\documentclass[pdflatex,sn-mathphys-num]{sn-jnl}

\usepackage{graphicx}%
\usepackage{multirow}%
\usepackage{amsmath,amssymb,amsfonts}%
\usepackage{amsthm}%
\usepackage{mathrsfs}%
\usepackage[title]{appendix}%
\usepackage{xcolor}%
\usepackage{textcomp}%
\usepackage{manyfoot}%
\usepackage{booktabs}%
\usepackage{algorithm}%
\usepackage{algorithmicx}%
\usepackage{algpseudocode}%
\usepackage{listings}%
\usepackage{float}
\usepackage{subcaption}
\usepackage{tikz}
\usetikzlibrary{arrows.meta, angles, quotes}
\begin{document}

\title[Article Title]{Energy Barriers for Reversible Chain Scission and Healing under Tension with Displacement Control}


\author[1]{\fnm{Mohammad A.} \sur{Ansari}}

\author[1]{\fnm{Kenneth M.} \sur{Liechti}}

\author[2]{\fnm{Dmitrii E.} \sur{Makarov}}

\author*[1]{\fnm{Rui} \sur{Huang}} \email{ruihuang@mail.utexas.edu}

\affil[1]{\orgdiv{Department of Aerospace Engineering and Engineering Mechanics}, \orgname{University of Texas}, \orgaddress{\city{Austin}, \state{TX}, \postcode{78712}, \country{USA}}}

\affil[2]{\orgdiv{Department of Chemistry and Oden Institute}, \orgname{University of Texas}, \orgaddress{\city{Austin}, \state{TX}, \postcode{78712}, \country{USA}}}

\abstract{
Polymer chain scission is a key mechanism for fracture of soft materials such as elastomers and hydrogels. It is well known from single-molecule force spectroscopy experiments that the critical condition for chain scission depends on the loading rate and other environmental effects (e.g., temperature and solvent). Common approaches to describing the kinetics of chain scission often assume force-controlled conditions, that is, when a polymer chain is stretched by a prescribed force. As a result of this assumption, chain scission is irreversible, excluding the possibility of healing. In many soft materials, however, self-healing has been observed after fracture, suggesting possibly reversible chain scission. Here, we show that reversible chain scission with healing is possible under displacement-controlled conditions, that is, when a polymer chain is stretched with a prescribed end-to-end distance. We present a breakable freely-jointed chain model, assuming that a polymer chain breaks when one of its links breaks while the other links remain nearly rigid. At a prescribed end-to-end distance, the free energy of the chain has two local minima and a local maximum (the transition state), giving rise to energy barriers for chain scission and healing. As the prescribed displacement increases, the energy barrier decreases for scission but increases for healing, depending on the chain length (number of links) and the potential energy of the link. With the energy barriers, we adopt a kinetic approach to predict the statistics and kinetics of a single polymer chain under tension, first by integrating the rate equation for the survival probability of the chain and then by kinetic Monte Carlo simulations. Notably, the present model predicts rate-dependent chain scission, with a lower bound for the rupture force that could be several orders of magnitude lower than the upper bound (which is close to the theoretical strength of the covalent bonds in the backbone).
}




\maketitle

\section{Introduction}\label{sec1}
Polymer chains are ubiquitous in natural and synthetic soft materials. Each polymer chain is a long molecule of many repeating units linked by covalent bonds. Many polymer chains aggregate in a condensed state, with relatively weak non-covalent interactions between the chains. The chains can be crosslinked by covalent bonds to form a three-dimensional polymer network.
Above a glass transition temperature, the polymer chains are flexible and mobile, and they readily slip relative to each other. Such a crosslinked polymer network is called an elastomer. Subject to forces, the elastomer can undergo large deformation. So long as the covalent bonds along the chains and the crosslinks do not break, the polymer network retains its connectivity, and the elastomer recovers its original shape when the applied forces are removed, that is, the elastomer undergoes reversible elastic deformation \cite{treloar1975physics, rubinstein2003polymer}. However, the weak interactions between the chains may induce irreversible viscoelastic effects \cite{ferry1980}. Both the elastic and viscoelastic properties of polymer networks can be traced to the chemistry and thermodynamics of polymer chains and junctions at the molecular level \cite{danielson2021molecular}.

Fracture of a polymer network necessarily requires chain scission by breaking some of the covalent bonds \citep{CS_2025}. The classic Lake-Thomas model \cite{Lake_Thomas_1967} assumes a perfect polymer network and localized chain scission along the crack plane. More realistically, in an imperfect polymer network, fracture often involves more distributed chain scission in a damage zone around the crack \cite{YYS_2019, Lin2020defects, Slootman_2020, Slootman_2022, Barney2022, Li2023_CrackTipKinematics}. 
It has been suggested that the kinetics of thermally activated chain scission could be the primary mechanism for temperature and rate-dependent fracture of elastomers \cite{Slootman_2020, Son2021_vitrimers, Ghareeb2021, Wang2024, Siavoshani2024, Wang2025_FreshConsiderations}.
Recently, dynamic covalent bonds have been exploited in transient networks such as vitrimers, where both chain scission and healing are involved in the deformation and fracture processes as well as self-healing via dynamic bond dissociation and re-association \cite{Stukalin2013selfhealing, yu2018dynamicbonds, Guo2020dynamicbonds, shen2020transientnetworks, lamont2021rate, Hui2021_dynamic, Zheng2021_dynamic}.  
More recently, it was demonstrated that force induced chain scission with weak bonds can be leveraged to achieve rapid self-strengthening of polymer networks by bond re-association to form new networks \cite{wang2025_self_strengthening}.
These developments call for a better understanding of chain scission and healing in polymer networks.

The classic freely jointed chain (FJC) model assumes purely entropic elasticity with rigid and unbreakable links \cite{Kuhn1942}, which does not predict chain scission. Many other polymer chain models have been developed to take into account the stiffness of bond stretching and/or bending such as the modified freely jointed chain (m-FJC) model \cite{Smith1996overstretching, Zhang2003, Buche2022} and the worm-like chain (WLC) model \cite{Marko1995DNA, Rosa2003, Kierfeld2004, Manca2012}, but they have largely focused on the elastic force-extension relations instead of chain scission. 

Mao et al. \cite{mao2017rupture} proposed a model of freely jointed chain with deformable links, including the internal energy due to bond stretching as part of the free energy. They postulated that chain scission occurs when the internal energy of each link reaches a critical value, similar to the assumption in the Lake-Thomas model \cite{Lake_Thomas_1967}. Such a chain model was incorporated in a phase field approach for modeling progressive damage and rupture of polymer networks \cite{Talamini2018}. A similar chain model was adopted by Lamont et al. \cite{lamont2021rate} in formulating a statistical transient network model for rate-dependent damage and self-healing of polymer networks with reversible bonds.
Lavoie et al. \cite{Lavoie2020} proposed a modified WLC model by taking into account the change in internal energy due to bond
stretching and bond rotation, and they adopted a kinetic model for chain scission \cite{Dudko2006rates, Schwaderer2008siloxane}.
Buche and Silberstein \cite{Buche2021chainbreaking} proposed a potential-supplemented freely jointed chain (uFJC) model that incorporates force-sensitive reversible bond breaking kinetics. The single chain model was incorporated in a general statistical mechanics framework, leading to a rate-dependent constitutive model for the macroscopic stress–strain behavior of elastomers. 
An extended and somewhat simplified version of the uFJC model was formulated by Mulderrig et al. \cite{mulderrig2023statistical}.

By assuming a typical breaking force of $f_{break} \approx 4.5$ nN \cite{Beyer2000} along with the m-FJC model for a typical polymer chain, Wang et al. \cite{wang2019quantitative} found that the stored energy per bond at chain scission could be well below the bond dissociation energy in the chain. Similar assumption of a fixed breaking force or strength of a polymer chain was adopted by others \cite{Lin2020defects, Ghareeb2020, Deng2023nonlocal, Hartquist2025scaling}.
However, it is well known from single-molecule force spectroscopy experiments \cite{Ghatak2000, Beyer2005, Makarov2016, Bowser2021} that the actual value of $f_{break}$ for a polymer chain is probabilistic and depends on the loading rate and other environmental effects (e.g., temperature and solvent) \cite{Ribas-Arino2012, Bao2020}.

Descriptions of kinetic chain scission go back to the works of Eyring \cite{Eyring}, Kramers \cite{Kramers1940}, Zhurkov \cite{zhurkov1965kinetic}, and Bell \cite{bell1978models}, based on theories of thermally activated barrier crossing such as transition state theory and Kramers theory  \cite{Hanggi1990, Peters2017, Elber2020}. Typically, it is  assumed that the rate coefficient of bond dissociation in a polymer chain follows the Arrhenius law for thermally activated processes:
\begin{equation}
\label{eq: Arrhenius}
     k(f)=\nu \exp{ \left[ -\frac{E(f)}{k_B T} \right] },
\end{equation}
where $E(f)$ is an energy barrier or activation energy, $k_BT$ is the temperature in the unit of energy, and $\nu$ is a frequency pre-factor. The energy barrier is often taken as a function of the force $f$ acting on the bond or the chain.
If both the energy barrier $E$ and the pre-factor $\nu$ are taken to be constant, the rate coefficient becomes a constant at a given temperature (assuming isothermal), independent of the force $f$ or deformation of the chain \cite{Guo2020dynamicbonds, shen2020transientnetworks, lamont2021rate, Hui2021_dynamic}.
More commonly, the energy barrier is assumed to decrease linearly with the force \cite{yu2018dynamicbonds, kothari2018, Ghareeb2021, Son2021_vitrimers}: $E(f)=E_0-fx$, where $E_0$ is the energy barrier in the absence of the force and $x$ is a length scale called the activation length. 
However, it has been shown that the linear dependence of the energy barrier on the applied force would not hold at large forces \cite{crist1984polymer, titinLi2003, Dudko2006rates, Konda2011, Makarov2016, Renn2018}.
Even at small forces, a recent model predicted a nonlinear dependence \cite{yang2020multiscale}.
Strictly, for a polymer chain subject to isothermal stretching, the energy barrier for chain scission is the free energy barrier, which is the change of free energy from the intact state of the chain to a transition state. Both the intact state and the transition state change with the force, and thus in general the free energy barrier is not a constant or a linear function of the force.

\begin{figure}[h]
    \centering
    \includegraphics[width=12cm]{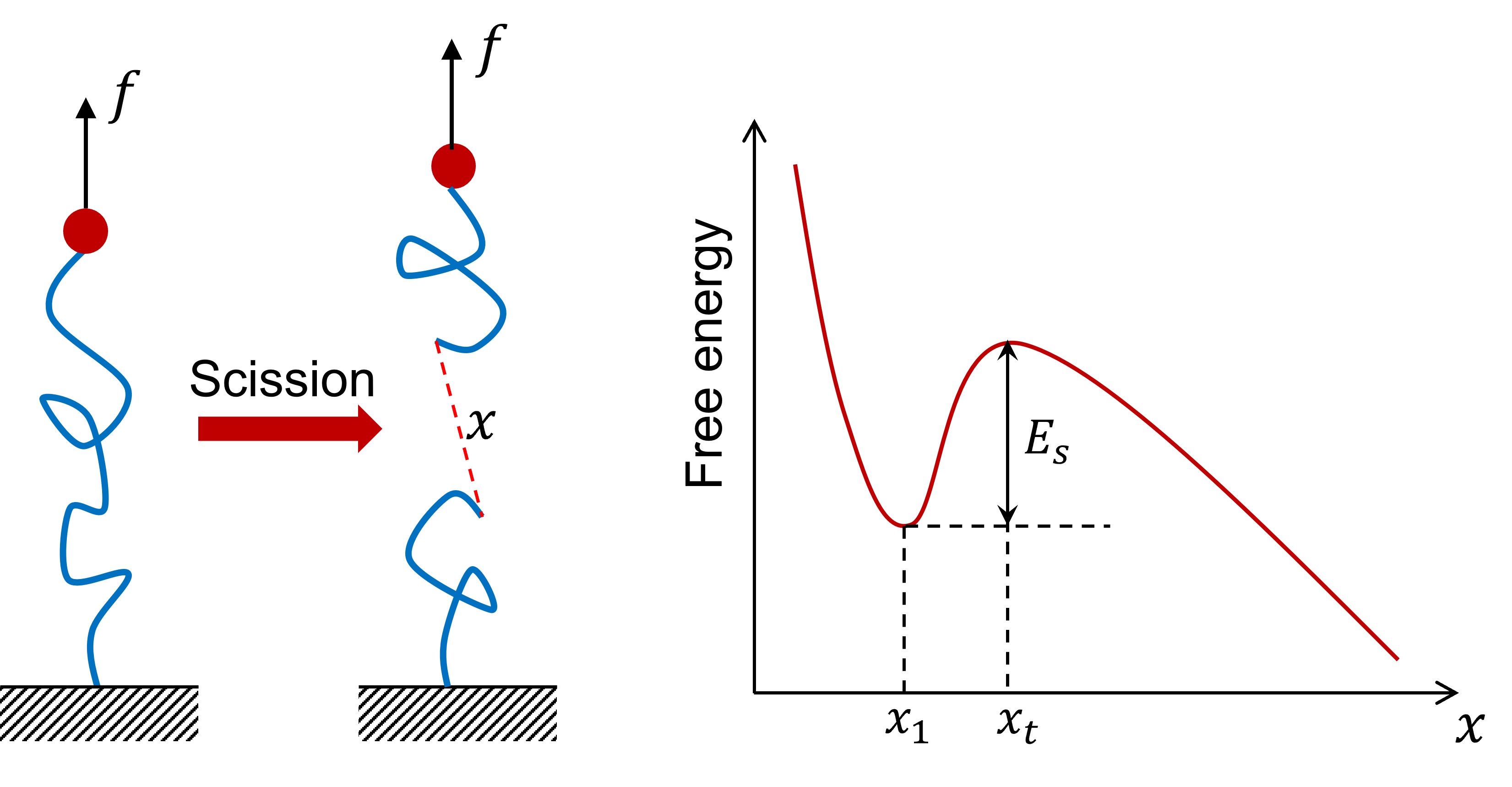}
    \caption*{(a)}
    
    \vspace{5pt} 
    
    \includegraphics[width=12cm]{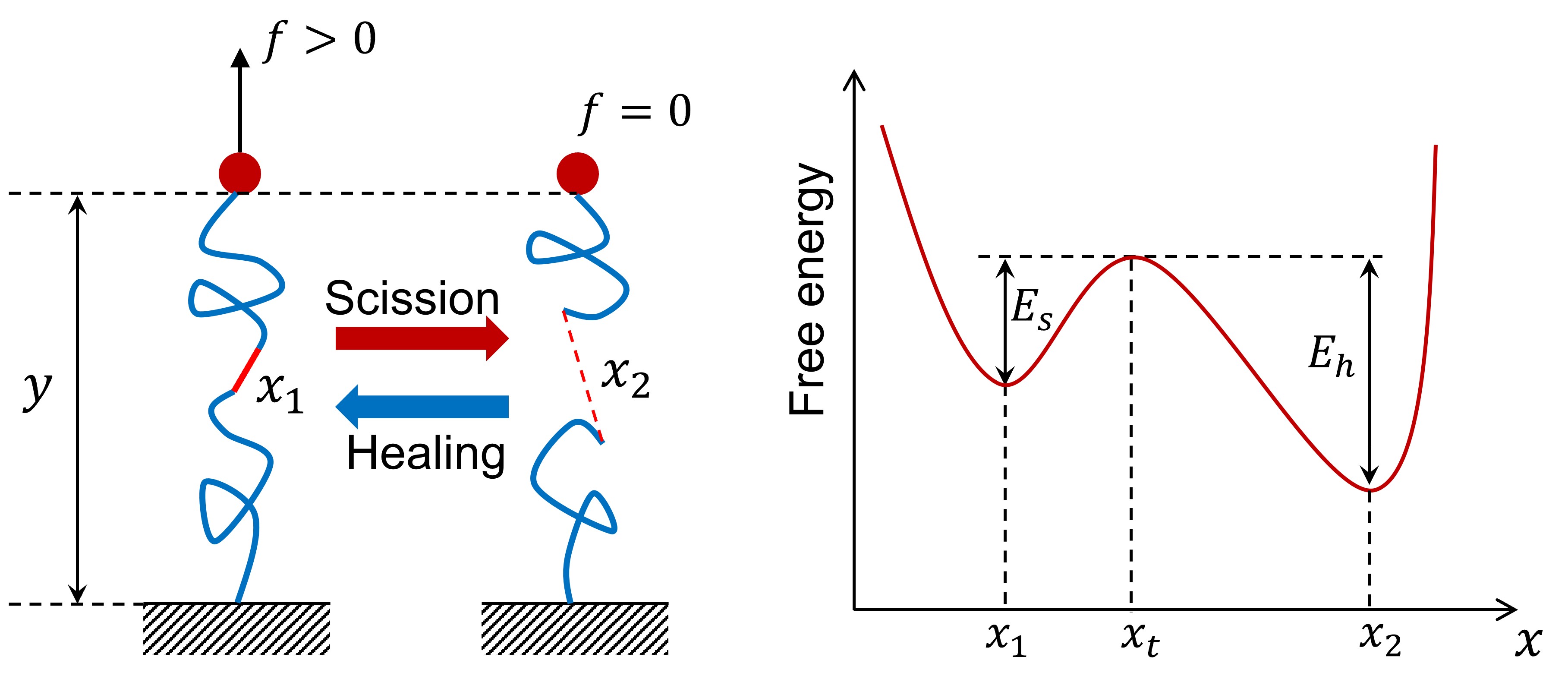}
    \caption*{(b)}
    
    \caption{(a) Force-controlled chain scission, where the free energy function for a prescribed force $f$ has one local minimum and the energy barrier ($E_s$) depends on the force. (b) Displacement-controlled chain scission and healing, where the free energy function for a prescribed end-to-end distance $y$ has two local minima so that the energy barriers for both chain scission ($E_s$) and healing ($E_h$) can be defined.}
    \label{fig:example}
\end{figure}

Healing of a broken chain by bond re-association has received less attention. Bell \cite{bell1978models} proposed a theoretical framework for cell adhesion considering both bond formation and rupture. 
Evidence of reversible chain scission and healing in polymer networks has been reported \cite{Yamaguchi2007, Kersey2007}. Motivated by experimental evidence, reversible chain scission with healing has been considered in modeling deformation and fracture polymer networks \cite{Stukalin2013selfhealing, yu2018dynamicbonds, lamont2021rate, Ghareeb2021, kothari2018, Guo2020dynamicbonds, shen2020transientnetworks, Hui2021_dynamic}. 
As in the case of bond dissociation, the rate of bond formation or re-association is assumed to follow the Arrhenius law in form of Eq. \eqref{eq: Arrhenius}, where the energy barrier is often taken to be a constant \cite{Guo2020dynamicbonds, shen2020transientnetworks, lamont2021rate, Hui2021_dynamic} or a linear function of the force \cite{yu2018dynamicbonds, kothari2018, Ghareeb2021}. In the latter case, the energy barrier for bond formation/re-association increases with the force, assuming a constant activation length, either equal to or different from the activation length for bond dissociation.
However, it was noted recently that, subject to a constant force, healing after chain scission is not possible because the free energy of the polymer chain has only one local minimum (Fig. \ref{fig:example}a) \cite{yang2020multiscale}. Consequently, chain scission is energetically favored at any force $f>0$. In contrast, when the end-to-end distance of the chain is prescribed, it was speculated that the free energy of the chain would have two local minima (Fig. \ref{fig:example}b) so that healing would be possible and chain scission is energetically favored only when the prescribed displacement exceeds a critical value.
Here we present such a model for a polymer chain being stretched under the displacement control, which predicts the energy barriers for both chain scission and healing as generally nonlinear functions of the end-to-end distance. 

As illustrated in Figure \ref{fig:example}(b), when the ends of a polymer chain are kept at a distance $y$, the state of the chain may be either intact or broken. 
Treat the chain as a thermal system, in equilibrium with a thermal reservoir of a fixed temperature $T$. The end-to-end distance $y$ is a control parameter. The chain consists of $n$ links (or Kuhn segments). Assume all links to be identical. Each link may be stretched and dissociated. Typically, chain scission occurs with only one link dissociated. To find the lowest energy barrier for chain scission, we take one arbitrary link to be stretchable and breakable, and treat all the other links as rigid. The multiplicity of chain scission pathways, each involving dissociation of a single link, is accounted for in estimating the rate coefficient of chain scission, Eq.~\eqref{eq: Arrhenius}, using the same energy barrier.
Take the length of the breakable link, $x$, as the reaction coordinate. At a given control parameter $y$, $x$ is an internal variable of the thermal system, and the free energy of the chain is a function of $x$, $A(x; y)$. The equilibrium configuration of the chain corresponds to the value of $x$ that minimizes the free energy. As the control parameter $y$ changes over time, $y=y(t)$, the free energy function changes, and the equilibrium configuration changes accordingly.     

Under the displacement control, it is found that the free energy function, $A(x; y)$ at a fixed value of $y$, has two local minima when $y$ is not too large (Fig. \ref{fig:example}b). In this case, two equilibrium configurations of the chain exist, corresponding to two different values of $x$. In the equilibrium configuration with the smaller value of $x$, say $x_1$, the chain remains intact. In the equilibrium configuration with the larger value of $x$, say $x_2$, the link is stretched significantly ($x_2 \gg x_1$) such that it is essentially dissociated. Transition between the two equilibrium configurations requires overcoming an energy barrier, because there exists a local maximum of the free energy function at $x=x_t$ between the two minima, $x_1 < x_t < x_2$, corresponding to the transition state of the chain scission/healing process. Define the energy barrier for chain scission by the difference between the free energy in the intact equilibrium configuration ($x=x_1$) and that of the transition state, $E_s (y) = A(x_t; y) - A(x_1; y)$. Similarly, define the energy barrier for chain healing by the difference between the free energy in the dissociated equilibrium configuration ($x=x_2$) and that of the transition state, $E_h (y) = A(x_t; y) - A(x_2; y)$. 
When $E_s(y)<E_h(y)$, chain scission is energetically favored, reducing the free energy by $\Delta E= E_h(y)-E_s(y)$. In contrast, when $E_s(y) > E_h(y)$, healing is energetically favored, also reducing the free energy.
Healing of the chain requires that the two dangling ends of the chain fragments meet and re-associate, which reduces the entropy of the chain thus entailing an energy barrier. This process is similar to diffusion controlled loop formation within a polymer, which can also be considered as crossing of an energy barrier \cite{SSS1980, Guo2009Association}.    

We emphasize that the existence of the second minimum of the free energy function corresponding to the state of a broken chain is specific to the displacement controlled stretching \cite{crist1984polymer, Manca2012}. If, instead, the two ends of the  chain are pulled by a prescribed force $f$ (acting in the $y$ direction), the free energy of the thermal system becomes $A(x,y)-fy$, including the potential energy of the force. In this case, the force $f$ is the control parameter, while both $x$ and $y$ are internal variables. At each prescribed force $f$, the free energy is a function of $x$ and $y$, and the free energy landscape becomes two dimensional. More generally, if all the $n$ links in the chain are stretchable, the free energy landscape becomes $n$-dimensional \cite{yang2020multiscale, Manca2012}. 
When the force is not too large, the free energy function would have one local minimum
where all links are equally stretched ($x=x_1$) and the chain is intact. Around the local minimum there are $n$ saddle points in the $n$-dimensional landscape, corresponding to $n$ transition states. In each transition state, one of the links is stretched more ($x_t > x_1$) while the other links remain equally stretched as in the intact state ($x=x_1$). Thus, by following a path with all the links stretched equally ($x=x_1$) except for one link with a varying stretch, we can find the transition state and the energy barrier for chain scission \cite{yang2020multiscale}. However, following the same path beyond the transition state, the free energy continues decreasing as $x$ increases (Fig. \ref{fig:example}a). The broken chain is not in equilibrium with the prescribed force $f>0$, thus not a minimum in the free energy landscape.
Consequently, under the force-controlled condition, healing by bond re-association is not possible, and chain scission would become irreversible \cite{mulderrig2023statistical,yang2020multiscale}. 
In a previous work \cite{yang2020multiscale}, irreversible chain scission was incorporated in a cohesive zone model to simulate rate-dependent fracture of a polymer interface, which agreed with experiments very well for the first cycle, but overestimated the amount of crack growth in the subsequent cycles under cyclic loading. It was suggested that reversible chain scission and healing could help improving the model for crack growth under cyclic loading.
 
In what follows we focus on a single polymer chain subject to displacement-controlled stretching. 
We first describe a breakable freely jointed chain model in Section 2, where the free energy function is derived by a statistical mechanics approach, including contributions from the potential energy of one link and the entropy of the chain fragments made of rigid links. By assuming the chain fragments to be Gaussian as an approximation in Section 3, the free energy function $A(x;y)$ is obtained in a closed form for the breakable chain model. Section 4 presents the breakable chain model with non-Gaussian fragments, which predicts an upper bound for the end-to-end distance and the corresponding force when the energy barrier for chain scission vanishes.
The energy barriers for both chain scission and healing are obtained from the breakable chain model as functions of the end-to-end displacement, $E_s(y)$ and $E_h(y)$, depending on the equilibrium bond energy and the number of links in the chain, as discussed in Section 5.
With the energy barriers, we adopt the kinetic approach to predict the statistics and kinetics of chain scission and healing for a single polymer chain, first by integrating the rate equation in terms of the survival probability of the chain (Section 6) and then by kinetic Monte Carlo simulations (Section 7). In conclusion, we summarize the findings and discuss the implications of the present results on rate dependent fracture of polymer networks.
For completeness, we present in Appendix a model for a freely jointed chain with rigid links under displacement control, which recovers the classical Gaussian and non-Gaussian models in limiting cases.

\section{A breakable freely jointed chain model}
\begin{figure}[htp]
    \centering
        \includegraphics[width=0.6\textwidth]{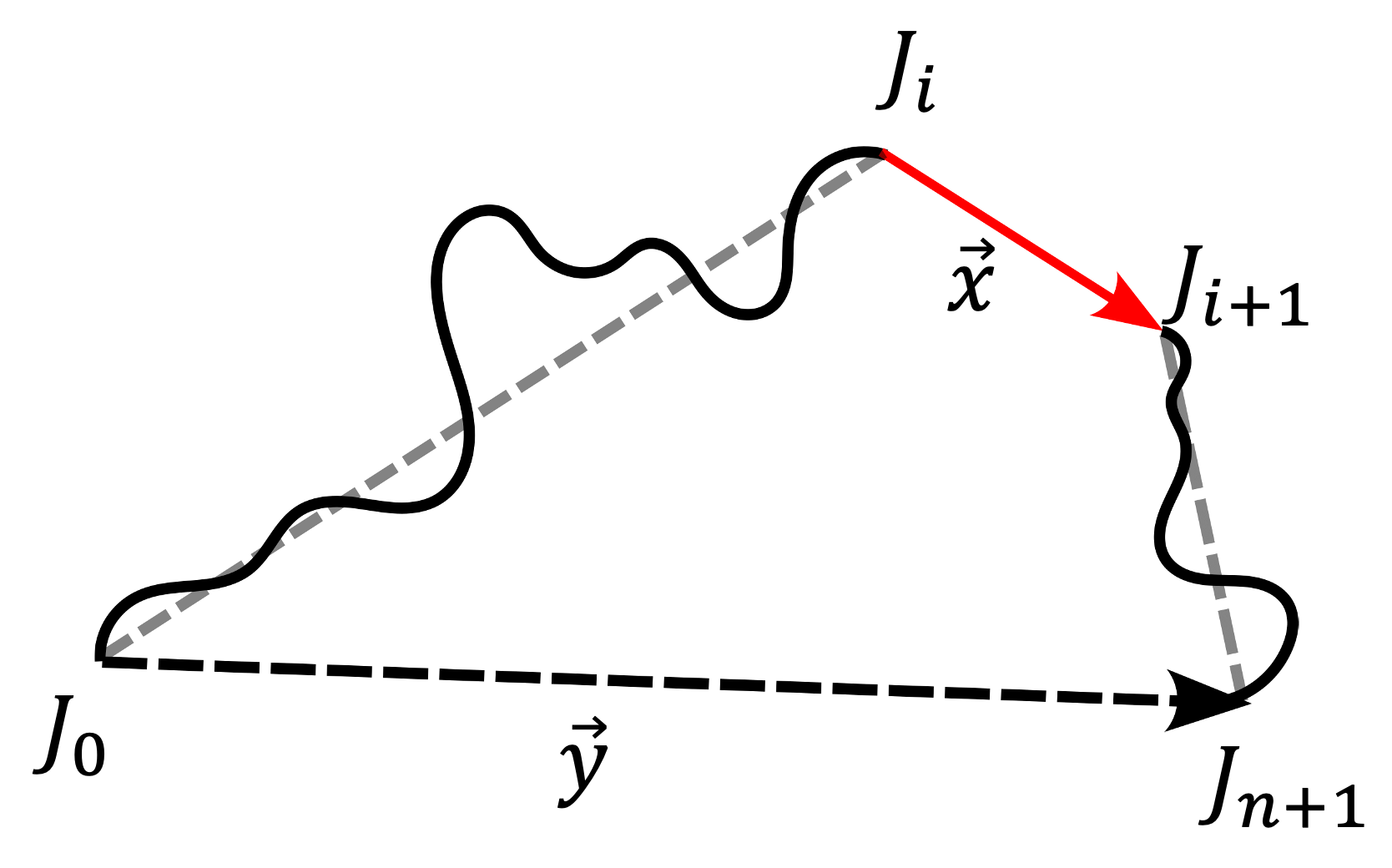}
    \caption{Schematic of a polymer chain with one arbitrary link (red) that is stretchable and breakable.}
    \label{fig:chain combined}
\end{figure}

Consider a polymer chain with $n+1$ identical links (Kuhn segments). The links are assumed to be freely jointed, with the joints marked continuously as $J_0, J_1, J_2, ..., J_{n+1}$, as shown in Figure \ref{fig:chain combined}. Fix one end of the chain, $J_0$, and pull the other end of the chain, $J_{n+1}$. Denote the end-to-end vector of the chain as $\vec{y}$. Among the $n+1$ links, only one link has to be broken for chain scission to occur, and it is statistically less likely to have more than one link simultaneously broken \cite{crist1984polymer, yang2020multiscale}.
Consider an arbitrary link that is to be broken, $J_iJ_{i+1}$ ($0 \leq i \leq n$). 
Let $\vec{x}$ be the vector for the link $J_iJ_{i+1}$, with a variable length $\left|\vec{x}\right| = x$, whereas all the other links remain intact and are treated as rigid links of a constant length $l$ (that is, the equilibrium bond length). 
Dissociation of the link $J_iJ_{i+1}$ would break the chain into two fragments, $J_0J_i$ and $J_{i+1}J_{n+1}$. 
The chain fragment $J_0J_i$ has $i$ links, and the chain fragment $J_{i+1}J_{n+1}$ has $n-i$ links. Let the end-to-end vectors of the two fragments be $\vec{r}_i$ and $\vec{r}_{n-i}$, respectively. Both the chain fragments follow the same statistics for a freely jointed chain of rigid links.
The joint probability for the chain to take the configuration with given $\vec{x}$ and $\vec{y}$ is  
\begin{equation}
    p(\Vec{x},\vec{y}) = p_1(\Vec{x}) \int  p_i(\vec{r}_i) p_{n-i}(\vec{r}_{n-i})
    \delta(\vec{r}_i +\vec{r}_{n-i} +\vec{x} -\vec{y})
    d\vec{r}_i d\vec{r}_{n-i} ,
\end{equation}
where $\delta(\vec{r})$ is the 3D Dirac delta function, $p_1(\vec{x})$ is the probability for the link $J_iJ_{i+1}$ to take the configuration with the vector $\vec{x}$, and 
$p_i(\vec{r}_i)$ is the probability for the chain fragment of $i$ rigid links to take the configuration with the end-to-end vector $\vec{r}_i$.

The above integral can be simplified first by integrating with respect to $\vec{r}_{n-i}$ with the property of the Dirac delta function, which yields
\begin{equation}
     p(\Vec{x},\vec{y})=p_1(\vec{x})\int 
     p_i(\vec{r}_i) p_{n-i}(\vec{y}-\vec{x}-\vec{r}_i)
     d\vec{r}_i .
\end{equation}

Integrating again with respect to $\vec{r}_i$, we obtain
\begin{equation}
    p(\Vec{x},\vec{y}) = p_1(\vec{x})
 p_{n}(\vec{y}-\vec{x}) .
 \label{eq:pxy}
\end{equation}
Here, $p_n(\vec{y}-\vec{x})$ is the probability for a chain of $n$ rigid links to take the configuration with the end-to-end vector $\vec{y} -\vec{x}$. Since both $p_1(\vec{x})$ and $p_n(\vec{y}-\vec{x})$ are independent of the specific location of the link $J_iJ_{i+1}$ along the chain, the joint probability $p(\Vec{x},\vec{y})$ is independent of the choice of the breakable link. 

Let us now assume a potential energy function for the breakable link, $V(x)$. The potential energy is independent of the orientation of the link for the freely jointed chain, and it is also independent of the specific choice of the link along the chain. In equilibrium with the thermal reservoir of temperature $T$, the probability for this link to take the configuration with the end-to-end vector $\vec{x}$ is (up to a constant factor):
\begin{equation}
    p_1 (\vec{x} ) \propto \exp (-\beta V(x) ) ,
    \label{eq:p1}
\end{equation}
where $x = \left|\vec{x}\right|$ and $\beta = 1/(k_BT)$.

The rest of the chain follows the statistics of a freely jointed chain with $n$ rigid links. Let the free energy of such a freely jointed chain be $g_n(\vec{r})$, where $\vec{r}$ is the end-to-end vector of the chain. Then, the probability for this chain to take the configuration of $\vec{r}$ is (up to a constant factor):
\begin{equation}
    p_n (\vec{r} ) \propto \exp (-\beta g_n(\vec{r})) .
    \label{eq:pn}
\end{equation}

Combining Eqs. \eqref{eq:pxy}-\eqref{eq:pn}, we obtain the joint probability for the breakable chain:
\begin{equation}
    p(\vec{x},\vec{y}) \propto \exp(-\beta V(x)) \exp(-\beta g_n(\vec{y}-\vec{x})) .
\end{equation}

To find the free energy as a function of the bond length $x$, we integrate the probability $p(\vec{x}, \vec{y})$ over a spherical surface defined by $\left| \vec{x} \right| = x$ and obtain
\begin{equation}
    p(x,\vec{y}) \propto \int_0^{\pi} 
    \exp(-\beta V(x)) \exp(-\beta g_n(\vec{y}-\vec{x}) ) 
    x^2 \sin{\phi} d\phi,
\end{equation}
where $\phi$ is the angle between vectors $\vec{x}$ and $\vec{y}$.

Given $\vec{y}$, $x$ is an internal variable, and the free energy of the chain is related to the probability (up to a constant) as:
\begin{equation}
    A(x,\Vec{y}) = -k_B T \log \left( p 
    (x,\vec{y}) \right),
    \label{eq:Axy}
\end{equation}
or, equivalently
\begin{equation}
     \exp (-\beta A(x,\vec{y})) \propto
     \int_0^{\pi} 
    \exp(-\beta V(x)) \exp(-\beta g_n(\vec{y}-\vec{x}) ) 
    x^2 \sin{\phi} d\phi.
    \label{eq:free_energy_integral}
\end{equation}
Thus, the free energy of the chain $A(x, \vec{y})$ is also independent of the specific location of the breakable link $J_iJ_{i+1}$ along the chain as long as all the links are identical.

As the end-to-end distance of the chain $y=\left| \vec{y} \right|$ increases, the entropy of the chain decreases by uncoiling of the freely jointed links, which increases the free energy. Meanwhile, the link $J_iJ_{i+1}$ may be stretched and dissociated, increasing the potential energy of the link and the free energy. Consequently, the free energy of the chain combines the contributions from the internal (potential) energy of the link (due to atomic interactions) and the conformational entropy (due to statistical fluctuations). 
The competition between the two contributions determines the stable configuration of the chain that minimizes the free energy. 

To be specific, we take the Lennard-Jones (LJ) potential for the breakable link so that
\begin{equation}
    V(x) = \epsilon_0 \left[ \left( \frac{l}{x} \right)^{12} - 2\left( \frac{l}{x} \right)^6 \right],
    \label{eq:LJ potential}
\end{equation}
where $\epsilon_0$ is the depth of the energy well (i.e., the equilibrium bond energy) at the equilibrium bond length, $x = l$. When $x \gg l$, the potential energy approaches zero, and the link is considered dissociated.

The specific form of the function $g_n(\vec{r})$ depends on the statistical model for the chain fragments with rigid links. 
Two approximate models are used in Sections 3 and 4. First, by assuming Gaussian statistics for the chain fragments, we obtain a closed-form free energy function for the breakable chain model. Then, the Kuhn-Grun approximation is adopted to account for the non-Gaussian statistics of a long freely jointed chain. The exact statistics for a freely jointed chain of rigid links is presented in Appendix, leading to an integral that is computationally challenging to calculate.

It is noted that, in many polymer systems, dynamic sacrificial bonds (including covalent and non-covalent) have been used for reversible chain scission and healing \cite{Kersey2007, Guo2009Association, yu2018dynamicbonds, Lavoie2020, wang2025_self_strengthening}. The present model can be readily adapted for such cases, by replacing the LJ potential with a specific potential energy function for the sacrificial bond.

\section{A breakable chain with Gaussian fragments}
Referring to Figure \ref{fig:chain combined}, we assume in this section the chain fragments are Gaussian except for the breakable link. In this case, the free energy function $g_n(\vec{y}-\vec{x})$ in Eq. \eqref{eq:free_energy_integral} takes the form (see \eqref{eq:gn Gaussian} in Appendix for details):
\begin{equation}
g_n(\vec{y} - \vec{x}) =\frac{3 k_B T}{2 nl^2} 
\left( y^2 + x^2 -2xy \cos{\phi} \right), 
\label{eq:gn_Gaussian2}
\end{equation}
where $\phi$ is the angle between two vectors $\vec{x}$ and $\vec{y}$, $x=\left| \vec{x} \right|$ and $y=\left| \vec{y} \right|$.

For a given $\vec{y}$, with Eqs. \eqref{eq:free_energy_integral} and \eqref{eq:gn_Gaussian2}, 
we obtain the free energy (up to a constant) of the breakable chain:
\begin{equation}
\label{eq:ideal_energy}
    \beta A(x,\vec{y})= \beta V(x) +\frac{3}{2n l^2} (x^2+y^2) -\log\left(\frac{x}{y} \sinh{\frac{3xy}{nl^2}} \right) .
\end{equation}

\begin{figure}[htp]
    \begin{center}
        \includegraphics[width=10cm]{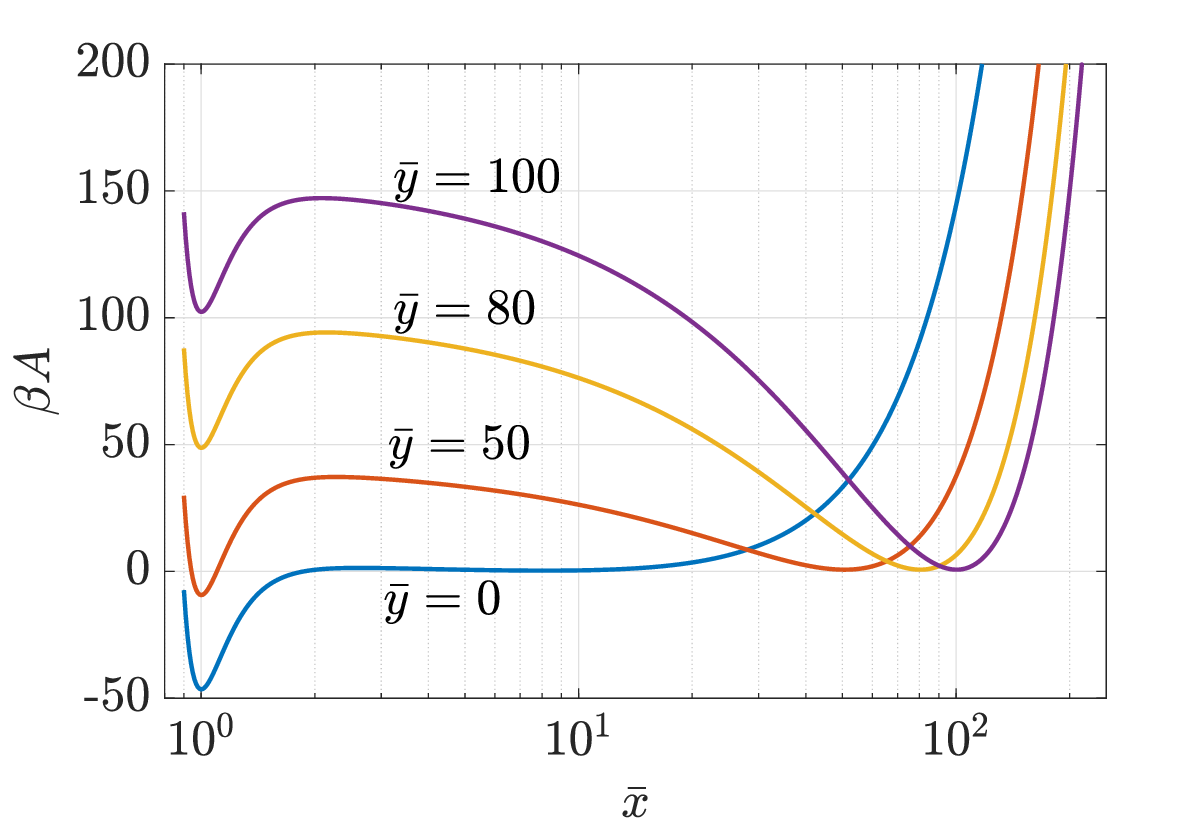}
    \caption{Normalized free energy as a function of the normalized bond length,  for a chain with $n=100$ and $\beta\epsilon_0=50$, when the chain fragments are approximated by Gaussian statistics.}
    \label{free_ideal}
    \end{center}
\end{figure}

The normalized free energy by Eq. \eqref{eq:ideal_energy} is plotted as a function of $\bar{x} = x/l$ in Figure \ref{free_ideal}, for a chain with $n = 100$ and $\beta\epsilon_0 = 50$, at four values of $\bar{y} = y/l$. The free energy of the chain includes the potential energy of one link that is stretchable and breakable, $V(x)$. The Lennard-Jones potential in Eq. \eqref{eq:LJ potential} is used here with $\beta\epsilon_0 = 50$. In addition, the free energy includes an entropic contribution. While the potential energy $V(x)$ has a minimum at $\bar{x} = 1$, the entropic contribution has a minimum at a much larger $\bar{x}$ that depends on $\bar{y}$. The combination of the two leads to a free energy function with two local minima, one near the equilibrium bond length, $\bar{x}_1 \approx 1$, and the other far from the equilibrium, $\bar{x}_2 \gg 1$. The first minimum corresponds to an equilibrium configuration of the intact chain, with a relatively low entropy. The second minimum corresponds to an configuration of a broken chain, with a relatively high entropy.
In between the two minima, there exists a local maximum of the free energy function for each value of $\bar{y}$ (so-called "transition state"). Thus, the energy barriers for both chain scission and healing can be determined as functions of $\bar{y}$.

We note that, under force-controlled condition, the free energy function is different, with one local minimum and one local maximum (Fig. \ref{fig:example}a) \cite{crist1984polymer, Dudko2006rates, mulderrig2023statistical, yang2020multiscale, Beyer2005}. Consequently, only the energy barrier for chain scission can be determined, and healing is not possible under force control. 

Interestingly, the location of the first local minimum ($\bar{x}_1$) is nearly independent of the applied stretch ($\bar{y}$). Thus, the length of the breakable link remains nearly constant, $x_1 \approx l$, and it is reasonable to assume that all links in the chain are rigid before chain scission. As $\bar{y}$ increases, the free energy of the intact chain at equilibrium ($x=x_1$) increases (due to decrease in entropy), and the energy barrier for chain scission decreases. Meanwhile, the free energy of the chain in the dissociated state ($x=x_2 \gg l$) remains nearly constant, $A(x_2, y) \approx 0$, and the energy barrier for healing increases with $\bar{y}$.

The corresponding force-displacement relation before chain scission can be obtained by setting $x=x_1$ in Eq. \eqref{eq:ideal_energy} and then taking derivative of the free energy with respect to $\vec{y}$. Note that, in principle, $x_1$ is a function of $y$. However, since $x_1 \approx l$, we obtain approximately

\begin{equation}
    \vec{f} (\vec{y}) \approx \frac{\partial A(l, \vec{y})}
    {\partial \vec{y}} ,
\end{equation}
which is a vector in the direction of $\vec{y}$, and the force magnitude is
\begin{equation}
    f(y) \approx \frac{k_BT}{l} \left( \frac{3y}{nl} 
    + \frac{l}{y} 
    -\frac{3}{n} \coth{\frac{3y}{nl}} \right).
    \label{eq:force Gaussian}
\end{equation}
The first term on the right hand side of Eq. \eqref{eq:force Gaussian} is identical to the linear force-displacement relation by Gaussian statistics for a chain of $n$ rigid links. When $y \ll nl$, Eq. \eqref{eq:force Gaussian} reduces to a linear relation. When $y$ is large, the Gaussian approximation breaks down.

\section{A breakable chain with non-Gaussian fragments}
While the Gaussian approximation yields a closed-form free energy function, Eq. \eqref{eq:ideal_energy}, it is expected to be inaccurate when the chain is highly stretched. In this section, we equip the breakable chain model with non-Gaussian statistics. 
As discussed in Appendix, for a freely jointed chain of rigid links, the free energy as a function of the end-to-end vector, $g_n(\vec{r})$, can be calculated by Eq. \eqref{eq:gn log}. However, the numerical calculation of the probability density by the integral in Eq. \eqref{eq: disp_prb_eq_lambda} is limited to relatively small stretches for a long chain. 
On the other hand, under a force controlled condition, the non-Gaussian statistics leads to a free energy function in terms of the applied force $f$ \cite{Kuhn1942, mao2017rupture}:
\begin{equation}
    w(f) = nk_B T \left( \frac{\beta fl}{\tanh{\beta fl}}
    +\log \frac{\beta fl}{\sinh{\beta fl}} \right) .
    \label{eq:free energy force}
\end{equation}
Under the force control, the end-to-end vector of the chain fluctuates, and the average end-to-end distance in the direction of the force is related to the force by
\begin{equation}
    r(f) = nl \left( \frac{1}{\tanh{\beta fl}}
    - \frac{1}{\beta fl} \right) .
    \label{eq:length force}
\end{equation}

Combining the two equations above, one obtains the free energy as a function of the average end-to-end distance, $w(r)$. It can be shown that, when $\beta fl \ll 1$, Eq. \eqref{eq:length force} recovers the linear force-displacement relation predicted by the Gaussian statistics, and Eq. \eqref{eq:free energy force} recovers the quadratic free energy function (up to a constant) by the Gaussian approximation under displacement control. Therefore, in the regime when $\beta fl \ll 1$ or $\lambda \ll n$, there is no difference between displacement control and force control, in terms of the free energy function and the force-displacement relation. However, when the applied force or stretch is large, the Gaussian approximation becomes invalid, and the difference between force and displacement control could be significant. Indeed, it is found that the difference between force and displacement control is significant only for short chains with small $n$, but the difference diminishes as $n$ increases \cite{Manca2012}. As shown in Figures \ref{fig:probability to force} (b) and (c), for $n = 100$, the free energy function and the force-displacement relation calculated under displacement control closely follow those obtained under force control, up to the point when the numerical results become inaccurate. Therefore, for long chains with large $n$, we may use Eqs. \eqref{eq:free energy force}-\eqref{eq:length force} to approximate the free energy function of the chain under displacement control, that is, $g_n(\vec{r}) \approx w(f)$ with $\left| \vec{r} \right| = r(f)$. 
This approximation is known as the Kuhn-Grun approximation \cite{Jernigan1969, Morovati2019}.
Here, we adopt this approximation for the rigid links in the breakable chain model.
Then, the free energy of the breakable chain can be obtained by Eq. \eqref{eq:free_energy_integral} as 
\begin{equation}
    \exp(-\beta A(x,\vec{y})) \propto \int_0^{\pi} \exp[-\beta V(x)] \exp[-\beta w(f) ] x^2 \sin{\phi} d\phi ,
    \label{eq:free enery exp}
\end{equation}
where $f$ is related to the length $r = \left| \vec{y}-\vec{x} \right|$ by Eq. \eqref{eq:length force}, namely
\begin{equation}
    \left| \vec{y}-\vec{x} \right| = nl \mathcal{L} (\bar{f}).
\end{equation}
Here, $\bar{f} = \beta fl$, and $\mathcal{L} (\bar{f}) = \coth{\bar{f}} - \bar{f}^{-1}$ is the Langevin function.

Re-write the above equation as
\begin{equation}
    \bar{f} = \mathcal{L}^{-1} \left( \frac{\left| \vec{y}-\vec{x} \right|}{nl} \right) = 
    \mathcal{L}^{-1} \left( \frac{1}{n} \sqrt{\bar{x}^2+\bar{y}^2-2\bar{x}\bar{y} \cos \phi} \right) ,
\end{equation}
where $\bar{x} = x/l$, $\bar{y}=y/l$, and $\phi$ is the angle between the two vectors, $\vec{x}$ and $\vec{y}$.
The inverse Langevin function can be calculated approximately by \cite{jedynak2015approximation}:
\begin{equation}
    \mathcal{L}^{-1}(\eta) \approx \frac{3\eta-\eta^3}{1-\eta^2} ,
\end{equation}
for $0 \leq \eta < 1$.

Re-write Eq. \eqref{eq:free energy force} as
\begin{equation}
    \beta w(f) = n h (\bar{f}) ,
\end{equation}
with
\begin{equation}
    h (\bar{f}) = \frac{\bar{f}}{\tanh{\bar{f}}}
    +\log \frac{\bar{f}}{\sinh{\bar{f}}} .
\end{equation}

Thus, by Eq. \eqref{eq:free enery exp}, we obtain
the free energy of the chain (up to a constant):
\begin{equation}
    \beta A(x,\vec{y}) = \beta V(x) - \log\left( x^2 
    I_n (x, \vec{y}) \right) ,
    \label{eq:free energy non Gaussian}
\end{equation}
where
\begin{equation}
    I_n (x, \vec{y}) = \int_{0}^{\pi} \exp \left[-nh \left( \mathcal{L}^{-1} \left( \frac{\left| \vec{y}-\vec{x} \right|}{nl} \right) \right)
    \right] 
    \sin \phi d\phi .
\end{equation}
This integral needs to be evaluated numerically.
For convenience, change the variable from $\phi$ to $c = \cos \phi$ for the integral:
\begin{equation}
    I_n (x, \vec{y}) = \int_{-1}^{1} \exp\left[-n h\left( \mathcal{L}^{-1}\left(\frac{1}{n} \sqrt{\bar{x}^2+\bar{y}^2-2\bar{x}\bar{y}c}\right)\right)\right] d c  .
    \label{eq:I integral}
\end{equation}
This integral is dimensionless, depending on $\bar{x}$, $\bar{y}$, and $n$.

Note that the length $\left| \vec{y} -\vec{x} \right|$ must be less than $nl$ due to the assumption of rigid links except for the breakable link in the chain. If $\left| \vec{y} -\vec{x} \right| \geq nl$, we have $\bar{f} \rightarrow \infty$ and $h(\bar{f}) \rightarrow \infty$ so that the integrand in \eqref{eq:I integral} becomes zero and can be excluded from the integral.  

\begin{figure}[htp]
    \begin{center}
        \includegraphics[width=10cm]{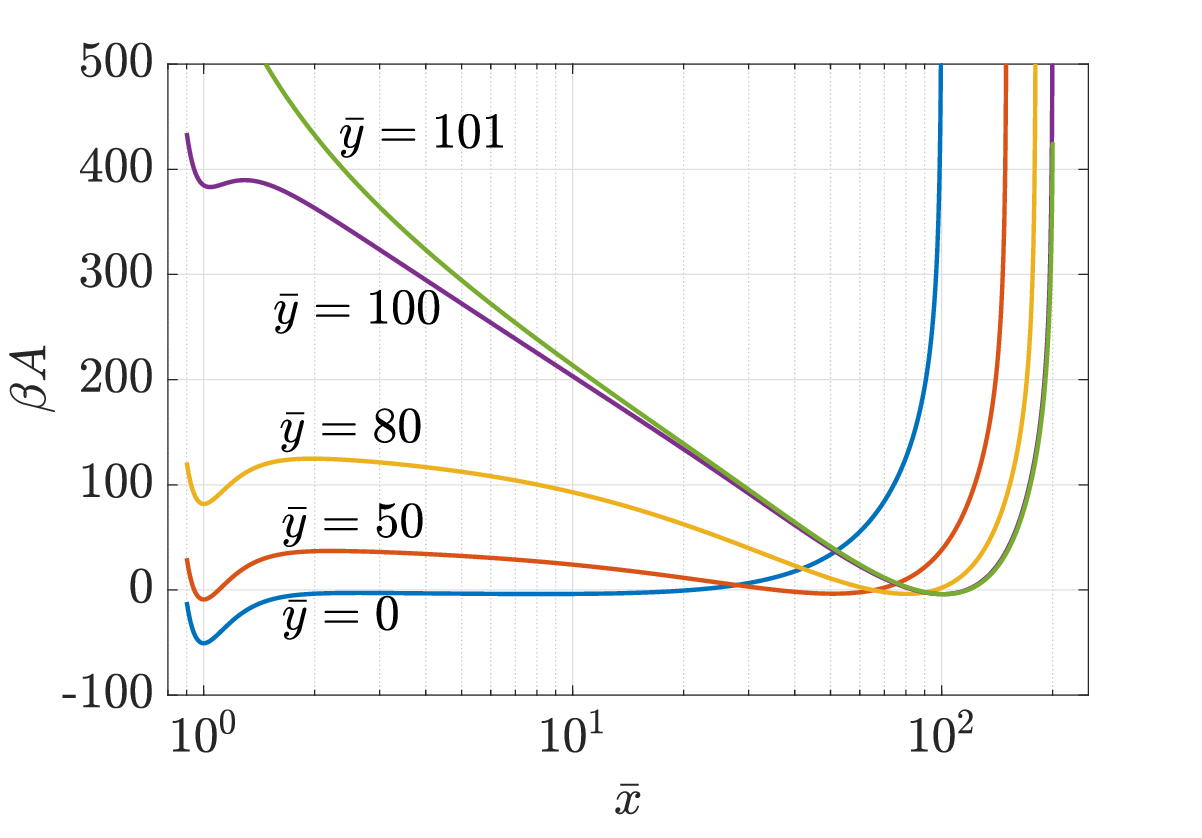}
    \caption{Normalized free energy as a function of the normalized bond length, for a chain of $n=100$ and  $\beta\epsilon_0=50$, when the chain fragments are non-Gaussian with rigid links.}
    \label{free_nonGaussian}
    \end{center}
\end{figure}

The normalized free energy in Eq. \eqref{eq:free energy non Gaussian} depends on four dimensionless parameters, $\bar{x}$, $\bar{y}$, $n$, and $\beta \epsilon_0$. For a particular chain at a constant temperature, $n$ and $\beta \epsilon_0$ are fixed.
At each prescribed end-to-end vector $\vec{y}$, $\bar{y}$ is fixed and the free energy $A(x,\vec{y})$ is calculated as a function of the normalized bond length $\bar{x}$, as shown in Figure \ref{free_nonGaussian} for $n=100$ and  $\beta\epsilon_0=50$. 
When $\bar{y}$ is relatively small, the free energy function is well predicted by the Gaussian approximation (Fig. \ref{free_ideal}). However, as $\bar{y}$ increases, the free energy function deviates from the Gaussian approximation, especially for $\bar{x} \approx 1$.
As $\bar{y} \rightarrow n$, the chain approaches its limit extension and the free energy increases dramatically at $\bar{x} \approx 1$. While the free energy function has two local minima when $\bar{y} \leq n$, the first minimum at $\bar{x} \approx 1$ disappears when $\bar{y} = n+1$. In this case, the energy barrier for chain scission becomes zero, and the chain breaks instantaneously.

\begin{figure}[htp]
     \begin{subfigure}[b]{0.48\textwidth}
         \centering        \includegraphics[width=1\textwidth]{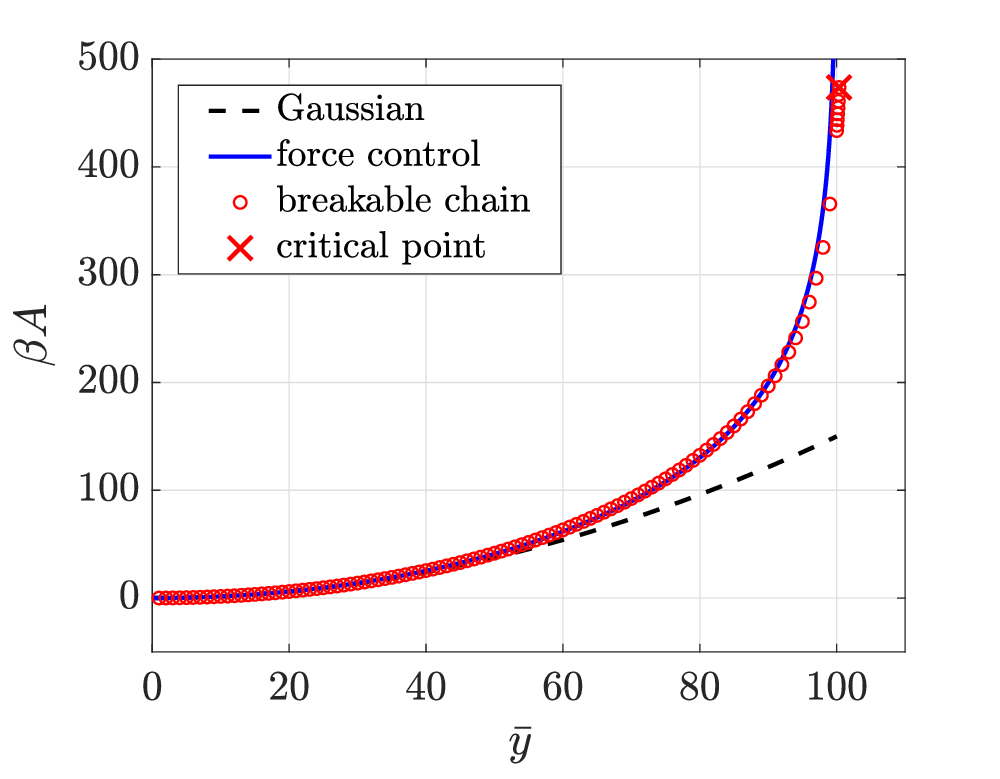}  \caption{\protect\label{}}
     \end{subfigure}
     \hfill
     \begin{subfigure}[b]{0.48\textwidth}
         \centering \includegraphics[width=1\textwidth]{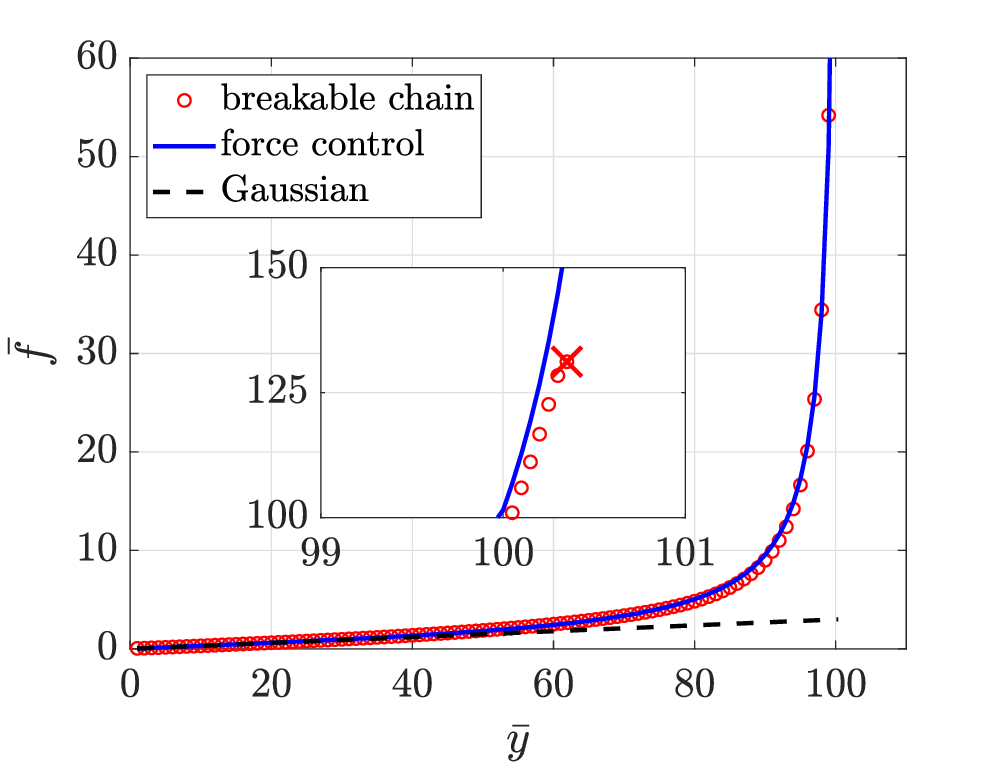} \caption{\protect\label{}}
     \end{subfigure}
    \caption{(a) Free energy at the first minimum ($\bar{x}=\bar{x}_1$) as a function of $\bar{y}$ for $n=100$ and $\beta\epsilon_0=50$, in comparison with the classical freely jointed chain models, and (b) comparison of the force-displacement curves obtained from the breakable chain model and the classical models.}
    \label{force_compare}
\end{figure}

Similar to the Gaussian approximation in Figure \ref{free_ideal}, the location of the first local minimum ($\bar{x}=\bar{x}_1$) of the free energy function in Figure \ref{free_nonGaussian} is nearly independent of the prescribed end-to-end distance ($\bar{y}$), and the length of the breakable link remains nearly constant, $\bar{x}_1 \approx 1$, before chain scission. As $\bar{y}$ increases, the free energy of the chain at $\bar{x}=\bar{x}_1$ increases, primarily due to decrease in entropy. As shown in Figure \ref{force_compare}(a), the free energy at $\bar{x}=\bar{x}_1$ is nearly identical to that by the classical freely jointed chain model with rigid links, whereas the Gaussian approximation is valid only when $\bar{y}$ is relatively small. However, as $\bar{y} \rightarrow n$, the present model for a breakable chain predicts a critical value, $\bar{y}=\bar{y}_c$, beyond which the first local minimum of the free energy function disappears so that the chain breaks instantaneously.
The corresponding force-displacement relation before chain scission can be obtained by setting $\bar{x}=\bar{x}_1$ in Eq. \eqref{eq:free energy non Gaussian} and then taking derivative of the free energy function with respect to $y$. 
Figure \ref{force_compare}(b) compares the force-displacement relation to Eqs. \eqref{eq:force Gaussian} and \eqref{eq:length force}. Since $\bar{x}_1 \approx 1$, the force-displacement relation by the breakable chain model is nearly identical to that by the classical freely jointed chain model with rigid links, Eq. \eqref{eq:length force}, whereas the Gaussian approximation in Eq. \eqref{eq:force Gaussian} is valid only in the linear regime when $\bar{y}$ is relatively small. However, as $\bar{y} \rightarrow n$, the breakable chain model predicts that the force-displacement curve terminates at the critical point when the first local minimum of the free energy function disappears. Thus, unlike the classical freely jointed chain model, the breakable chain model predicts that the force acting on the chain has an upper bound at the critical end-to-end distance, $\bar{f}_c = \bar{f}(\bar{y}_c)$. It is found that the critical force $\bar{f}_c$ is slightly lower than the maximum force (bond strength) dictated by the Lennard-Jones potential used for the breakable link, $\bar{f}_{max} \approx 2.69\beta\epsilon_0$. While the latter is a property of the link (Kuhn segment), the former is a property of the chain and weakly depends on the chain length $n$. 
In contrast, the critical end-to-end distance $\bar{y}_c \approx n$ is insensitive to the bond energy.
It should be noted that chain scission may occur at a force well below this upper bound ($\bar{f}_c$ or $\bar{f}_{max}$) when the energy barrier ($E_s$) becomes sufficiently small.


Moreover, as shown in Figure \ref{free_nonGaussian}, the location of the second minimum ($\bar{x}=\bar{x}_2$) changes with $\bar{y}$, but the free energy at $\bar{x}=\bar{x}_2$ remains constant, $A(\bar{x}_2, \bar{y}) \approx 0$. Thus, the corresponding force is zero, and the chain is in the dissociated state. It can be seen that the dissociated state becomes energetically favored only when $\bar{y}$ is greater than a threshold value.

\section{Energy barriers for chain scission and healing}
\begin{figure}[htp]
     \begin{subfigure}[b]{0.48\textwidth}
         \centering        \includegraphics[width=1\textwidth]{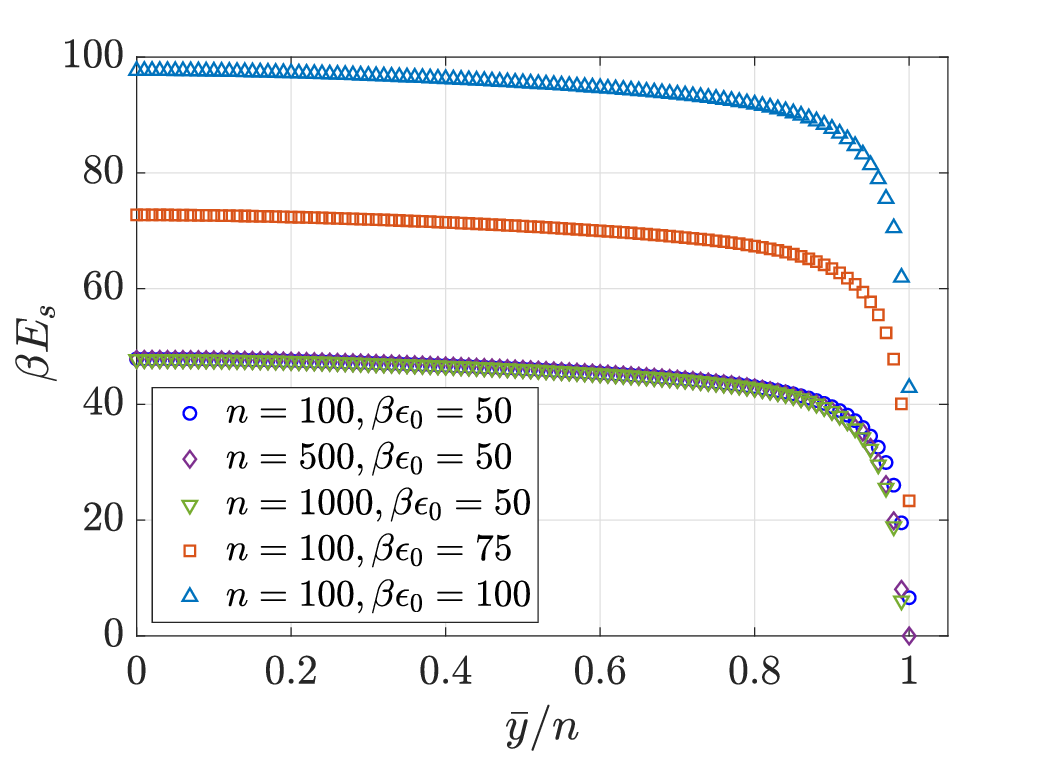}  \caption{\protect\label{}}
     \end{subfigure}
     \hfill
     \begin{subfigure}[b]{0.48\textwidth}
         \centering \includegraphics[width=1\textwidth]{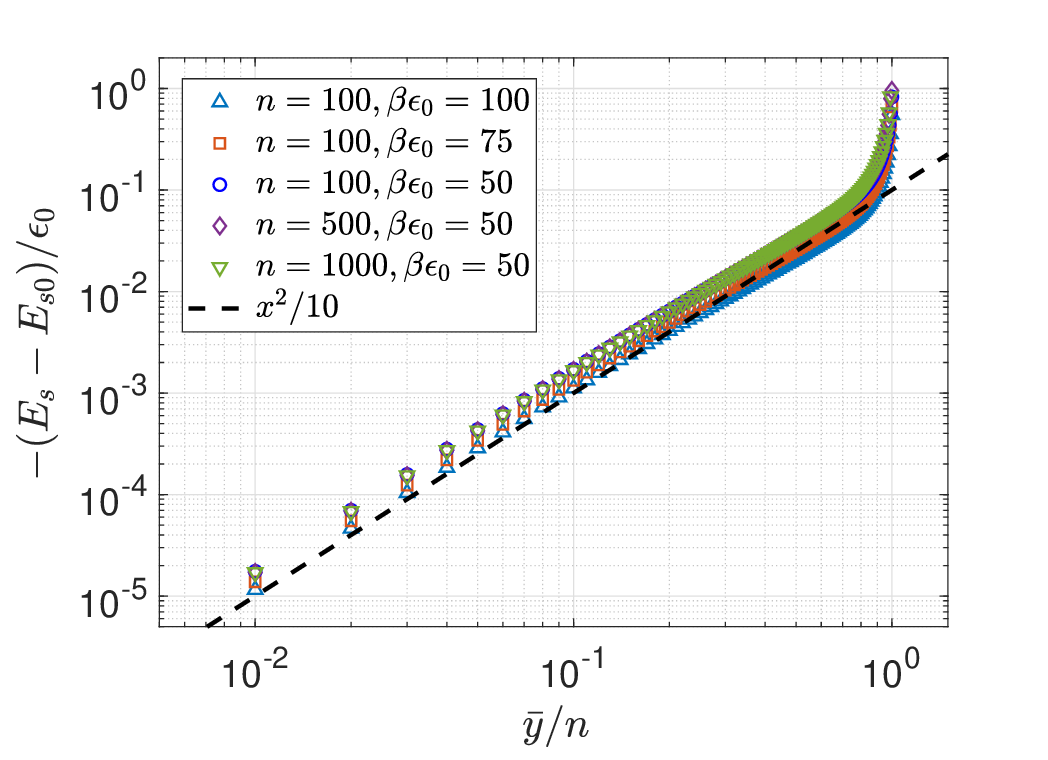} \caption{\protect\label{}}
     \end{subfigure}
\caption{(a) Normalized energy barrier for chain scission, with various values of $\beta \epsilon_0$ and $n$, and (b) Normalized change in the energy barrier for chain scission. The dashed line shows a quadratic function.}
\label{fig:scission barrier new}
\end{figure}

\begin{figure}[htp]
     \begin{subfigure}[b]{0.48\textwidth}
         \centering        \includegraphics[width=1\textwidth]{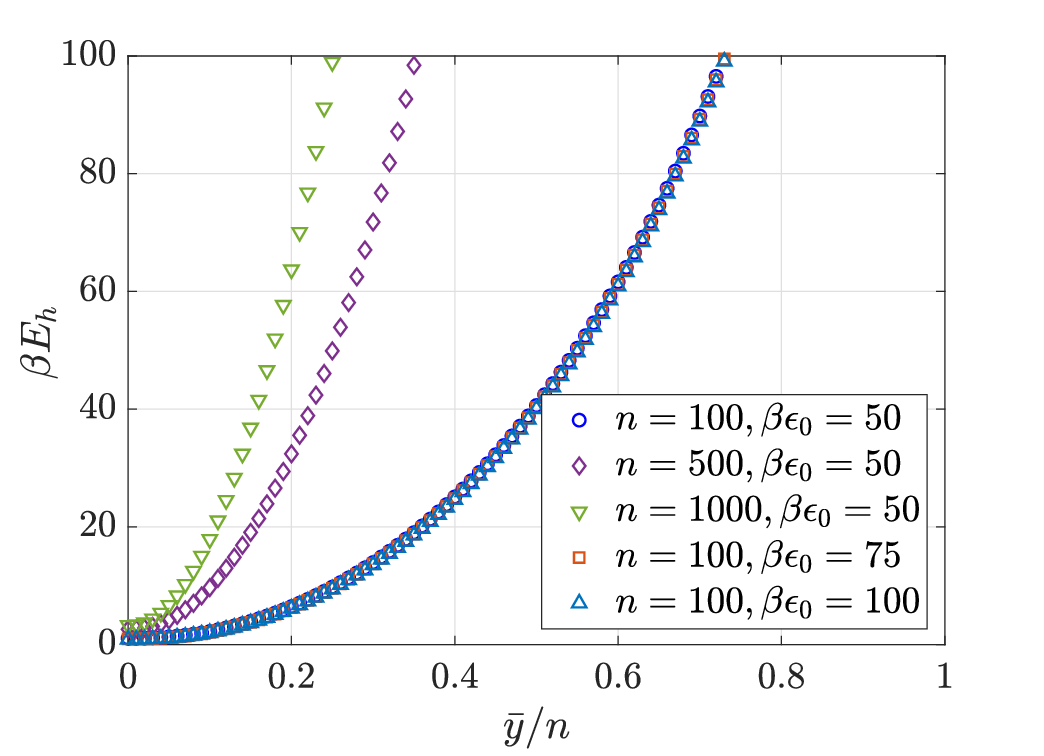}  \caption{\protect\label{}}
     \end{subfigure}
     \hfill
     \begin{subfigure}[b]{0.48\textwidth}
         \centering \includegraphics[width=1\textwidth]{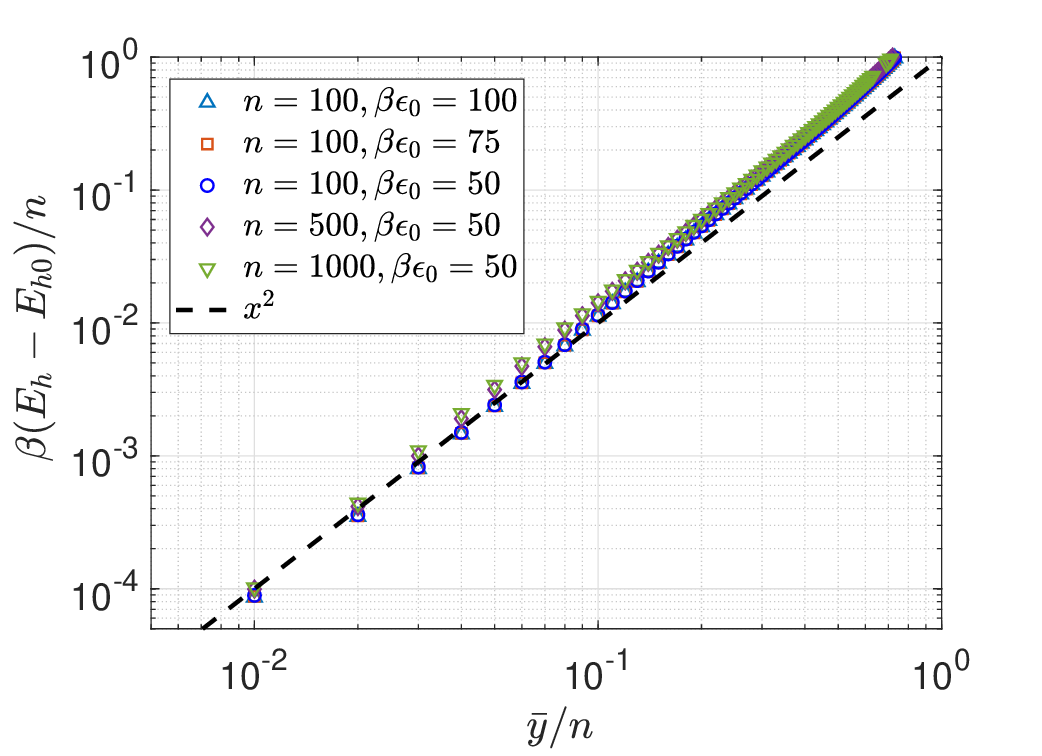} \caption{\protect\label{}}
     \end{subfigure}
\caption{(a) Normalized energy barrier for chain healing, with various values of $\beta \epsilon_0$ and $n$, and (b) Normalized change in the energy barrier for chain healing. The dashed line shows a quadratic function.}
\label{fig:healing barrier new}
\end{figure}

With the free energy function $A(x, \vec{y})$, we determine the energy barriers for chain scission and healing for each $\vec{y}$, that is, under the displacement control. 
As shown in Figure \ref{free_nonGaussian}, the free energy function has two local minima when $\bar{y} \leq n$, and there is a local maximum between the two minima. Let the two minimum free energy values be $A_1$ and $A_2$, corresponding to the two equilibrium states with $\bar{x}_1 \approx 1$ and $\bar{x}_2 \gg 1$, respectively. Let the local maximum free energy be $A_t$ (the transition state). Then,
the energy barrier for chain scission is 
$E_s = A_t-A_1$. Similarly, the energy barrier for healing is $E_h = A_t-A_2$. When the first minimum at $\bar{x}_1 \approx 1$ disappears for $\bar{y}>n$, $E_s =0$ and $E_h \rightarrow \infty$, in which case chain scission would be instantaneous and healing would be impossible. 
For a given chain, with $\beta \epsilon_0$ and $n$ fixed, both the energy barriers are functions of $\bar{y}$, as shown in Figures \ref{fig:scission barrier new} and \ref{fig:healing barrier new}. 

Figure \ref{fig:scission barrier new}(a) plots the normalized energy barrier for chain scission versus $\bar{y}/n$ for various values of $\beta \epsilon_0$ and $n$. At $\bar{y}=0$,  the energy barrier $E_s$ is slightly less than the bond energy $\epsilon_0$, due to the entropic contribution to the free energy function. As shown in Eq. \eqref{eq:ideal_energy} by the Gaussian approximation, besides the purely energetic Lenard-Jones term, the entropic terms raise the free energy and lower the depth of the energy well at $\bar{x} \approx 1$. Let $E_{s0}$ be the energy barrier for scission at $\bar{y} = 0$, and $\Delta E_{s0} = \epsilon_0 - E_{s0}$ be the reduction due to the entropic effect. It is found that $\beta \Delta E_{s0}$ increases slightly from $2.15$ to $2.23$ as $n$ increases from $100$ to $1000$ for $\beta \epsilon_0=50$. For $n = 100$, $\beta \Delta E_{s0}$ increases slightly from $2.15$ to $2.35$ as $\beta \epsilon_0$ increases from $50$ to $100$. Thus, the entropic effect $\Delta E_{s0}$ is relatively small compared to the bond energy $\epsilon_0$, and $\beta \Delta E_{s0} \approx 2.25$ may be used as an approximation for the ranges of $n$ and $\beta \epsilon_0$ considered here.

As $\bar{y}$ increases, the energy barrier $E_s$ decreases. For relatively small $\bar{y}/n$, the energy barrier for chain scission may be obtained from Eq. \eqref{eq:ideal_energy} by the Gaussian approximation. Interestingly, the leading order approximation for the change in the energy barrier is quadratic, namely, $E_s - E_{s0} \propto (\bar{y}/n)^2$, as shown by plotting $E_s - E_{s0}$ versus $\bar{y}/n$ in the log-log scales (Fig. \ref{fig:scission barrier new}b). This result is in contrast with the commonly assumed linear relation between the energy barrier and the applied force \cite{yu2018dynamicbonds, kothari2018, Ghareeb2021, Son2021_vitrimers}, given that the force is linearly proportional to $y$ in the small stretch limit ($\bar{y} \ll n$). Here, the energy barrier for chain scission under the displacement control is also different from that in a previous study \citep{yang2020multiscale} where force control was assumed.

For large $\bar{y}$, the energy barrier $E_s$ rapidly approaches zero as $\bar{y}$ approaches $n$. The results for $\beta\epsilon_0 = 50$ and three values of $n$ in Figure \ref{fig:scission barrier new}(a) nearly collapse onto one curve, suggesting that the energy barrier for chain scission depends on $n$ primarily through $\bar{y}/n$. For a fixed $\bar{y}/n$, the energy barrier $E_s$ increases with the bond energy $\epsilon_0$. 
Figure \ref{fig:scission barrier new}(b) shows that all results 
nearly collapse onto one curve when plotting the normalized change, $-(E_s - E_{s0})/\epsilon_0$ versus $\bar{y}/n$. Thus, 
the energy barrier for chain scission may be approximated by the form, 
\begin{equation}
E_s = \epsilon_0 - \Delta E_{s0} - \epsilon_0 \phi_s(\bar{y}/n),
\end{equation}
where $\Delta E_{s0} \approx 2.25 k_BT$. As shown in Figure \ref{fig:scission barrier new}(b), the function $\phi_s(\bar{y}/n)$ is approximately quadratic when $\bar{y}/n \ll 1$ and approaches $1$ as $\bar{y}/n \rightarrow 1$.

We next turn our attention to the energy barrier for healing in Figure \ref{fig:healing barrier new}. Figure \ref{fig:healing barrier new}(a) plots the normalized energy barrier for healing versus $\bar{y}/n$ for various values of $n$ and $\beta\epsilon_0$.  At $\bar{y}=0$, the energy barrier for healing is nearly zero, but not exactly due to the entropic terms in the free energy function. Let $E_{h0}$ be the energy barrier for healing at $\bar{y}=0$. It is found that $\beta E_{h0}$ decreases slightly from $1.07$ to $0.86$ as $\beta\epsilon_0$ increases from $50$ to $100$ for $n=100$. For $\beta\epsilon_0 = 50$, $\beta E_{h0}$ increases from $1.07$ to $3.28$ as $n$ increases from $100$ to $1000$. Thus, while $E_{h0}$ is relatively small compared to the bond energy $\epsilon_0$, it increases with the chain length but nearly independent of the bond energy.

For given $n$ and $\beta\epsilon_0$, the energy barrier $E_h$ increases with $\bar{y}/n$. At each $\bar{y}/n$, the energy barrier $E_h$ increases with $n$. A longer chain would make it more difficult for the two broken fragments to come together to re-associate, and thus the higher energy barrier for healing. On the other hand, the energy barrier $E_h$ is nearly independent of the bond energy $\epsilon_0$. 
Figure \ref{fig:healing barrier new}(b) plots the normalized change in the energy barrier for healing, $\beta (E_h-E_{h0})/n$, versus $\bar{y}/n$. The results in Figure \ref{fig:healing barrier new}(a) nearly collapse onto one curve in Figure \ref{fig:healing barrier new}(b), approximately following a quadratic relation:
\begin{equation}
    \beta (E_h-E_{h0})/n \approx (\bar{y}/n)^2 .
\end{equation}

\begin{figure}[htbp]
    \begin{center}
        \includegraphics[width=10cm]{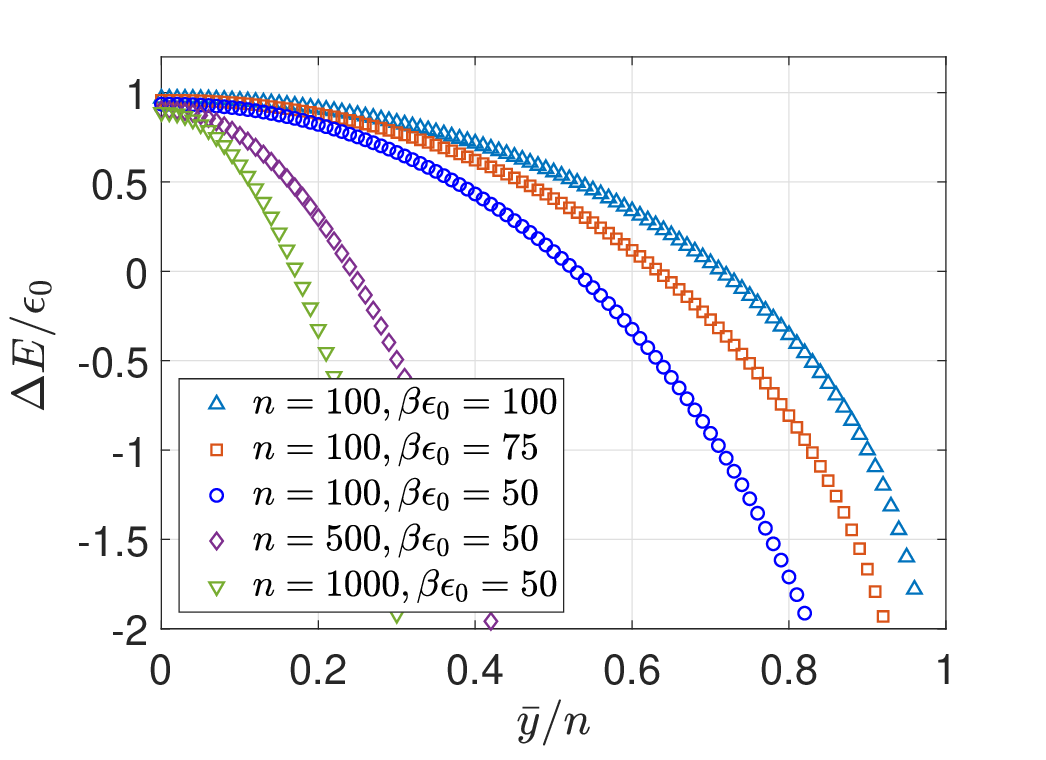}
    \caption{Difference between the two energy barriers, $\Delta E = E_s-E_h$, normalized by the bond energy $\epsilon_0$, versus $\bar{y}/n$ for various values of $n$ and $\beta\epsilon_0$. }
    \label{fig:barrier difference}
    \end{center}
\end{figure}

Comparing Figure \ref{fig:scission barrier new} and Figure \ref{fig:healing barrier new}, we can conclude that the energy barrier for chain scission is dominated by the energetics of the individual link, while the energy barrier for healing is highly entropic in nature. Given $\beta \epsilon_0$ and $n$ for a polymer chain, the energy barrier for scission is higher than that for healing when $y$ is small, while the opposite is true when $y$ is large. Figure \ref{fig:barrier difference} shows the difference between the two energy barriers, $\Delta E = E_s-E_h$, as a function of $\bar{y}/n$ for various values of $n$ and $\beta\epsilon_0$. The difference $\Delta E$ decreases monotonically with $\bar{y}/n$ for each chain. 
When $\Delta E<0$, chain scission is energetically favored. In contrast, when $\Delta E > 0$, healing is energetically favored. In both cases, the free energy of the chain reduces, resulting in dissipation of the free energy.
For each chain, the two energy barriers are equal at a particular value of $\bar{y}/n$, corresponding to $\Delta E = 0$. Moreover, $\Delta E/\epsilon_0$ increases with the normalized bond energy $\beta\epsilon_0$ (due to increasing $E_s$) but decreases with the chain length $n$ (due to increasing $E_h$).

\section{Statistics and kinetics of a single chain under tension}
We now consider a single polymer chain under tension (Fig. \ref{fig:example}), a setup similar to single molecule force spectroscopy experiments using atomic force microscopy (AFM) \cite{Zhang2003, Dudko2006rates, Schwaderer2008siloxane, Grandbois1999, Dudko2008, Suzuki2010, Suzuki2013, Makarov2015}. When the loading device such as the AFM cantilever is much stiffer than the polymer chain, the end-to-end distance of the chain can be varied by controlling the AFM to achieve the displacement-controlled condition. In contrast, if the polymer chain is much stiffer than the loading device (AFM), a force-controlled condition is realized \cite{Manca2012}. In practice, however, the AFM experiment may not be perfectly displacement control or force control. 
The displacement control can also be achieved by using an extension clamp with a feedback mechanism as in the ultrastable AFM \cite{King2009, Suzuki2013} or the interfacial force microscope (IFM) \cite{Joyce1991}. 
Here, we assume the displacement-controlled condition, where a polymer chain is subject to a prescribed end-to-end displacement history, $y(t)$. 
The probability for a polymer chain of $n+1$ identical links to be intact is a function of time, $P(t)$, governed by the first-order kinetics \cite{Makarov2015, bell1978models}:  
\begin{equation}
    \frac{\mathrm{d}P}{\mathrm{d}t}=-(n+1) P \nu_s \exp{(-\beta E_s)} + (1-P) \nu_h \exp{(-\beta E_h)} .
    \label{eq:rate1}
\end{equation}
Here, $E_s$ and $E_h$ are the energy barriers for chain scission and healing, respectively, both dependent on the end-to-end displacement $y(t)$, and $\nu_s$ is the microscopic ``attempt frequency'' for chain scission and $\nu_h$ is the microscopic frequency for healing. The frequency $\nu_s$ is often taken as the natural frequency of atomic oscillation, $\nu_s \approx 10^{13}$ s$^{-1}$, following the classic transition-state theory \cite{Eyring, Makarov2015}. However, for processes in condensed phases it has been noted that $\nu_s$ could be considerably lower \cite{Kramers1940, Hanggi1990, Evans1998}, and Kramers theory \cite{Kramers1940} has been used to estimate this pre-factor \cite{titinLi2003,Dudko2006rates,Serdal2005, Thirumalai2016,Woodside2016}. The frequency $\nu_h$ is largely unknown. Here, we assume that $\nu_h = \nu_s = \nu$ and take $\nu = 10^{13}$ s$^{-1}$ as the upper bound estimate of the frequencies.
Since any one of the $n+1$ links of the chain may break for chain scission, the first term on the right side of \eqref{eq:rate1}  includes a factor of $n+1$ for the rate of chain scission.
We note that, if reversible chain scission is facilitated by dynamic sacrificial bonds \cite{Kersey2007, Guo2009Association}, the factor $n+1$ should be replaced by $1$ or the number of sacrificial bonds per chain.

For a given displacement history $y(t)$, the probability for a chain to be intact at time $t$ can be obtained from a large number of repeated experiments. At any time $t$, the chain is intact in some of the experiments but broken in the other experiments. The probability $P(t)$ is defined by the number of experiments when the chain is intact divided by the total number of experiments. 

In an infinitesimal time interval between $t$ and $t+dt$, the probability for a chain to break is $P(t)k_s(y(t))dt$, where $k_s(y) = (n+1)\nu_s \exp (-\beta E_s(y))$ is the rate coefficient of chain scission and the energy barrier $E_s$ depends on $y$. Denote this probability as $\rho_s(t) dt$, where $\rho_s(t) = P(t)k_s(y(t))$ is the probability density for chain scission. At a given end-to-end distance $y$, the rate coefficient of chain scission is a constant $k_s(y)$. For chain scission to occur between $t$ and $t+dt$, the chain must be intact at time $t$.
Thus, the probability density for chain scission depends on both the time-dependent probability $P(t)$ and the rate coefficient $k_s(y)$.
Similarly, the probability density for healing is: $\rho_h(t)=(1-P(t))k_h(y(t))$, where $k_h(y)=\nu_h \exp \left( -\beta E_h(y) \right)$ is the healing rate coefficient that depends on $y$. As such, the rate equation \eqref{eq:rate1} can be rewritten as 
\begin{equation}
\begin{aligned}
    \frac{\mathrm{d}P}{\mathrm{d}t} &= -k_s(y) P + k_h(y)(1-P) \\
    & = -\rho_s(t)+\rho_h(t) .
    \end{aligned}
    \label{eq:rate1_rho}
\end{equation}
This equation relates the time-dependent probability $P(t)$ to the probability densities for scission $\rho_s(t)$ and for healing $\rho_h(t)$.

\subsection{Equilibrium statistics}
At each prescribed end-to-end distance $y$, there exists an equilibrium probability for the chain to be intact. It is found by setting $dP/dt = 0$ in Eq. \eqref{eq:rate1_rho}:
\begin{equation}
    P_{eq}(y) = \frac{k_h(y)}{k_s(y)+k_h(y)} .
    \label{eq:Peq}
\end{equation}

\begin{figure}[h!]
    \centering

    \begin{subfigure}[t]{0.45\textwidth}
        \centering
        \includegraphics[width=\textwidth]
        {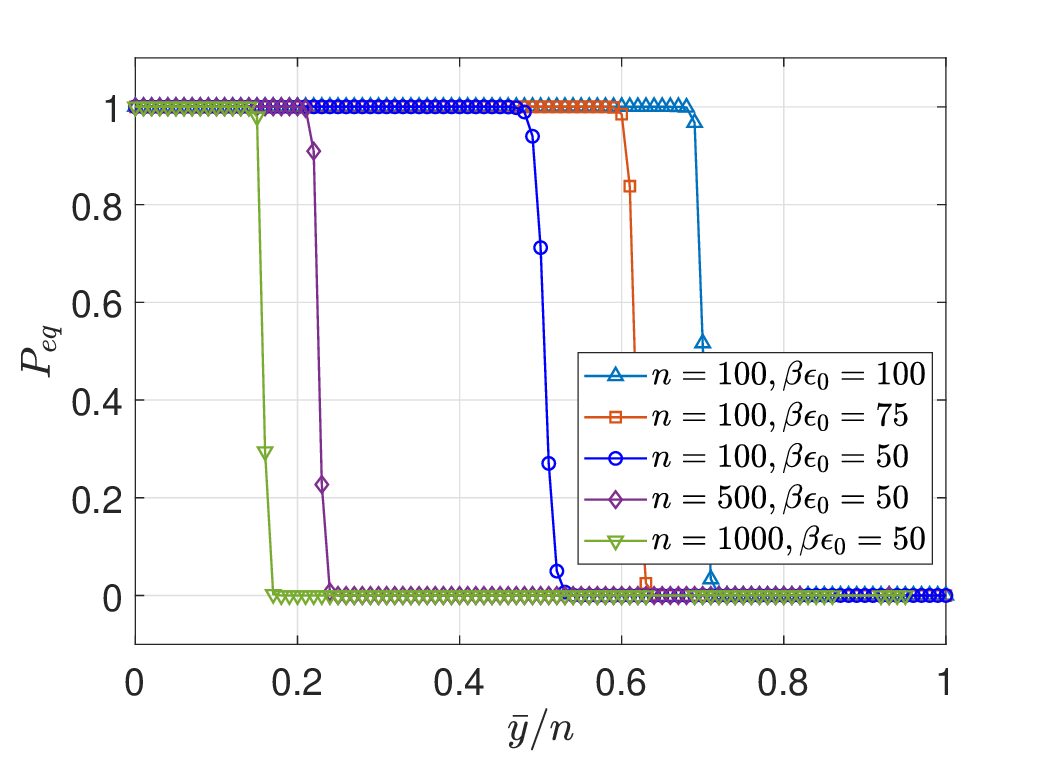}
        \caption*{(a)} 
    \end{subfigure}
    \hfill
    \begin{subfigure}[t]{0.45\textwidth}
        \centering
        \includegraphics[width=\textwidth]
        {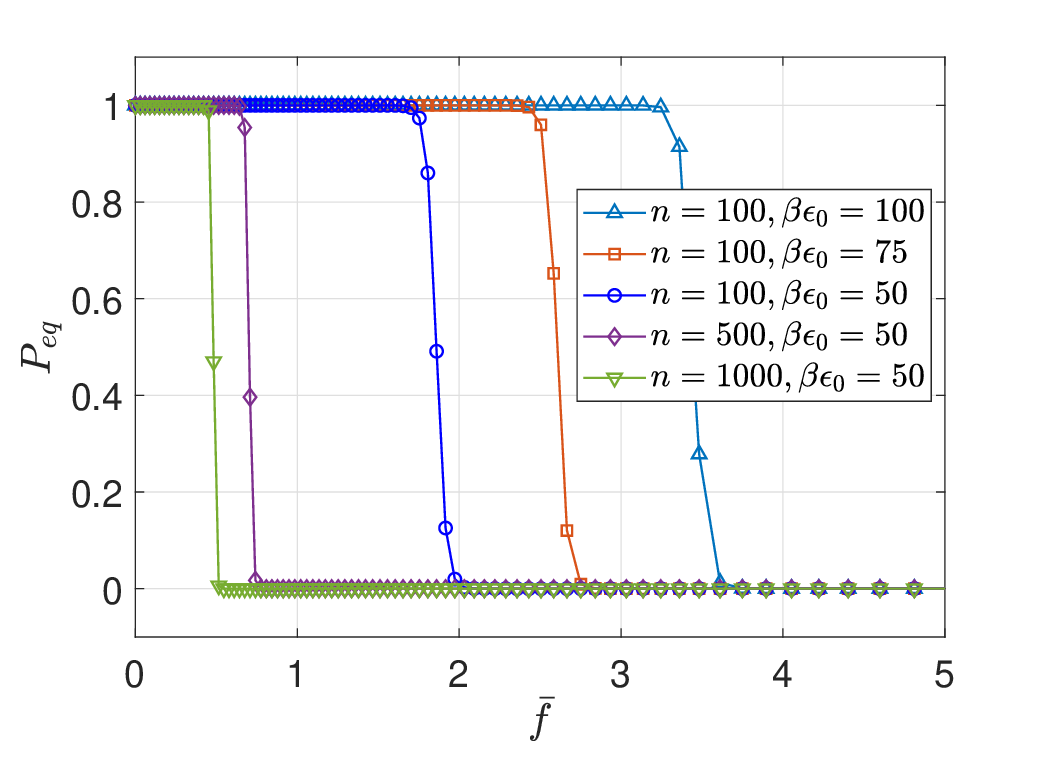}
        \caption*{(b)}
    \end{subfigure}

    \vspace{0.1cm}

    \begin{subfigure}[t]{0.45\textwidth}
        \centering
        \includegraphics[width=\textwidth]{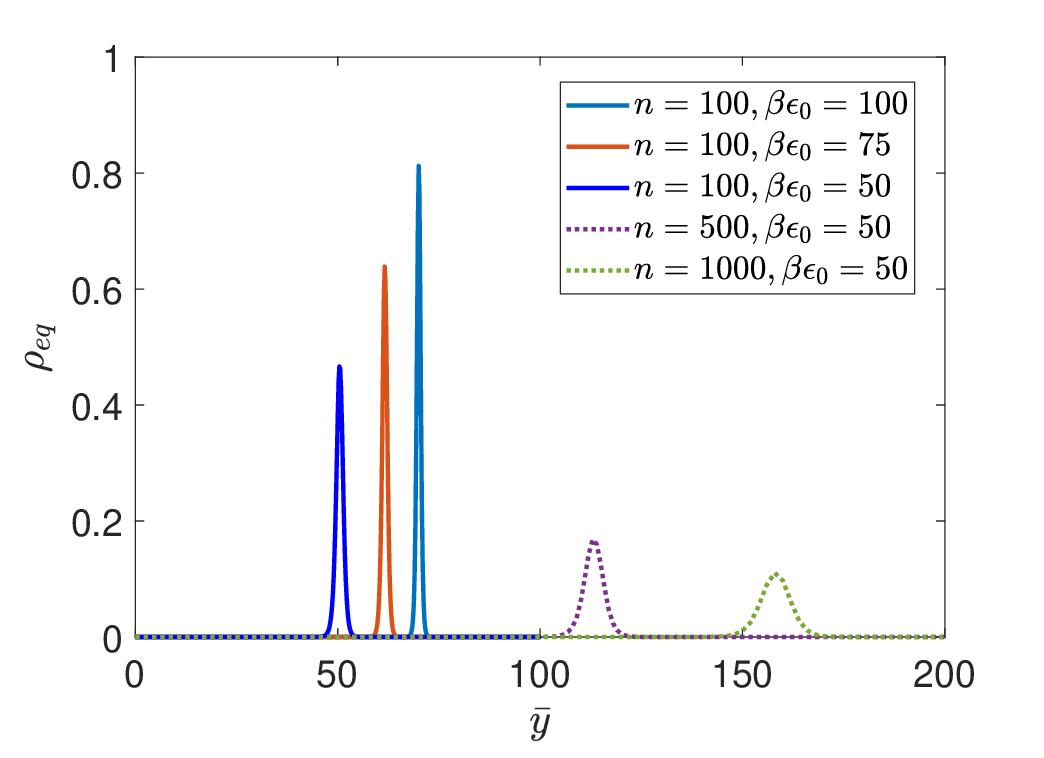}
        \caption*{(c)}
    \end{subfigure}
    \hfill
    \begin{subfigure}[t]{0.45\textwidth}
        \centering
        \includegraphics[width=\textwidth]{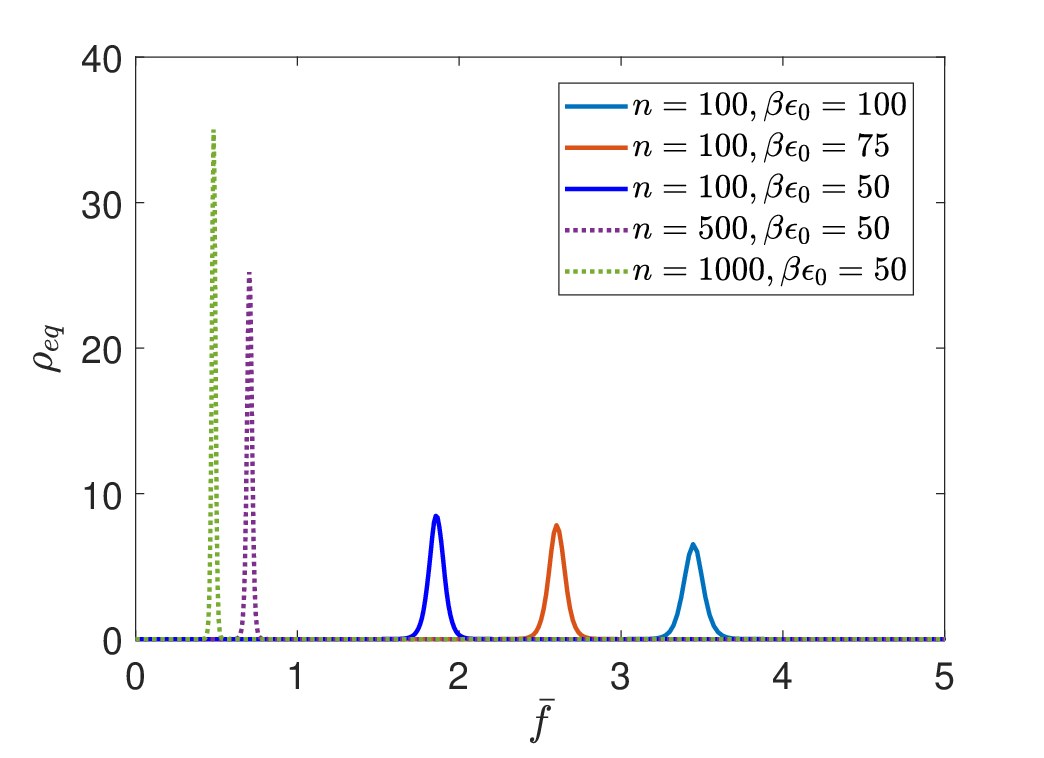}
        \caption*{(d)}
    \end{subfigure}

    \caption{(a-b) Equilibrium probability as a function of $\bar{y}=y/l$ and $\bar{f}=\beta fl$. (c-d) Corresponding equilibrium probability density functions.}
    \label{eq_probs}
\end{figure}

The equilibrium probability depends on the ratio, $k_s(y)/k_h(y)$, which in turn depends on the chain length $n+1$ and the difference between the two energy barriers, $\Delta E = E_s-E_h$. When $E_s > E_h$ and $\beta \Delta E \gg 1$, we obtain $k_h(y) \gg k_s(y)$ and $P_{eq} \approx 1$, corresponding to an intact chain in equilibrium. When $E_s < E_h$ and $-\beta \Delta E \gg 1$, we obtain $k_h(y) \ll k_s(y)$ and $P_{eq} \approx 0$, corresponding to a broken chain in equilibrium. When $E_s=E_h$, we have $\Delta E = 0$ and $P_{eq}= 1/(n+2)$, which is close to zero when $n \gg 1$ for a long chain. 
Figure \ref{eq_probs}(a) shows the equilibrium probability as a function of $\bar{y}/n$ for various values of $n$ and $\beta\epsilon_0$.
For each chain with given $n$ and $\beta\epsilon_0$, $\Delta E$ decreases monotonically with $y$ (Fig. \ref{fig:barrier difference}), and thus the equilibrium probability is a decreasing function of $y$, $P_{eq}(y)$. 
Recall that the force pulling each chain is an increasing function of $y$ (Fig. \ref{force_compare}b), $f(y)$. The two functions combine to give the equilibrium probability as a decreasing function of the force: $P_{eq}(f)$, as shown in Figure \ref{eq_probs}(b).

In both Figures \ref{eq_probs} (a) and (b), the equilibrium probability transitions sharply from $1$ to $0$ as $y$ or $f$ increases for each chain. The corresponding probability density functions can be obtained from the derivatives of the equilibrium probability with respect to $y$ and $f$, that is, $\rho_{eq}(y) =-dP_{eq}(y)/dy$ and $\rho_{eq}(f) = -dP_{eq}(f)/df$, as shown in 
Figures \ref{eq_probs} (c) and (d) for various values of $n$ and $\beta \epsilon_0$.
The equilibrium probability density function, $\rho_{eq}(y)$, has a sharp peak for a relatively short chain ($n = 100$) and the peak shifts right as the bond energy $\beta\epsilon_0$ increases. For a longer chain, the equilibrium distribution $\rho_{eq}(y)$ is broader with a peak located at a larger $\bar{y}$. Thus, a longer chain is more likely to survive at a larger end-to-end distance. 
In contrast, the equilibrium probability density function, $\rho_{eq}(f)$, has a higher peak at a lower force for a longer chain, suggesting that a shorter chain is more likely to survive a larger force. With the same chain length, the peak shifts right with increasing bond energy $\beta\epsilon_0$. In all cases considered,
the peak of the equilibrium distribution $\rho_{eq}(f)$ is located at a force far below the upper bound of the chain force as noted in Figure \ref{force_compare}b.

As shown in Section 6.3, the equilibrium probability sets the lower bound for the probability of the chain to be intact if pulled by a constant rate. At an extremely slow rate, the state of the chain approaches equilibrium at each step. The probability density functions, $\rho_{eq}(y)$ and $\rho_{eq}(f)$, result from the difference between the probability density for chain scission ($\rho_s$) and that for healing ($\rho_h$).

\subsection{Kinetics}
Chain scission and healing are competing kinetic processes. Consider an experiment where a chain is held at a constant end-to-end distance $y$. In this case, both the energy barriers are fixed, independent of time, so that Eq. \eqref{eq:rate1_rho} can be rewritten as 
\begin{equation}
\frac{\mathrm{d}P}{\mathrm{d}t} =-\frac{P}{\tau(y)}+k_h(y) ,
\label{eq:rate2}
\end{equation}
where $\tau(y)=\left[k_s(y)+k_h(y)\right]^{-1}$ is a time scale. The two terms on the right hand side of \eqref{eq:rate2} compete to determine if the probability decreases or increases over time.
The solution to the rate equation \eqref{eq:rate2} takes the form 
\begin{equation}
    P(t) = (P_0-P_{eq}) \exp{\left(-\frac{t}{\tau}\right) + P_{eq}},
    \label{eq:P evolution}
\end{equation}
where $P_0$ is the probability at $t=0$ and $P_{eq}$ is the equilibrium probability that depends on $y$. 

\begin{figure}[h!]
    \centering

    \begin{subfigure}[t]{0.45\textwidth}
        \centering
        \includegraphics[width=\textwidth]
        {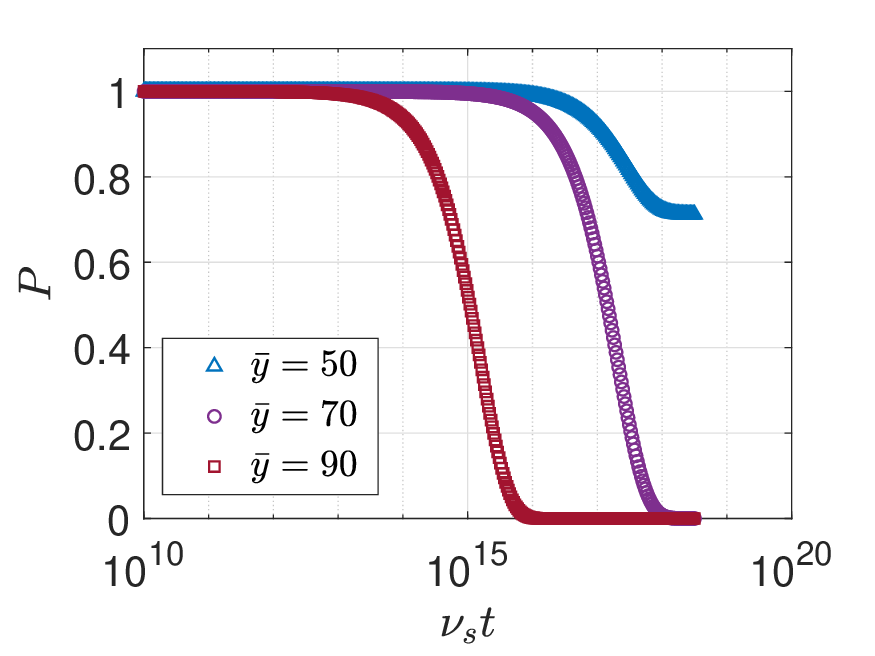}
        \caption*{(a)}
    \end{subfigure}
    \hfill
    \begin{subfigure}[t]{0.45\textwidth}
        \centering
        \includegraphics[width=\textwidth]
        {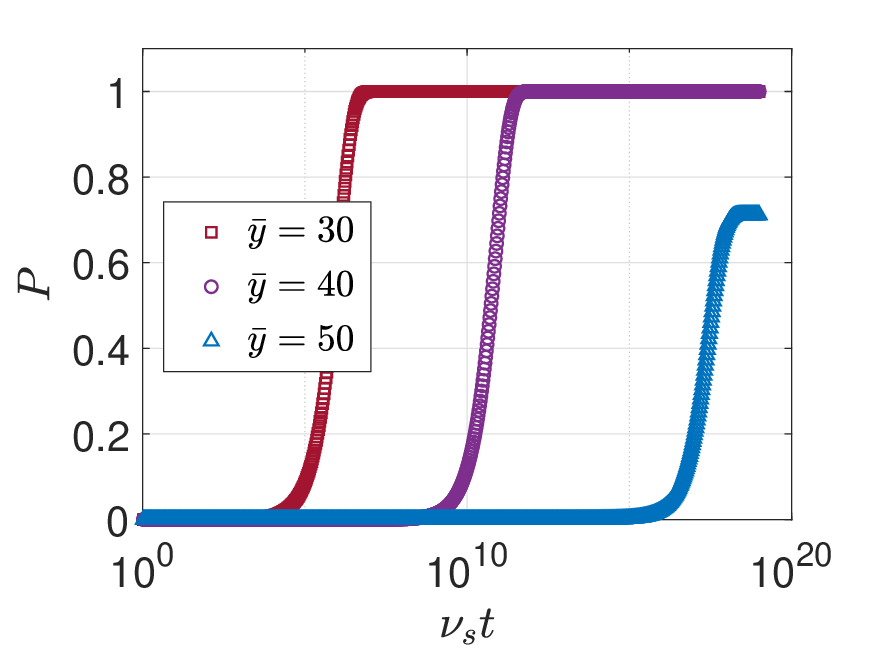}
        \caption*{(b)}
    \end{subfigure}

    \caption{Probability for a chain to be intact as a function of time with (a) $P_0=1$ and (b) $P_0=0$, for a chain with $n=100$ and $\beta\epsilon_0=50$ held at various values of $\bar{y}$.}
    \label{probs_hold_y}
\end{figure}

If the chain is intact at $t=0$, we have $P_0=1$, in which case the probability $P(t)$ decreases over time and approaches $P_{eq}$ as $t\rightarrow \infty$. 
Figure \ref{probs_hold_y}(a) shows the evolution of the probability $P(t)$ for a chain with $n=100$ and $\beta\epsilon_0=50$, subject to various end-to-end distances $\bar{y}$. When $\bar{y}$ is relatively small, the equilibrium probability is close to 1, and thus the probability $P(t)$ stays close to 1 for all time and the chain is unlikely to break. 
In contrast, when $\bar{y}$ is relatively large, the equilibrium probability is close to 0, and $P(t)$ approaches 0 after a finite time. That is, the chain is likely to break after a certain time that depends on the end-to-end distance.

Alternatively, if we start with a broken chain, $P_0=0$, the probability $P(t)$ increases over time, that is, the probability for the chain to heal increases with time, as shown in Figure \ref{probs_hold_y}(b) for $n=100$ and $\beta\epsilon_0=50$. When $\bar{y}$ is relatively large, the equilibrium probability is close to 0, and thus the probability $P(t)$ stays close to 0 for all time and the chain is unlikely to heal. In contrast, when $\bar{y}$ is relatively small, the equilibrium probability is close to 1, and $P(t)$ approaches 1 after a finite time. That is, the broken chain is likely to heal after a certain time that depends on the end-to-end distance. 
It should be noted that healing of a broken chain requires that the ends of both segments remain reactive with each other as free radicals. If the radicals react with other molecules in the surrounding, the chain would not be able to heal as described here.

\begin{figure}[htbp]
    \begin{center}
        \includegraphics[width=10cm]{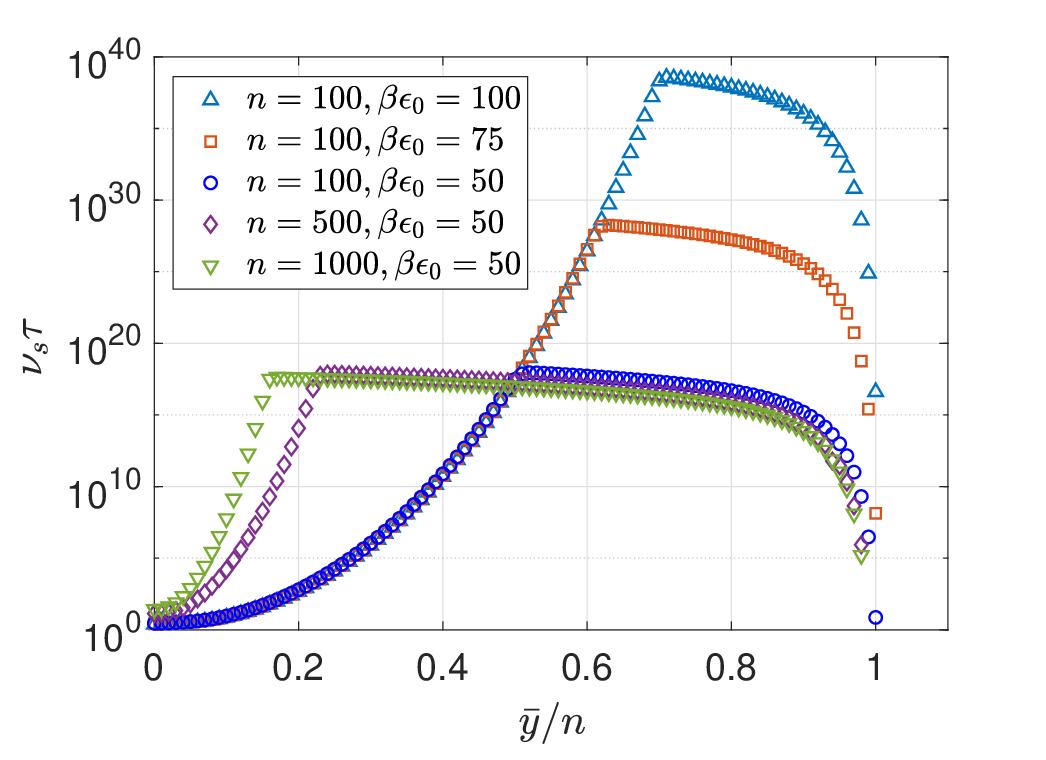}
    \caption{Normalized characteristic time scale for chain scission and healing, $\nu_s\tau$, as a function of $\bar{y}/n$ for chains with various values of $n$ and $\beta\epsilon_0$. }
    \label{fig:mean time}
    \end{center}
\end{figure}

In both cases, the characteristic time scale is set by $\tau(y)$, which depends on the two energy barriers $E_s(y)$ and $E_h(y)$ as well as the chain length $n+1$. 
As shown in Figure \ref{fig:mean time}, for each chain with given $n$ and $\beta\epsilon_0$, the characteristic time first increases with $\bar{y}/n$ and then decreases. When $\bar{y}/n$ is small, healing dominates, and the time scale for healing increases with $\bar{y}/n$. When $\bar{y}/n$ is large, chain scission dominates, and the time scale for scission decreases with $\bar{y}/n$. 
A transition occurs at the peak $\tau$. While the peak value of $\tau$ depends primarily on the bond energy $\beta\epsilon_0$, the location of the transition in terms of $\bar{y}/n$ also depends on the chain length: the transition occurs at a smaller $\bar{y}/n$ for a longer chain, similar to the transition of the equilibrium probability $P_{eq}(y)$ in Figure \ref{eq_probs}(a).

Practically, if the normalized time scale $\nu_s\tau$ is greater than $10^{21}$, it may be considered to be infinite ($> 3$ years). By this consideration, for the case of a relatively large bond energy (e.g., $\beta\epsilon_0 = 75$ and $100$ in Fig. \ref{fig:mean time}), there exists a range of intermediate values of $\bar{y}/n$, for which $\nu_s\tau > 10^{21}$ and the time scale is practically infinite for both chain scission and healing.
Also practically, if the normalized time scale is less than $10^{10}$, it may be considered to be nearly zero ($< 10^{-3}$ second).
Thus, for each chain, there exist two critical values, $\bar{y}_{1}$ and $\bar{y}_{2}$. When $\bar{y}<\bar{y}_{1}$, healing is practically instantaneous. When $\bar{y}>\bar{y}_{2}$, chain scission is practically instantaneous. 
Figure \ref{fig:mean time} shows that $\bar{y}_{1}/n$ for instantaneous healing decreases as the chain length increases, independent of the bond energy, whereas $\bar{y}_{2}/n$ for instantaneous chain scission is close to $1$, depending weakly on the chain length and the bond energy.

To measure the kinetics of chain scission and healing of a polymer chain, one may repeat the same experiment for many times under the same condition. At a given time, the state of the chain may be either intact or broken in each experiment, and the probability for the chain to be intact can then be obtained from a sufficiently large number of experiments.
For each end-to-end distance $y$, the measured probability as a function of time, $P(t)$, may be compared to Eq. \eqref{eq:P evolution} to determine the time scale $\tau$ and the equilibrium probability $P_{eq}$. Then, the two energy barriers, $E_s$ and $E_h$, can be determined, assuming that the chain length $n+1$ is known. Repeating the process for different values of $y$, one obtains the energy barriers as functions of $y$. Alternatively, the energy barriers can also be determined from a single experiment for each value of $y$, as discussed in Section 7.

\subsection{Rate dependent chain scission}
Next we consider pulling a chain at a constant rate under the displacement control, that is, $y(t)=\gamma t$, where $\gamma$ is the pulling rate.
The kinetic processes of the thermally activated chain scission and healing as described by the rate equation \eqref{eq:rate1} naturally leads to rate-dependent chain scission as commonly observed in single-molecule force spectroscopy experiments \cite{Rief1997, Bustamante1997, Zhang2003, Dudko2006rates, Schwaderer2008siloxane, Grandbois1999, Dudko2008, Suzuki2010, Makarov2016}. Moreover, a statistical distribution of the rupture forces at the final chain scission can be predicted, which depends on the pulling rate as well.

\begin{figure}[h!]
    \centering

    \begin{subfigure}[t]{0.45\textwidth}
        \centering
        \includegraphics[width=\textwidth]
        {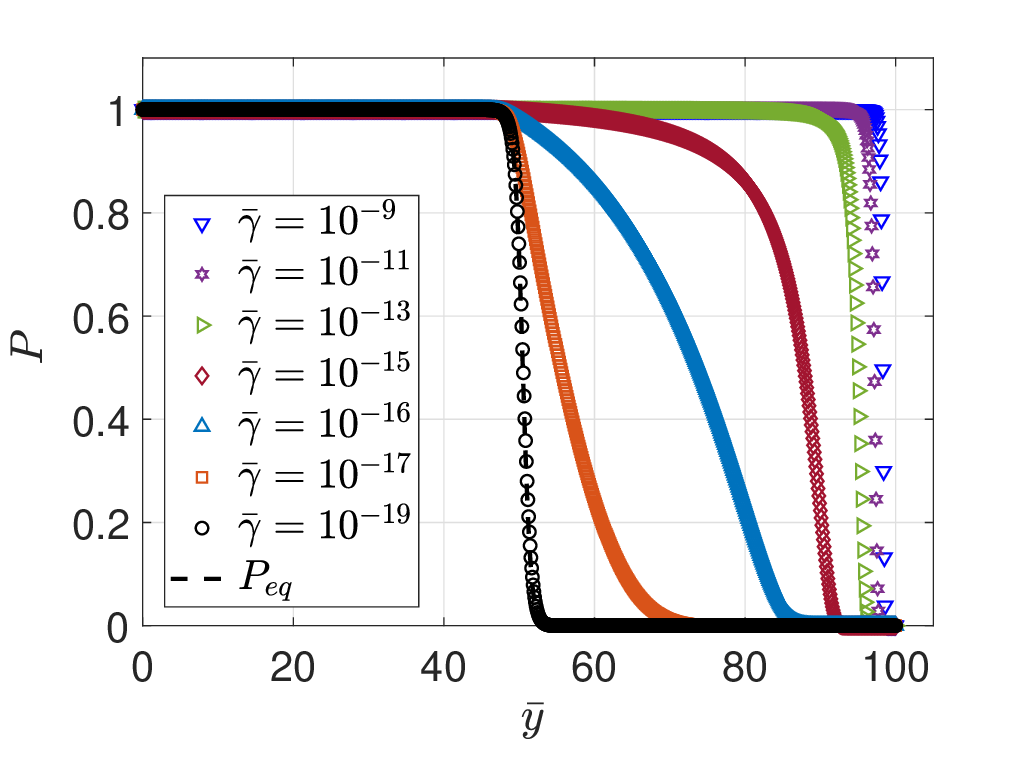}
        \caption*{(a)}
    \end{subfigure}
    \hfill
    \begin{subfigure}[t]{0.45\textwidth}
        \centering
        \includegraphics[width=\textwidth]
        {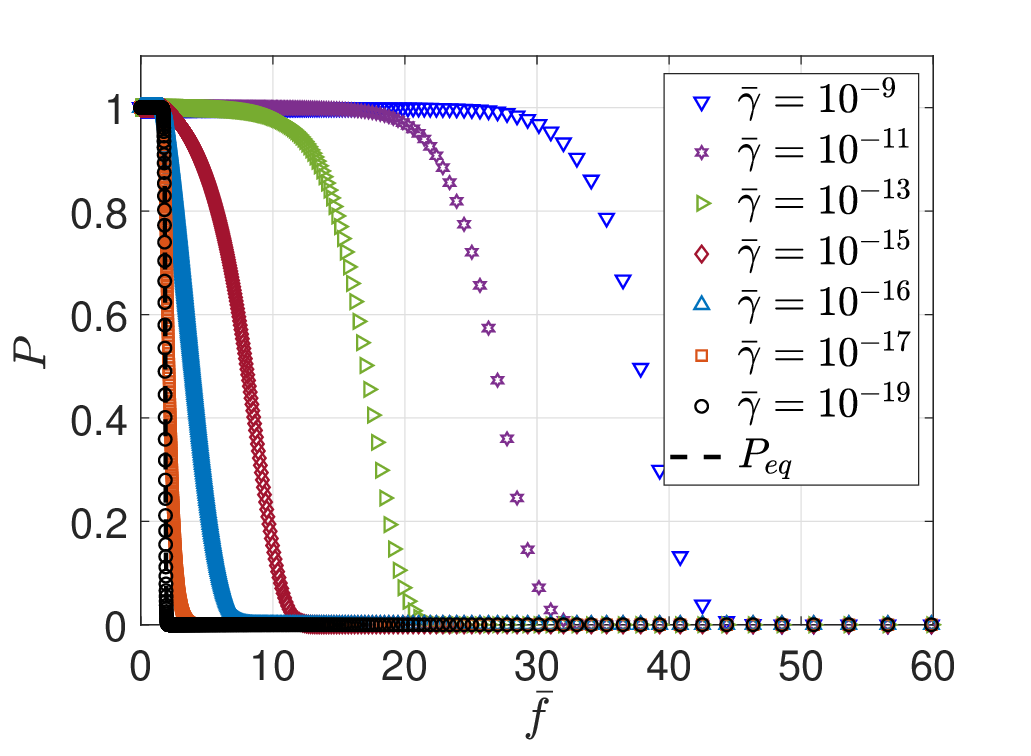}
        \caption*{(b)}
    \end{subfigure}

    \vspace{0.1cm}

    \begin{subfigure}[t]{0.45\textwidth}
        \centering
        \includegraphics[width=\textwidth]{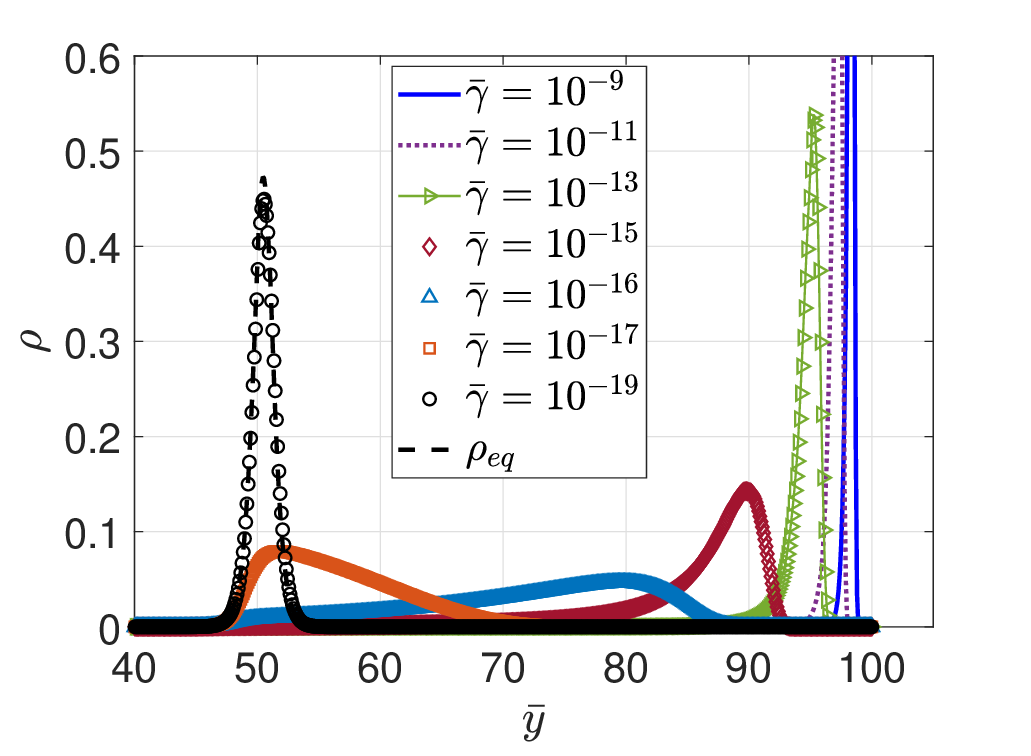}
        \caption*{(c)}
    \end{subfigure}
    \hfill
    \begin{subfigure}[t]{0.45\textwidth}
        \centering
        \includegraphics[width=\textwidth]{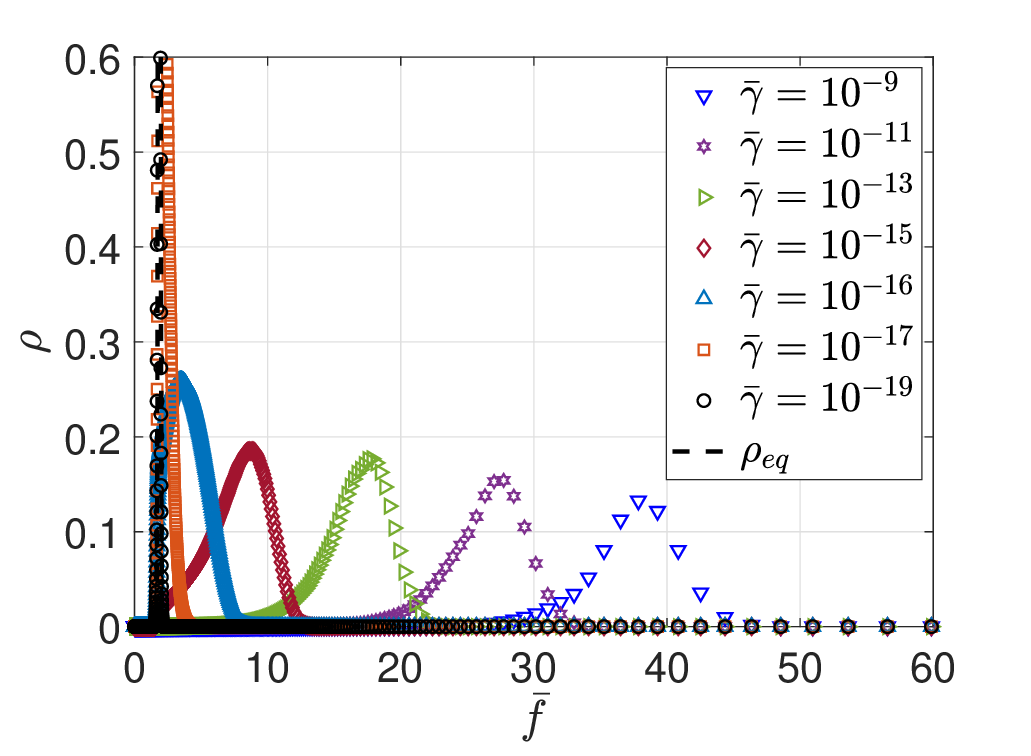}
        \caption*{(d)}
    \end{subfigure}

    \vspace{0.1cm}



    \caption{Probability for a chain ($n=100,\beta\epsilon_0=50$) to be intact when pulled at various normalized rates versus (a) $\bar{y}=y/l$ and (b) $\bar{f}=\beta fl$. (c–d) Corresponding probability density functions at various pulling rates. }
    \label{rate_probs}
\end{figure}

Typical pulling rates in AFM-based single-chain experiments range from 10 to $10^4$ nm/s \citep{Dudko2008}, whereas much higher pulling rates (up to 300 m/s) are often applied in molecular dynamics (MD) simulations \cite{schulten1998, schulten2001, Rohrig2001, Lupton2005}. Assuming $l\approx 1$ nm for each link and $\nu = 10^{13}$ s$^{-1}$, the normalized pulling rate, $\bar{\gamma} = \gamma/(\nu l)$, ranges from $10^{-12}$ to $10^{-9}$ in experiments but can be much higher (up to $10^{-2}$) in MD simulations. 
By integrating Eq. \eqref{eq:rate1} with a constant rate $\bar{\gamma}$, we obtain the probability as a function of $\bar{y}$, as shown in Figure \ref{rate_probs}(a) for $n=100$ and $\beta\epsilon_0=50$. 
This probability can be obtained from a large number of repeated experiments for a given rate $\gamma$. At any time $t$, $y = \gamma t$, the chain is intact in some of the experiments but broken in the other experiments. The probability $P(y; \gamma)$ is defined by the number of experiments when the chain is intact divided by the total number of experiments. 
For very slow pulling rates ($\bar{\gamma} < 10^{-19}$), the probability $P(y; \gamma)$ follows the equilibrium probability $P_{eq}(y)$ in \eqref{eq:Peq} (Fig. \ref{eq_probs}a), which is the lower bound to the probabilities at higher pulling rates.
As the pulling rate increases, the probability increases at increasingly larger $\bar{y}$. At a relatively high rate (e.g., $\bar{\gamma}=10^{-9}$), the probability remains nearly $1$ until $\bar{y}$ is close to $n$.
The same probability is plotted versus the normalized force $\bar{f}=\beta fl$ in Figure \ref{rate_probs}(b). The normalized force increases monotonically with $\bar{y}$ for an intact chain (Fig. \ref{force_compare}b). Again, at very low rates, the probability follows the equilibrium probability shown in Figure \ref{eq_probs}(b). As the pulling rate increases, the probability for the chain to be intact increases at larger forces.

At a constant pulling rate $\gamma$, the probability for a chain to break in an infinitesimal interval between $y$ and $y+dy$ is $P(y)k_s(y)dy/\gamma$. Thus, the probability density for chain scission as a function of the end-to-end distance is:
\begin{equation}
    \rho_s(y) = P(y)k_s(y)/\gamma .
\end{equation}
Similarly, the probability density for healing of a broken chain is:
\begin{equation}
    \rho_h(y) = (1-P(y))k_h(y)/\gamma .
\end{equation}
Then, Eq. \eqref{eq:rate1_rho} can be re-written as
\begin{equation}
    \rho(y) \equiv-\frac{dP}{dy} = \rho_s(y)-\rho_h(y).    
\label{eq:rhoy}
\end{equation}
Here, $\rho(y)$ is defined as the combined probability density function due to the difference between the probability density functions for chain scission and healing.
As shown in Figure \ref{rate_probs}(c), the probability density function $\rho(y)$ has a sharp peak at very low rates, nearly identical to the equilibrium probability density function in Figure \ref{eq_probs}(c). As the pulling rate increases, the peak of $\rho(y)$ first decreases and then increases, whereas the location of the peak shifts right towards $\bar{y} = n$ at relatively high rates.

In terms of the force, the combined probability density function is:
\begin{equation}
    \rho(f) \equiv -\frac{dP}{df} = \rho(y) \left( \frac{df}{dy} \right)^{-1}.
\label{eq:rhof}
\end{equation}
Here, $df/dy$ is the tangent stiffness of the chain. The tangent stiffness is a constant in the linear regime when the force is small, but becomes increasingly higher as the force increases beyond the linear regime (see Fig. \ref{force_compare}b). 
As shown in Figure \ref{rate_probs}(d), the peak of the probability density function $\rho(f)$ decreases monotonically with the pulling rate up to $\bar{\gamma}=10^{-9}$. The apparently different statistical distributions in $\rho(y)$ and $\rho(f)$ can be understood with reference to the force-displacement curve in Figure \ref{force_compare}(b). At a relatively low rate (e.g., $\bar{\gamma} = 10^{-16}$), $\rho(y)$ has a low peak at $\bar{y} \approx 80$ with a fairly wide distribution in the range $45<\bar{y}<90$. Within this range, the stiffness of chain increases mildly, the normalized force $\bar{f}$ varies from $1.6$ to $9.0$, and $\rho(f)$ has a peak at $\bar{f} \approx 3.5$. In contrast, at a relatively high rate (e.g., $\bar{\gamma} = 10^{-11}$), $\rho(y)$ has a sharp peak at $\bar{y} \approx 97$ with a narrow distribution in the range $93<\bar{y}<98$. However, as shown in Figure \ref{force_compare}(b), the stiffness of chain is very large in this range, as the normalized force $\bar{f}$ rises sharply from $12$ to $34$, and thus $\rho(f)$ has a wider distribution and a lower peak compared to $\rho(y)$. 

In single-chain experiments, the rupture forces are often reported in histograms from a large number of repeated experiments. When the pulling rate is slow, a chain may break more than once, followed by healing, until the final scission event. The statistical distribution for the rupture forces at the final scission event can be predicted as follows.
Let $\Phi(t_1, t_2)$ be the probability for the chain not to heal in the time interval $t_1 < t < t_2$, conditional upon being broken at time $t_1$.
Then, the probability for the chain to break in an infinitesimal interval between $t$ and $t+dt$ and not to heal thereafter is
\begin{equation}
    \rho_{s, last} (t) dt = P(t) k_s(y(t))dt \times \Phi(t, \infty) .
\end{equation}

It can be shown that the probability $\Phi(t_1, t_2)$ satisfies a rate equation
\begin{equation}
    \frac{\partial \Phi(t_1, t_2)}{\partial t_2} = - k_h(y(t_2)) \Phi(t_1, t_2) ,
\end{equation}
and $\Phi(t_1, t_1)=1$. Thus, we obtain
\begin{equation}
    \Phi(t_1, t_2) = \exp \left( -\int_{t_1}^{t_2} k_h(y(t)) dt \right) .
\end{equation}
Therefore, the probability density for the final scission event is
\begin{equation}
    \rho_{s, last} (t) = P(t) k_s(y(t)) \times \exp \left( -\int_{t}^{\infty} k_h(y(t)) dt \right) .
\end{equation}

At a constant pulling rate $\gamma$, we have $dy = \gamma dt$ and $\rho_{s,last}(t)dt = \rho_{s,last}(y)dy$. Thus, 
we obtain
\begin{equation}
\begin{aligned}
    \rho_{s, last} (y) &= \frac{1}{\gamma} P(y) k_s(y) \times \exp \left( - \frac{1}{\gamma} \int_{y}^{\infty} k_h(y) dy \right) \\
    & = \rho_s(y) \times \exp \left( - \frac{1}{\gamma} \int_{y}^{\infty} k_h(y) dy \right) .
\end{aligned}
\label{eq:rhoy_last}
\end{equation}
It can be shown that
$\int_0^\infty \rho_{s,last}(y)dy = 1$, 
because no chain could survive as $y \rightarrow \infty$. 
Note that $\rho_s(y)$ is the probability density for chain scission to occur, including all scission events before the final one. Thus, $\rho_s(y) \geq \rho_{s,last}(y)$ and $\int_0^\infty \rho_{s}(y)dy \geq 1$, because it is possible to have more than one scission event.

In terms of the force, the probability density for the final scission event is
\begin{equation}
\begin{aligned}
    \rho_{s,last}(f) &= \rho_{s, last} (y) \left( \frac{df}{dy} \right)^{-1} \\
    & = \rho_s(f) \times \exp \left( - \frac{1}{\gamma} \int_{y(f)}^{\infty} k_h(y) dy \right) ,
\end{aligned}
\label{eq:rhof_last}
\end{equation}
where $\rho_s(f) = \rho_s(y) \left( \frac{df}{dy} \right)^{-1}$.

\begin{figure}[h!]
    \centering

    \begin{subfigure}[t]{0.45\textwidth}
        \centering
        \includegraphics[width=\textwidth]
        {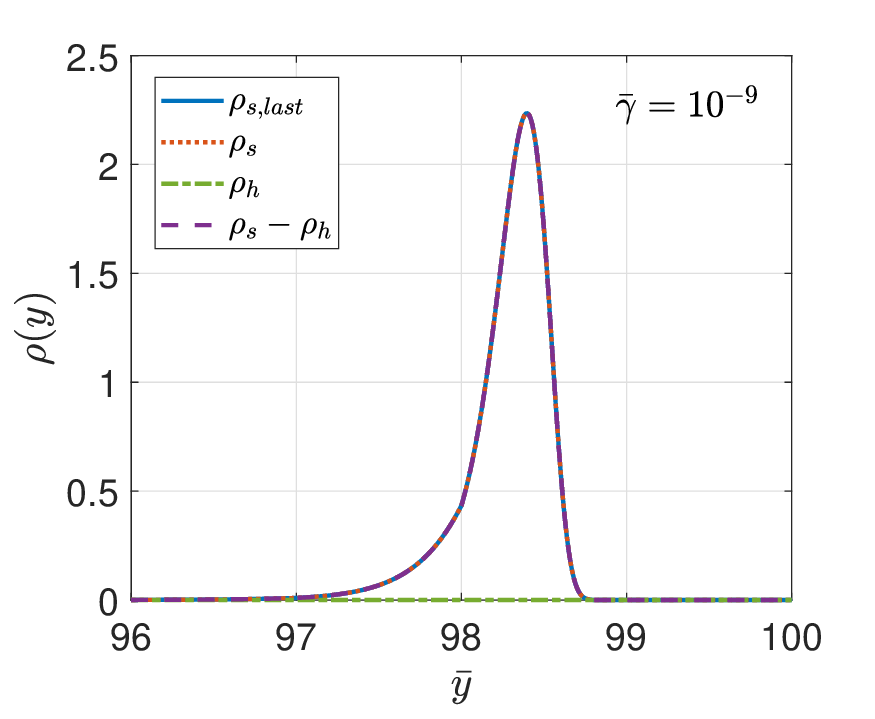}
        \caption*{(a)}
    \end{subfigure}
    \hfill
    \begin{subfigure}[t]{0.45\textwidth}
        \centering
        \includegraphics[width=\textwidth]
        {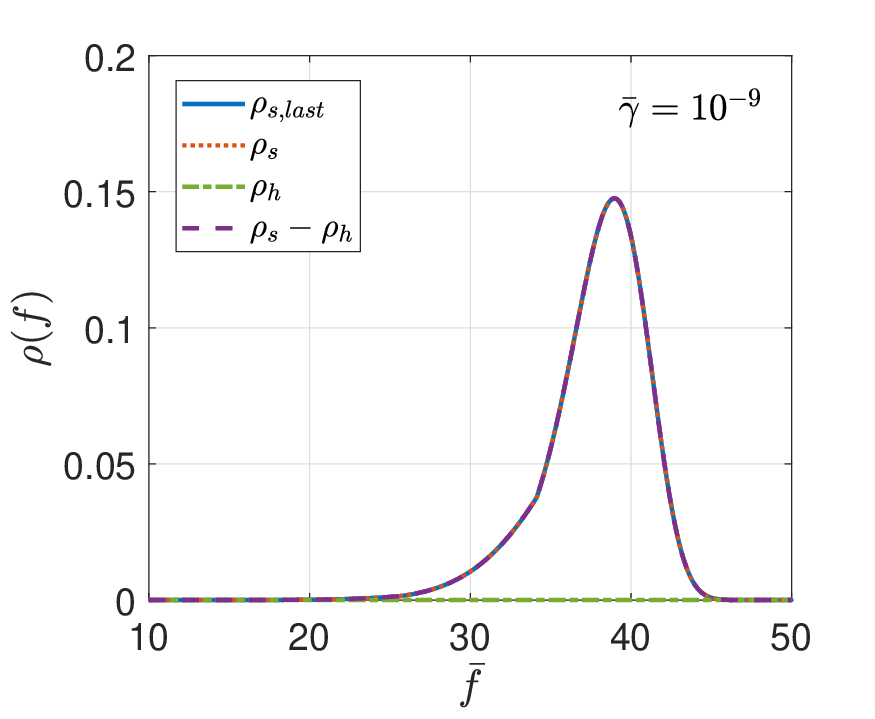}
        \caption*{(b)}
    \end{subfigure}

    \vspace{0.1cm}

    \begin{subfigure}[t]{0.45\textwidth}
        \centering
        \includegraphics[width=\textwidth]{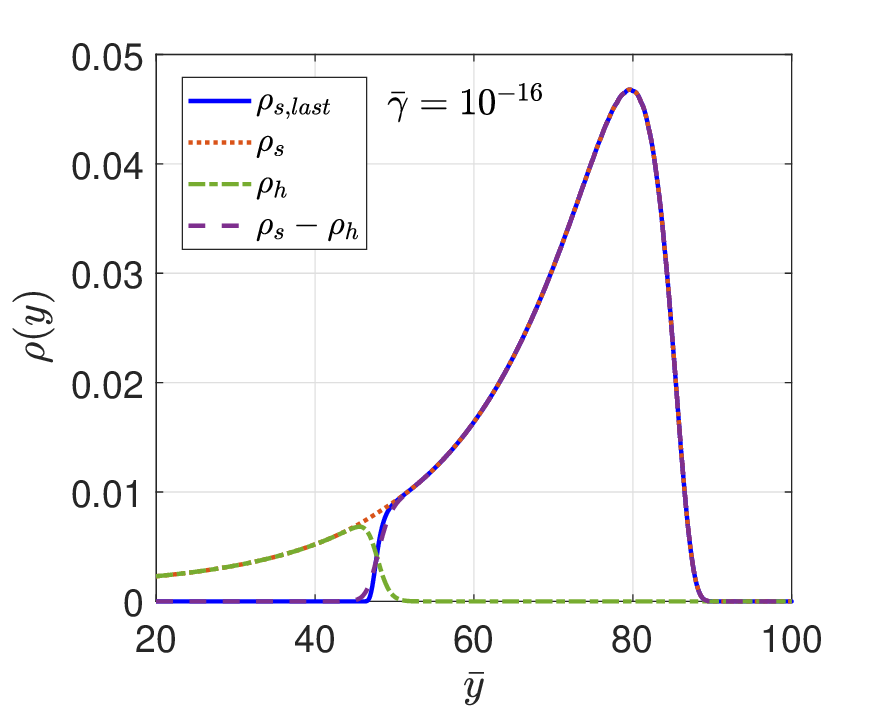}
        \caption*{(c)}
    \end{subfigure}
    \hfill
    \begin{subfigure}[t]{0.45\textwidth}
        \centering
        \includegraphics[width=\textwidth]{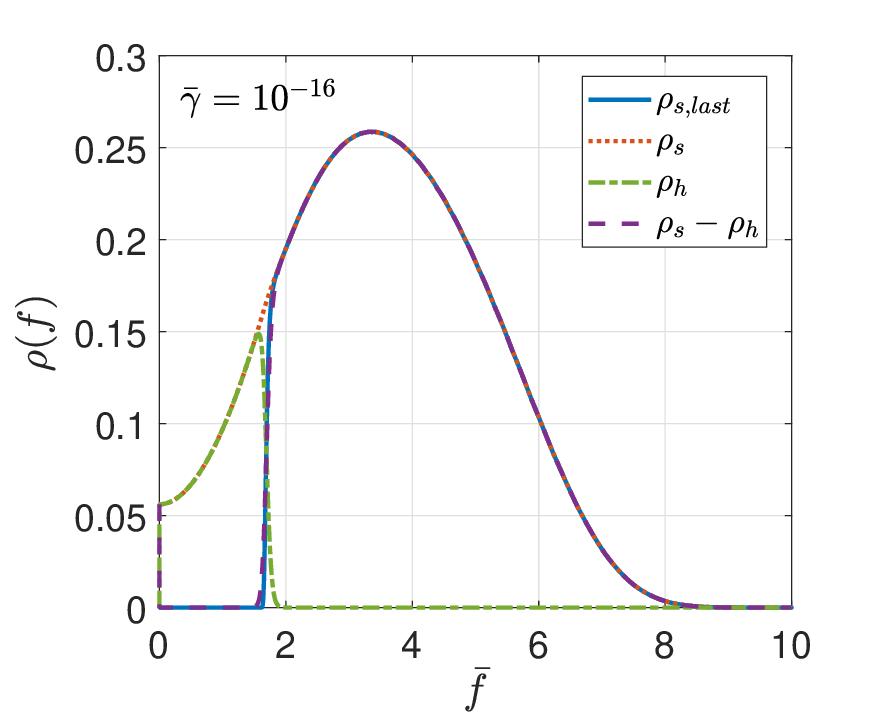}
        \caption*{(d)}
    \end{subfigure}

    \vspace{0.1cm}

    \begin{subfigure}[t]{0.45\textwidth}
        \centering
        \includegraphics[width=\textwidth]{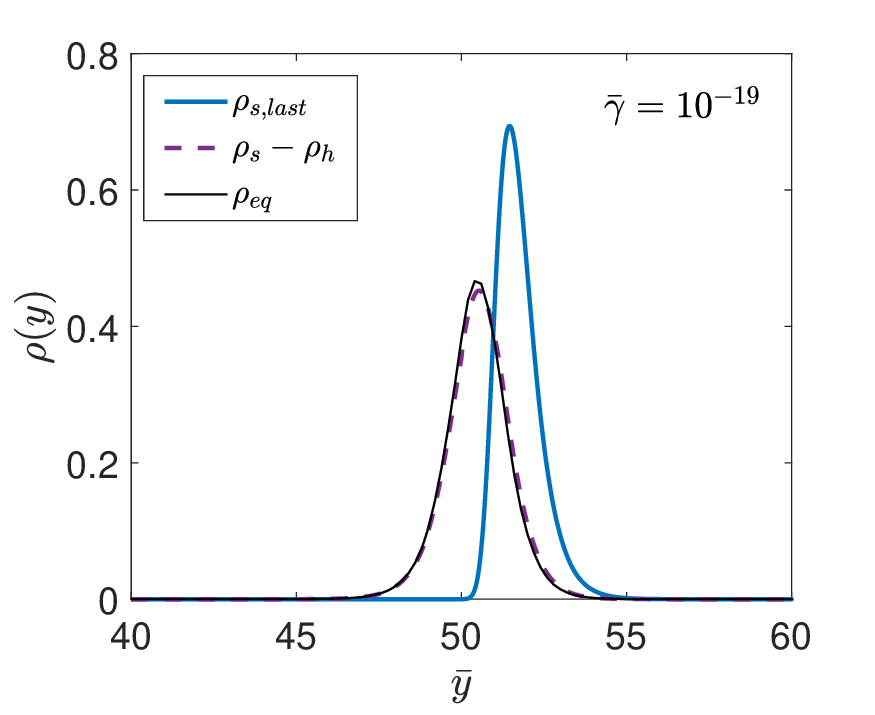}
        \caption*{(e)}
    \end{subfigure}
    \hfill
    \begin{subfigure}[t]{0.45\textwidth}
        \centering
        \includegraphics[width=\textwidth]{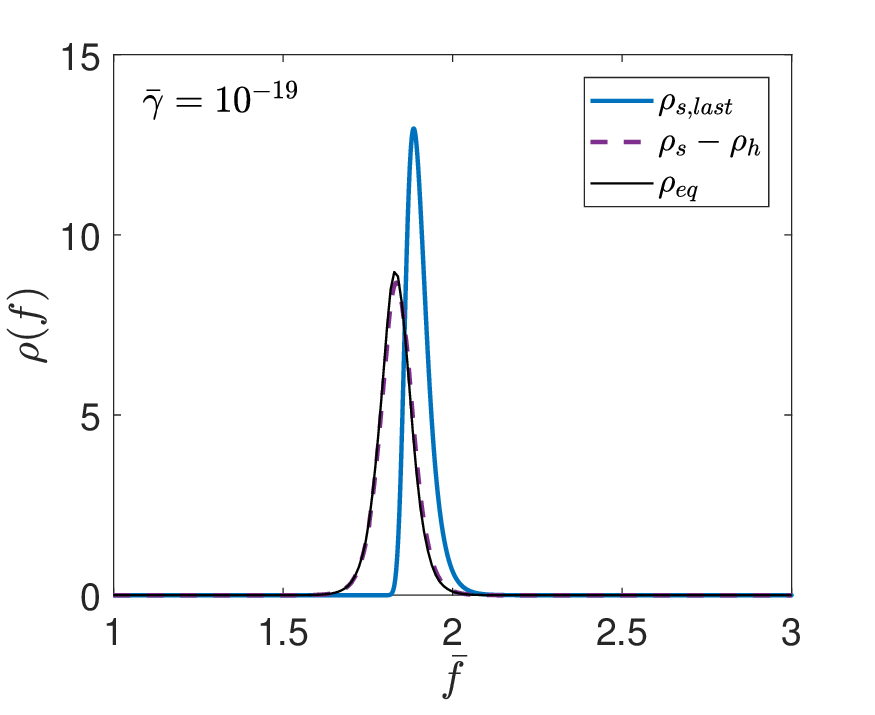}
        \caption*{(f)}
    \end{subfigure}

    \caption{Probability density functions for the last chain scission event of a polymer chain ($n=100, \beta \epsilon_0=50$) subject to different pulling rates. (a,b) $\bar{\gamma}=10^{-9}$ (a high rate), (c,d) $\bar{\gamma}=10^{-16}$ (a low rate), and (e,f) $\bar{\gamma}=10^{-19}$ (an extremely low rate).}
    \label{rho_comp}
\end{figure}

Both $\rho_{s,last}(y)$ and $\rho_{s,last}(f)$ are rate dependent. Figures \ref{rho_comp} compare $\rho_{s,last}$ to $\rho=\rho_s-\rho_h$.
For a relatively high pulling rate ($\bar{\gamma} > 10^{-15}$), the effect of healing is negligible. As a result, both $\rho_{s,last}$ and $\rho$ are approximately equal to $\rho_s$, as shown in Figures \ref{rho_comp}(a,b) for $\bar{\gamma} = 10^{-9}$. 
In this case, healing is unlikely after scission, so that the probability density functions, $\rho(y)$ and $\rho(f)$ defined in Eqs. \eqref{eq:rhoy} and \eqref{eq:rhof} are nearly identical to the probability density functions for the final scission event, $\rho_{s,last}(y)$ and $\rho_{s,last}(f)$.

Subject to a low pulling rate, the effect of healing becomes noticeable. As shown in Figure \ref{rho_comp}(c) for $\bar{\gamma} = 10^{-16}$, when $\bar{y} < 45$, $\rho_h \approx \rho_s$ so that both $\rho_{s,last}$ and $\rho = \rho_s-\rho_h$ are close to zero.
When $\bar{y} > 50$, $\rho_h \ll \rho_s$ so that both $\rho_{s,last}$ and $\rho$ are approximately equal to $\rho_s$. When $45<\bar{y}<50$, the probability density for healing $\rho_h$ decreases, while $\rho_s$ continues increasing. As a result, both $\rho_{s,last}$ and $\rho$ undergo a transition from nearly zero to $0.01$. Correspondingly, Figure \ref{rho_comp}(d) shows a similar but even sharper transition for the probability density function $\rho_{s,last}(f)$ in terms of the force,  between $\bar{f} = 1.5$ to $\bar{f} = 1.8$. Interestingly, the combined probability density $\rho = \rho_s-\rho_h$ differs slightly from $\rho_{s,last}$ in the transition region only.

At an extremely low pulling rate, both $\rho_s$ and $\rho_h$ are large, and the combined probability density $\rho = \rho_s-\rho_h$ approaches the equilibrium probability density $\rho_{eq}$ in Figures \ref{eq_probs}(c,d). In contrast, the probability density for the final scission event, $\rho_{s,last}$, differs from $\rho_{eq}$, as shown in Figures \ref{rho_comp}(e,f) for $\bar{\gamma} = 10^{-19}$. As the pulling rate decreases further, the peak of $\rho_{s,last}$ shifts right, opposite to the shift at higher pulling rates. At such extremely slow pulling rates, the probability of late healing and thus a late scission event increase as the rate decreases. 

The probability density $\rho_{s,last}$ may be compared to the histograms of the critical displacements and rupture forces for the final chain scission events in single-chain experiments. In particular, one may be interested in the average end-to-end distance ($\left< y_s \right>$) and the average rupture force ($\left< f_s \right>$),
which can be calculated as
\begin{equation}
    \left< y_s \right> =\int_0^\infty y\rho_{s,last}(y)dy ,
\end{equation}
\begin{equation}
    \left< f_s \right> =\int_0^\infty f\rho_{s,last}(f)df .
\end{equation}

\begin{figure}[h!]
    \centering

    \begin{subfigure}[t]{0.45\textwidth}
        \centering
        \includegraphics[width=\textwidth]{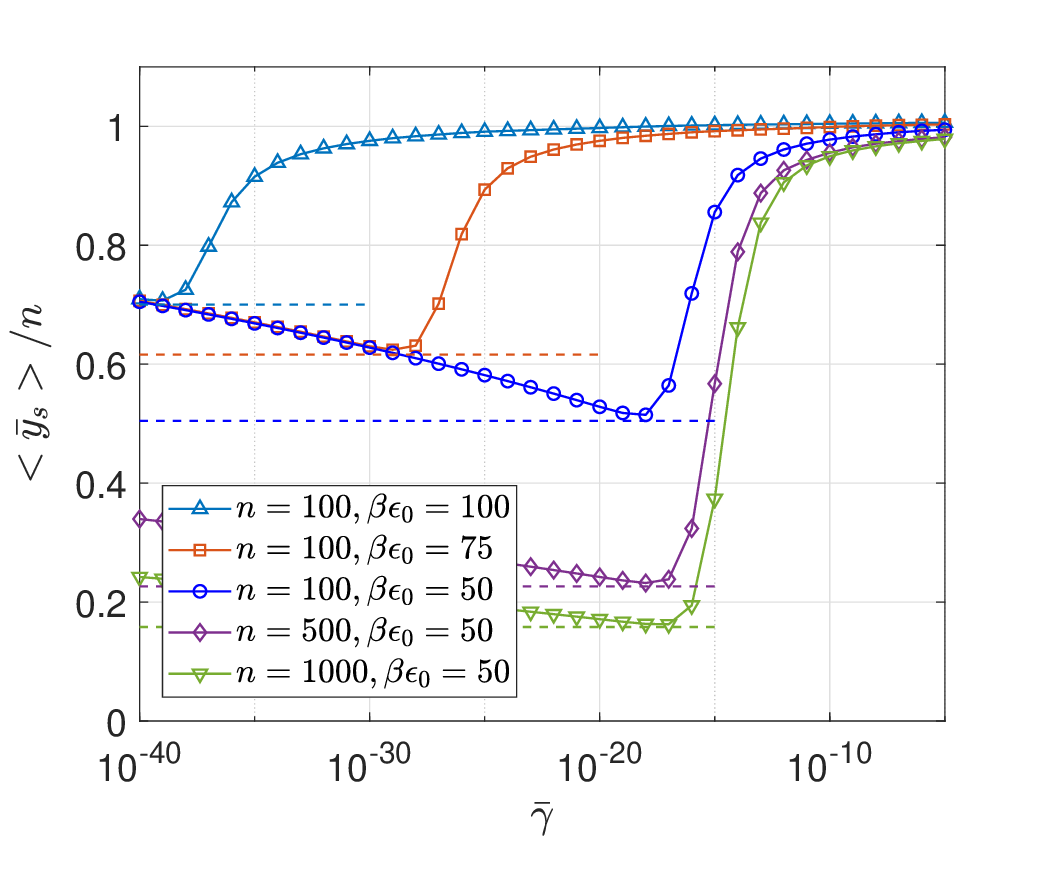}
        \caption*{(a)}
    \end{subfigure}
    \hfill
    \begin{subfigure}[t]{0.45\textwidth}
        \centering
        \includegraphics[width=\textwidth]{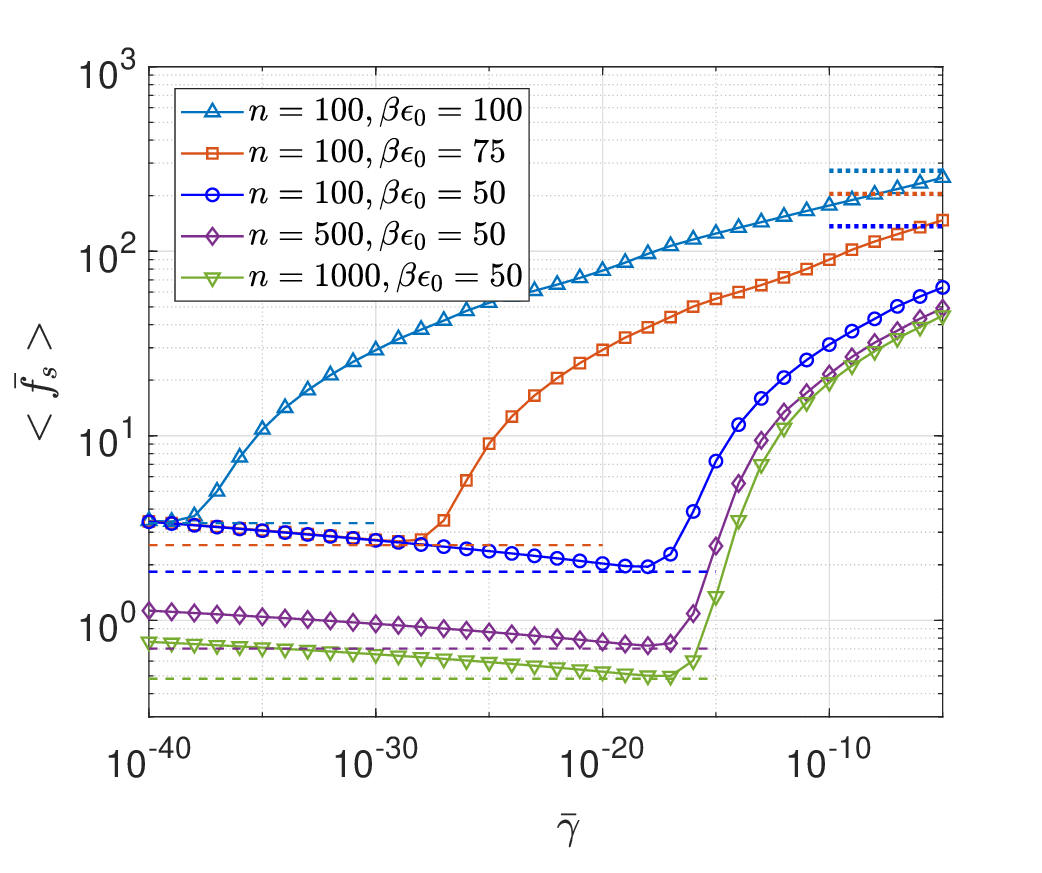}
        \caption*{(b)}
    \end{subfigure}

    \caption{Rate dependent chain scission: (a) Average end-to-end distance $\left< \bar{y}_s \right>/n$ and (b) average rupture force $\left< \bar{f}_s \right>$, at the final chain scission events for various values of $n$ and $\beta \epsilon_0$. For each chain, the lower bounds for the end-to-end distance and the rupture force are shown as dashed lines. The upper bound for the rupture force is shown as a dotted line, and it is independent of the chain length. }
    \label{mean_values}
\end{figure}

Figures \ref{mean_values} (a,b) show the average end-to-end distance and the average rupture force as functions of the pulling rate for various values of chain length and bond energy. For given $n$ and $\beta\epsilon_0$, the average rupture force increases with the pulling rate except for extremely low rates. The minimum is close to that set by the equilibrium probability density function $\rho_{eq}$:
\begin{equation}
    \left< y_s \right>_{\min}=\int_0^\infty y\rho_{eq}(y)dy ,
\end{equation}
\begin{equation}
    \left< f_s \right>_{\min}=\int_0^\infty f\rho_{eq}(f)df .
\end{equation}
As noted above, $\rho_{s,last} \approx \rho_s-\rho_h$ except for extremely low pulling rates. The combined probability density function, $\rho = \rho_s-\rho_h$, approaches the equilibrium limit $\rho_{eq}$ as the pulling rate decreases ($\bar{\gamma}\rightarrow 0$). 
In contrast, $\rho_{s,last}$ does not have an equilibrium limit, but is bounded by the minimum. 

At the limit of extremely high pulling rate, it is expected that chain scission occurs at the upper bound set by the zero energy barrier, $E_s(y)=0$, corresponding to the critical point in Figure \ref{force_compare}(b). However, up to a very high pulling rate ($\bar{\gamma}=10^{-3}$), the normalized average rupture force $\left< \bar{f}_s \right>$ remains considerably below the upper bound, while $\left< \bar{y}_s \right>/n$ is close to $1$. The upper bound for the rupture force is approximately equal to the strength of each link (covalent bonds), $\bar{f}_{max} \approx 2.69 \beta\epsilon_0$ by the LJ potential in the present model. However,
to reach this theoretical limit for the rupture force, an unrealistically high pulling rate would be required.
Therefore, the strength of a polymer chain is rate dependent and could be considerably lower than the strength of the covalent bonds in the backbone of the chain.

The effects of chain length and bond energy on the rate-dependent chain scission are shown in Figure \ref{mean_values}. As the chain length increases, the average end-to-end distance and the average rupture force for the final chain scission decreases, and the rate dependence becomes more significant in the range $10^{-15} < \bar{\gamma} < 10^{-9}$. The lower bound for the average rupture force decreases with increasing chain length, but the upper bound is independent of the chain length. Both the lower bound and the upper bound for the average rupture force increase with the bond energy. In all cases, 
the lower bound for the average rupture force is $2-3$ orders of magnitude lower than the upper bound.

\section{Kinetic Monte Carlo simulations}
As an complementary  approach, we conduct kinetic Monte Carlo (KMC) simulations \cite{Hansma2001} to  provide a direct visualization of single-chain experiments.  The same energy barriers obtained by the breakable chain model are used in the KMC simulations to examine the kinetics and statistics of chain scission and healing events in single-chain experiments. The state of a polymer chain is either intact (healed) or broken, described by a time-dependent state variable $S(t)$ taking the value of 1 or 0, respectively. Start with an intact chain ($S = 1$) at $t=0$. Hold the end-to-end distance of the chain $\bar{y}$ constant. 
For each time increment $\Delta t$, a random number $\xi$ is generated from the standard uniform distribution in the interval $(0,1)$, which is used to determine the state of the chain at $t+\Delta t$. If $S(t) = 1$, the random number $\xi$ is compared to the probability of chain scission, $P_s=(n+1)\nu_s\Delta t \exp(-\beta E_s)$, where $E_s$ is the energy barrier for chain scission at the prescribed end-to-end distance $\bar{y}$. If $\xi < P_s$, the chain breaks and the state of the chain is set to 0 at $t+\Delta t$; otherwise the chain remains intact. Similarly, if $S(t) = 0$, we compare the random number $\xi$ to the probability of healing, $P_h = \nu_h\Delta t\exp(-\beta E_h)$, where $E_h$ is the energy barrier for healing at $\bar{y}$. If $\xi < P_h$, the chain heals and the state of the chain is set to 1 at $t+\Delta t$; otherwise the chain remains broken. Repeat the process for each time step until a sufficiently long time $t_f$ to reach equilibrium ($t_f \gg \tau$). For the KMC simulation to converge, the time step $\Delta t$ must be sufficiently small. When $S(t)=1$, we take $\nu_s \Delta t = \exp{(\beta E_s)}/(100(n+1))$. When $S(t)=0$, we take $\nu_h \Delta t = \exp{(\beta E_h)}/100$.

\begin{figure}[h!]
  \centering

  \begin{subfigure}{0.8\textwidth}
    \centering
    \includegraphics[width=\linewidth]{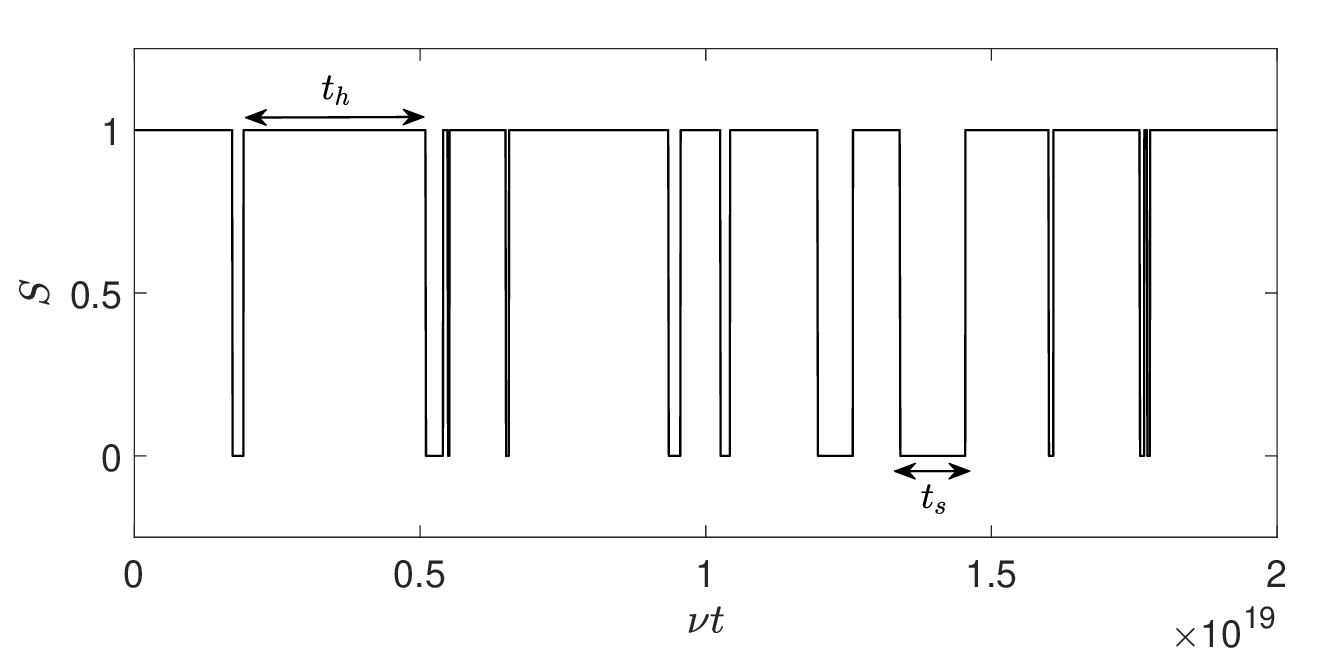}
    \caption{}
  \end{subfigure}\hfill

  \vspace{0.1cm}

  \begin{subfigure}{0.32\textwidth}
    \centering
    \includegraphics[width=\linewidth]{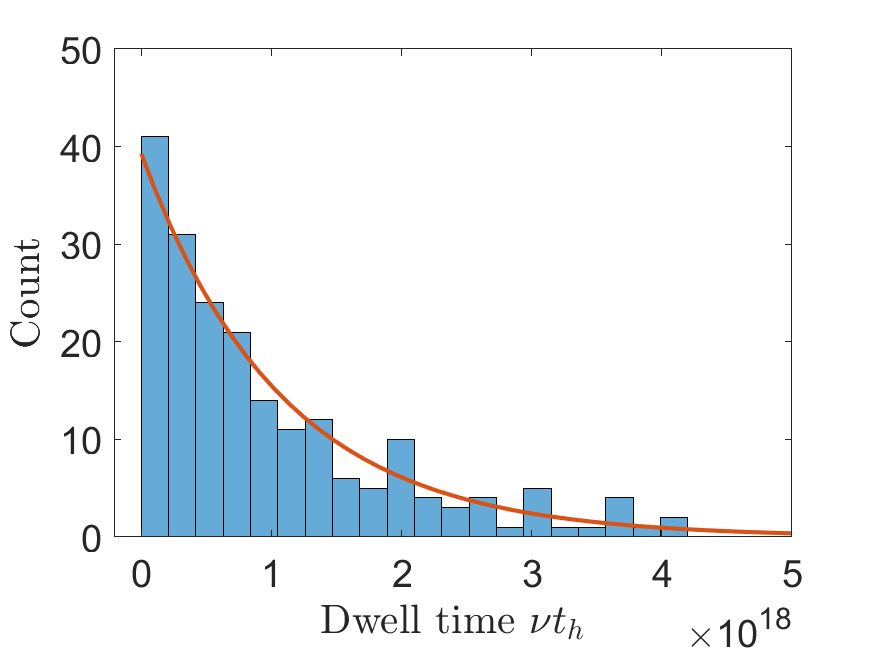}
    \caption{}
  \end{subfigure}\hfill
  \begin{subfigure}{0.32\textwidth}
    \centering
    \includegraphics[width=\linewidth]{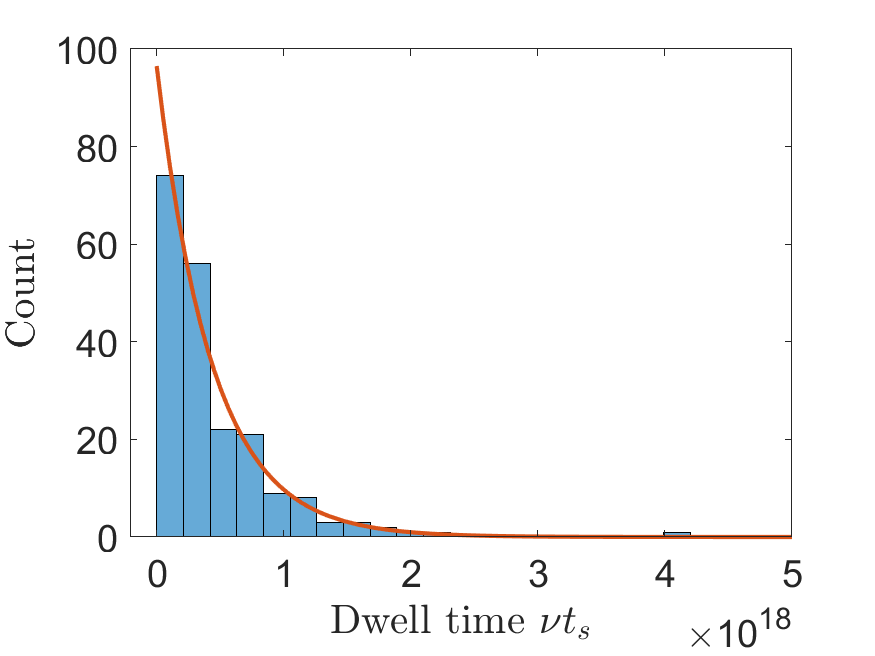}
    \caption{}
  \end{subfigure}
    \begin{subfigure}{0.32\textwidth}
    \centering
    \includegraphics[width=\linewidth]{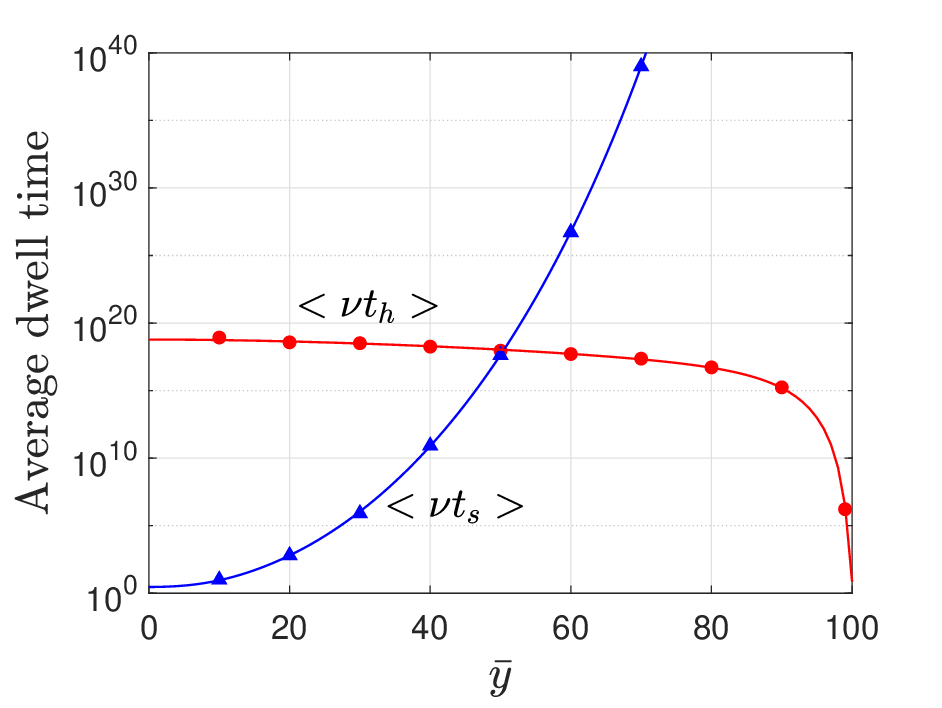}
    \caption{}
  \end{subfigure}

  \caption{(a) The state of a polymer chain with $n=100$ and $\beta\epsilon_0=50$ when the end-to-end distance is held at $\bar{y} = 50$, obtained by a kinetic Monte Carlo (KMC) simulation. (b,c) Histograms of the dwell times, $t_h$ and $t_s$, from the KMC simulation, in comparison with the exponential distributions (solid lines). (d) The average dwell times versus $\bar{y}$ from KMC simulations (symbols) and from the expected exponential distributions (lines). }
  \label{fig:KMC1}
\end{figure}

Each KMC simulation describes one possible scenario in a single experiment. Figure \ref{fig:KMC1}(a) shows the state variable $S(t)$ obtained from one KMC simulation for a chain with $n=100$ and $\beta\epsilon_0 = 50$ when the end-to-end distance is held constant, $\bar{y} = 50$. The statistics of chain scission and healing can be shown as histograms of the dwell time in each state, $t_h$ for $S=1$ and $t_s$ for $S=0$ in Figures \ref{fig:KMC1} (b) and (c), respectively. After sufficiently long time, the histograms of both dwell times approach the exponential distributions, and the average dwell times approach the expected values, $\left< \nu t_h \right> = \exp{(\beta E_s)}/(n+1)$ and $\left< \nu t_s \right> =\exp{(\beta E_h)}$. 
The equilibrium probability for the chain to be intact is then obtained as $P_{eq} = \left< t_h \right> /(\left< t_h \right> + \left< t_s \right>$), as predicted by Eq. \eqref{eq:Peq}. Therefore, the energy barriers for both chain scission and healing can be determined by measuring the average dwell times in one such experiment for a prescribed $\bar{y}$. Repeating the experiments at different $\bar{y}$ could determine the energy barriers as functions of $\bar{y}$. As shown in
Figure \ref{fig:KMC1}(d), the average dwell times obtained from the KMC simulations compare closely with the expected values by the exponential distributions.
A similar approach has been used to measure the rates of protein folding and unfolding by optical tweezers based single-molecule force spectroscopy \cite{Makarov2016}. Both the force-induced protein folding/unfolding and reversible chain scission/healing are examples of conformational transitions by single-molecule mechanochemstry \cite{Ribas-Arino2012, Bao2020}.

Note that, when the end-to-end distance of a chain is held constant, the rates of chain scission ($k_s = (n+1) \nu_s \exp (-\beta E_s)$) and healing ($k_h = \nu_h \exp (-\beta E_h)$) are time-independent. In this case, a more efficient algorithm can be used to simulate the single-chain experiments without requiring discretization of time in short intervals \cite{Elber2020}. However, such time discretization is necessary when the rates become time-dependent, for example, when the end-to-end distance of the chain is a function of time, $y(t)$. 

\begin{figure}[h!]
    \centering

    \begin{subfigure}[t]{0.45\textwidth}
        \centering
        \includegraphics[width=\textwidth]{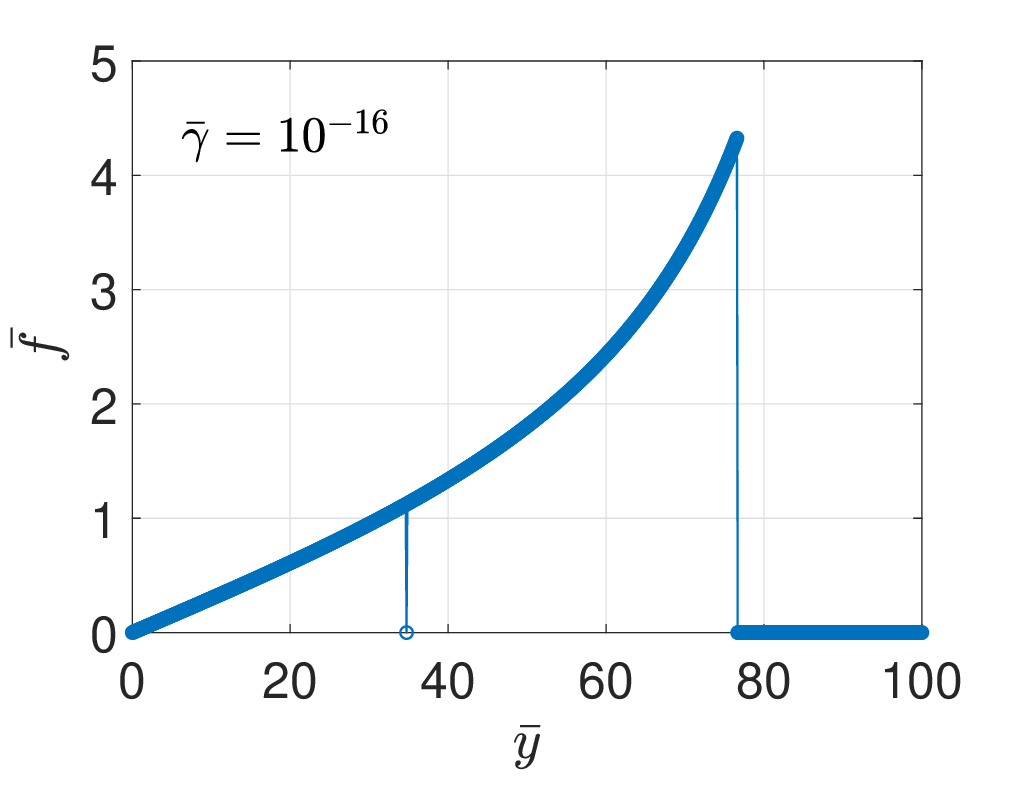}
        \caption*{(a)}
    \end{subfigure}
    \hfill
    \begin{subfigure}[t]{0.45\textwidth}
        \centering
        \includegraphics[width=\textwidth]{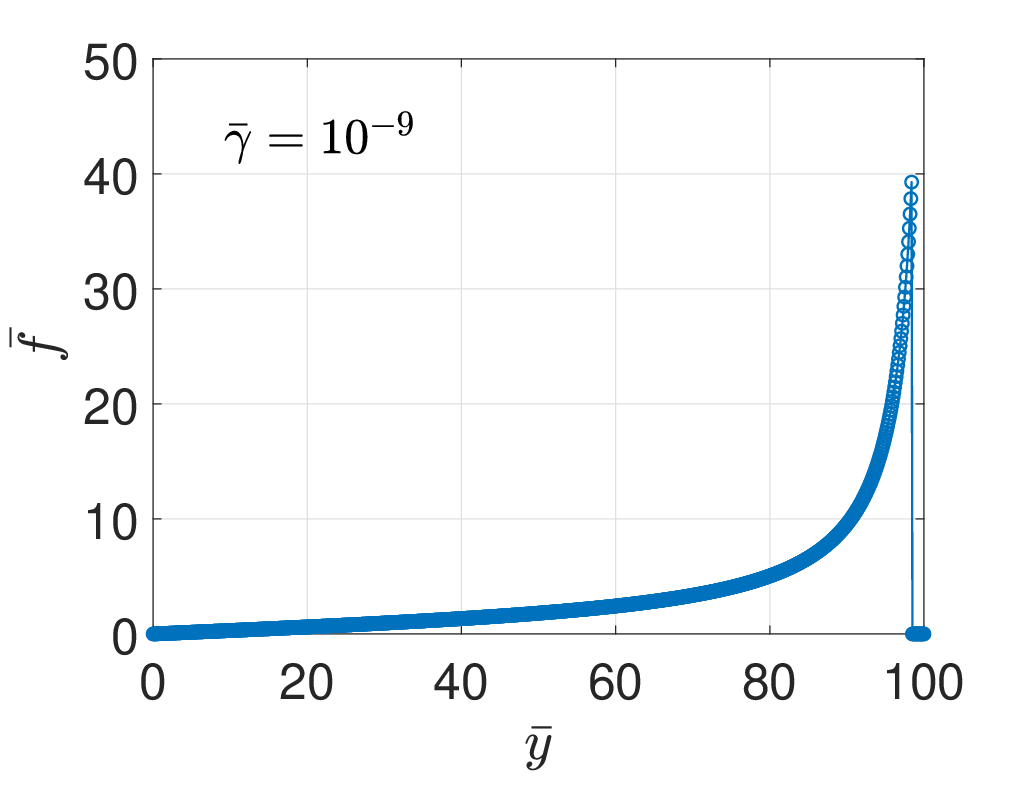}
        \caption*{(b)}
    \end{subfigure}

    \caption{Normalized force–displacement curves for a polymer chain with $n=100$ and $\beta\epsilon_0=50$, obtained from KMC simulations for two pulling rates, (a) $\bar{\gamma}=10^{-16}$ and (b) $\bar{\gamma}=10^{-9}$.}
    \label{force_MC}
\end{figure}

Next, we conduct similar KMC simulations of pulling a polymer chain at a constant rate, $y(t)=\gamma t$. 
Start with an intact chain, $S = 1$ at $t = 0$.
As the end-to-end distance changes, the energy barriers change. 
For each small time increment from $t$ to $t+\Delta t$, the energy barriers at $y(t)$ are used to calculate the probability of chain scission or healing, and then the state of the chain at $t+\Delta t$ is determined.
When $S(t) = 1$, the corresponding force $f(t)$ can be calculated approximately by Eq. \eqref{eq:length force} with $r = y(t)$.
When $S(t)=0$, the force is set to be zero. Therefore, each KMC simulation produces one possible force-displacement curve in a single experiment, which follows the same curve in Figure \ref{force_compare}(b) except for the points where the chain is broken. 
Figure \ref{force_MC} shows two force-displacement curves, obtained from the KMC simulations at two pulling rates. At a low rate, $\bar{\gamma}=10^{-16}$, the final chain scission occurs at a relatively low force, $\bar{f}_s \approx 4.2$, after which the chain remains in the broken state. At a higher rate, $\bar{\gamma} = 10^{-9}$, the chain stiffens significantly before chain scission, resulting in a much higher rupture force, $\bar{f}_s \approx 40$, but still well below the upper bound set by zero energy barrier in Figure \ref{force_compare}(b). Interestingly, at the low rate, chain scission may occur more than once at a very low force, followed immediately by healing. The same scission/healing event is less likely at the higher rate.

\begin{figure}[h!]
    \centering

        \begin{subfigure}[t]{0.45\textwidth}
        \centering
        \includegraphics[width=\textwidth]{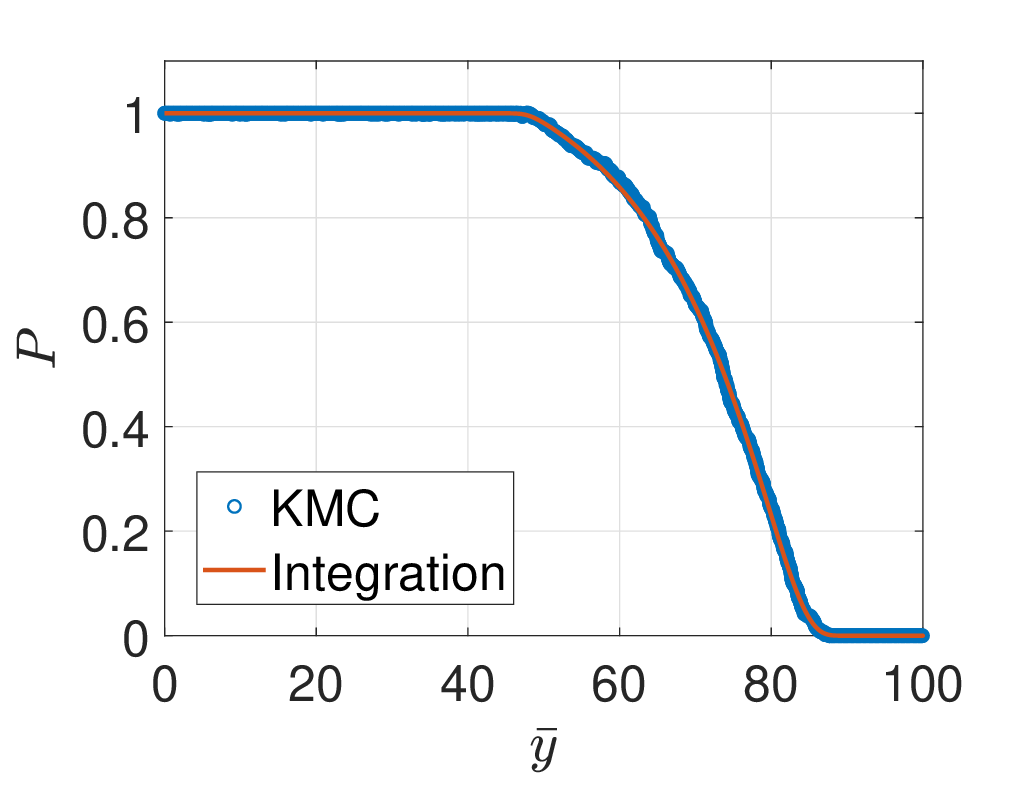}
        \caption*{(a)}
    \end{subfigure}
    \hfill
    \begin{subfigure}[t]{0.45\textwidth}
        \centering
        \includegraphics[width=\textwidth]{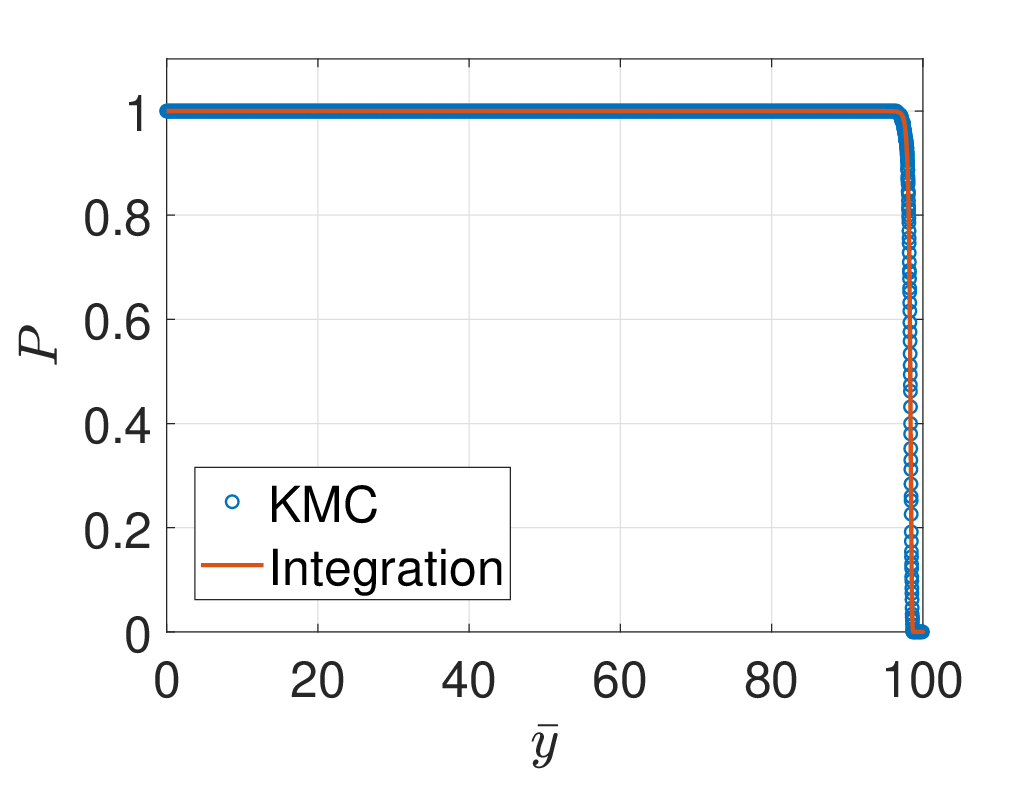}
        \caption*{(b)}
    \end{subfigure}
    
    \caption{Probability for a polymer chain ($n=100$ and $\beta\epsilon_0 = 50$) to be intact when pulled at a constant pulling rate, (a) $\bar{\gamma}=10^{-16}$ and (b) $\bar{\gamma} = 10^{-9}$, obtained from 500 KMC simulations (symbols) and from integration of the rate equation \eqref{eq:rate1} (lines).}
    \label{rate_effect_single_chain}
\end{figure}

\begin{figure}[htp]
    \centering
    \begin{subfigure}[b]{0.48\textwidth}
        \centering
        \includegraphics[width=\textwidth]{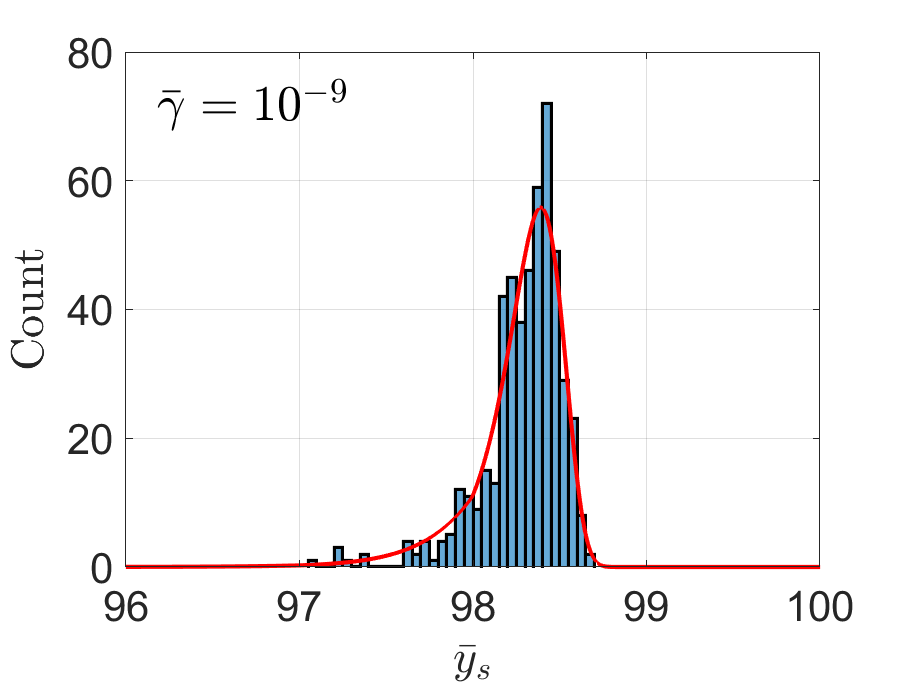}
        \caption{}
    \end{subfigure}
    \hfill
    \begin{subfigure}[b]{0.48\textwidth}
        \centering
        \includegraphics[width=\textwidth]{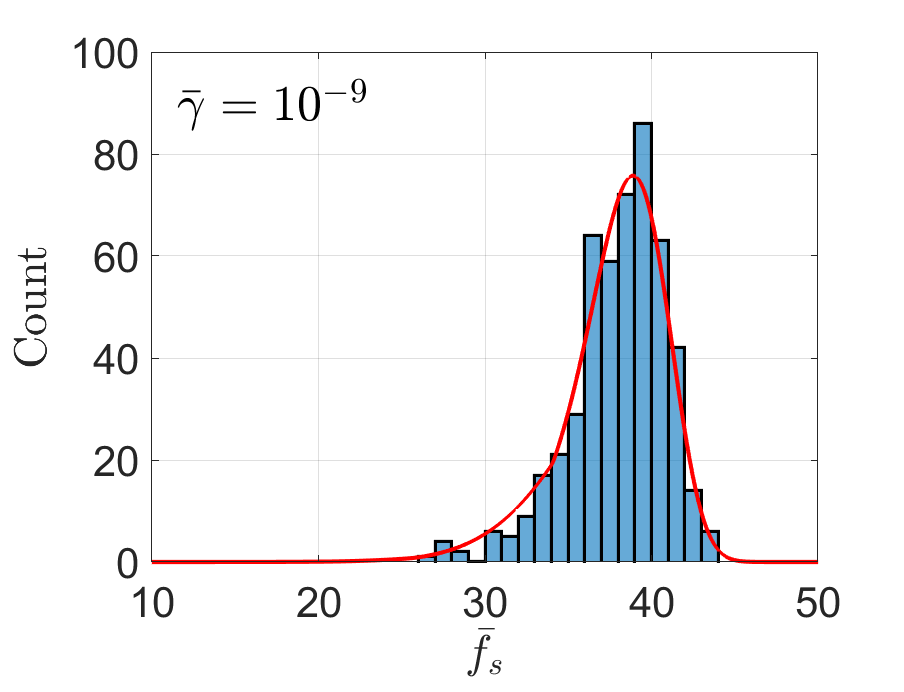}
        \caption{}
    \end{subfigure}

    \vspace{0.1cm} 

        \begin{subfigure}[b]{0.48\textwidth}
        \centering
        \includegraphics[width=\textwidth]{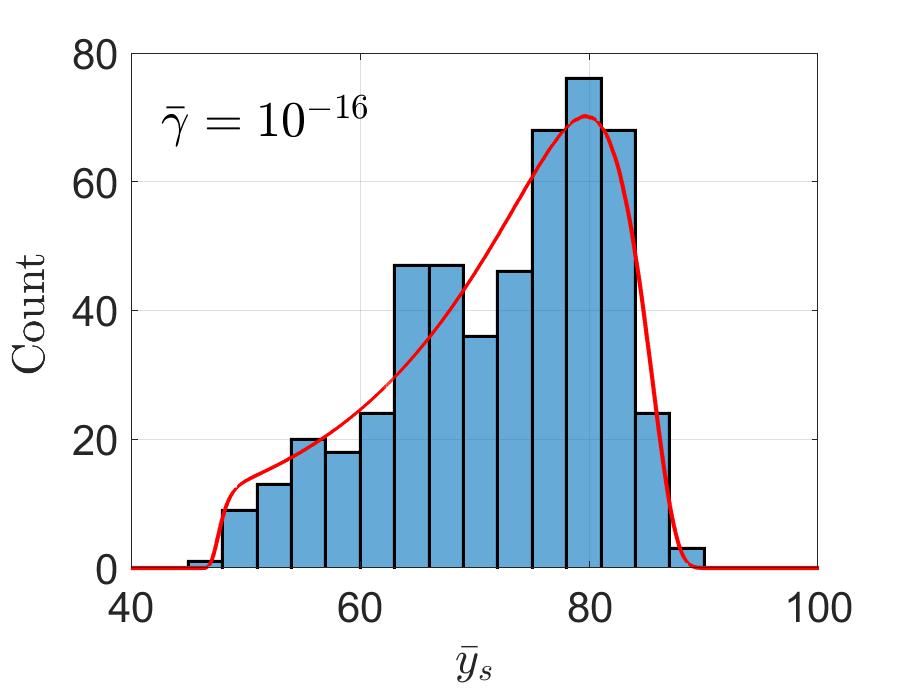}
        \caption{}
    \end{subfigure}
    \hfill
    \begin{subfigure}[b]{0.48\textwidth}
        \centering
        \includegraphics[width=\textwidth]{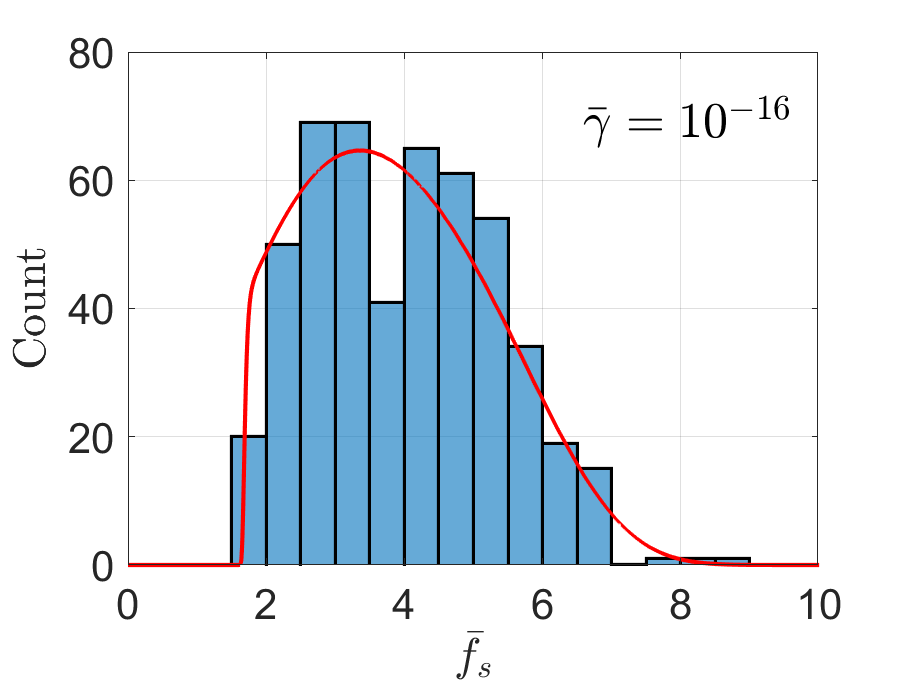}
        \caption{}
    \end{subfigure}

    \caption{Histograms of the normalized end-to-end distance $\bar{y}$ and force $\bar{f}$ for the last chain scission from 500 KMC simulations for each pulling rate, (a, b) $\bar{\gamma}=10^{-9}$ and (c, d) $\bar{\gamma}=10^{-16}$. The red line in each figure is obtained by multiplying the probability density $\rho_{s,last}(y)$ or $\rho_{s,last}(f)$ in Figure \ref{rho_comp} by $500 \Delta\bar{y}$ or $500\Delta\bar{f}$, where $\Delta\bar{y}$ or $\Delta\bar{f}$ is the bar width of the histogram.}
    \label{fig:mc_histograms}
\end{figure}

For each pulling rate, the strength of the chain $f_s$ at which the final chain scission occurs varies from simulation to simulation. 
By conducting a large number of KMC simulations, we may determine the statistics of the final chain scission and compare to that predicted by the rate equation \eqref{eq:rate1}. 
Figure \ref{rate_effect_single_chain} shows the probability for the chain to be intact obtained from 500 KMC simulations for each of the two pulling rates, $\bar{\gamma}=10^{-16}$ and $10^{-9}$, both in close agreement with those obtained by integrating Eq. \eqref{eq:rate1}. Moreover, 
Figure \ref{fig:mc_histograms} shows the histograms for the values of $\bar{y_s}$ and $\bar{f_s}$ corresponding to the final chain scission in the KMC simulations. In cases when there are multiple scission/healing events in one
simulation (possible when the pulling rate is low), only the final scission event is recorded for the
histograms. These histograms compare closely with the probability density for the final chain scission in Figure \ref{rho_comp} (a-d), as predicted by the rate equation \eqref{eq:rate1}. The KMC simulations confirm that the strength of a polymer chain in terms of either the rupture force or the end-to-end distance is stochastic and rate dependent.
In particular, at the relatively high rate ($\bar{\gamma} = 10^{-9}$), while the distribution of $\bar{y}_s$ is fairly narrow, the distribution of $\bar{f}_s$ is wider, and the average strength in terms of the rupture force is well below the upper bound set by zero energy barrier for chain scission in Figure \ref{force_compare}(b).

\section{Summary}
In summary, we present a breakable polymer chain model to calculate the free energy barriers for chain scission and healing. With the assumption of a freely jointed chain and the Lennard-Jones potential for the breakable link, the free energy of the polymer chain is obtained as a function of the prescribed end-to-end distance and an internal variable corresponding to the length of the breakable link. At a given end-to-end distance, the free energy has two local minima, corresponding to the two equilibrium states, one for the intact state and the other for a broken chain. Consequently, the energy barriers for both chain scission and healing are determined as functions of the end-to-end distance, depending on the properties of the polymer chain characterized by two parameters, the bond energy for each link and the number of links in the chain. We note the following features for the energy barriers predicted by the present model.

\begin{itemize}
    \item The energy barrier for chain scission starts at a level close to the bond energy when the end-to-end distance is zero. As the end-to-end distance increases, the energy barrier for chain scission decreases, approximately quadratically until the end-to-end distance is close to the contour length of the chain.
    \item The energy barrier for chain scission becomes zero at a critical end-to-end distance, which sets the upper bound for the distance and correspondingly the critical force for instantaneous chain scission. This critical force is close to the strength of the breakable link, which is proportional to the bond energy.
    \item The energy barrier for healing or re-association of a broken chain starts at a low level comparable to $k_BT$ when the end-to-end distance is zero. As the end-to-end distance increases, the energy barrier for healing increases, approximately quadratically as well.
    The energy barrier for healing depends on the chain length but is insensitive to the bond energy of individual links in the chain.
\end{itemize}

With the energy barriers, we examine the statistics and kinetics of a single polymer chain under tension, first by integrating the rate equation based on the first-order kinetics and then by kinetic Monte Carlo (KMC) simulations. The two approaches are equivalent, predicting the same equilibrium probability for a chain to be intact at a prescribed end-to-end distance. The characteristic time scale for the chain to reach equilibrium depends on both energy barriers. For each KMC simulation at a prescribed end-to-end distance, the dwell time in each state (intact or broken) follows an exponential distribution, with the average dwell time depending on one of the energy barriers. Therefore, it is possible to determine both energy barriers at a given end-to-end distance from one such experiment. 

When pulling a polymer chain at a constant rate under displacement control, the probability density for chain scission is rate dependent in general. Each KMC simulation produces a force-displacement curve, with the point of the final chain scission following the rate-dependent statistical distribution. Therefore, the strength of a polymer chain is statistical and rate dependent in nature. Notably, even at a practically high pulling rate, the average strength of a polymer chain in terms of the rupture force is well below the upper bound set by zero energy barrier for scission or the bond strength of each link. At extremely low pulling rates, the average rupture force has a lower bound that is $2-3$ orders of magnitude lower than the upper bound.

While this work has focused on a single polymer chain under tension, the breakable freely jointed chain model may be incorporated in a polymer network model to simulate statistically distributed chain scission near a crack tip. Such a model would be naturally rate dependent. 
In contrast to the classical Lake-Thomas model \cite{Lake_Thomas_1967}, the present model describes chain scission as a statistical and kinetic process. As a result, a distributed chain scission around a crack tip is expected, even for a perfect polymer network. Moreover, the dissipation by chain scission would be rate dependent, with a lower bound as the quasistatic limit at extremely low rates. Compared to the Lake-Thomas model, the dissipation due to chain scission would be much less for each chain, but the number of polymer strands that undergo chain scission as the crack grows in a polymer network could be much larger. By incorporating the statistics and kinetics of chain scission and healing, such a polymer network model would help elucidate the molecular origin of crack resistance in polymer networks.

\section*{Appendix: A freely jointed chain of rigid links}
To determine the free energy function $g_n(\vec{r})$ for the chain fragments, this appendix presents a statistical analysis of a freely jointed chain of rigid links under the displacement controlled condition.
To this end, we represent a chain of $n$ identical rigid links by $n$ vectors, $\vec{l}_i$ ($i = 1,...,n$). Each link has a constant length, $l = \left| \vec{l}_i \right|$. Fix one end of the chain at the origin of a coordinate, and let the other end fluctuate freely. Write the probability for the free end to be at a particular location $\vec{r}$ and within a small volume element, $d\vec{r} = dxdydz$, as $p(\vec{r})d\vec{r}$. The probability density (up to a constant factor) can be written as
\begin{equation}
p(\vec{r}) \propto \left< \delta(\Sigma \vec{l}_i-\vec{r}) \right>, 
\end{equation}
where $\left< . \right>$ represents the ensemble average of a random variable.
To evaluate the probability density, we use the representation of the Dirac delta function in Fourier space: 
\begin{equation}
\delta(\vec{x})=\frac{1}{(2\pi)^3}\int \exp(i \vec{k} \cdot\vec{x})\mathrm{{d}}\vec{k}. 
\end{equation}
The probability density becomes 
\begin{equation}
    p(\vec{r}) \propto \int \left< \exp\left(i\vec{k}\cdot (\Sigma\vec{l}_i-\vec{r})\right) \right> \mathrm{d}\vec{k} .
\end{equation}

Since all the rigid links are identical and freely jointed, they follow the same statistics, with the ensemble average
\begin{equation}
    \left< \exp(i\vec{k}\cdot\vec{l}_i) \right> =\frac{\int_0^{\pi}\exp(ikl\cos \phi)\sin\phi \mathrm{d}\phi}{\int_0^{\pi}\sin\phi \mathrm{d}\phi}=\frac{\sin kl}{kl} ,
\end{equation}
where $k = \left| \vec{k} \right|$. Hence, we obtain the probability density:
\begin{equation}
    p(\vec{r}) \propto \int_0^{\infty} \left[ k^2 \left(\frac{\sin kl}{kl}\right)^n \int_0^{\pi}\exp(-ikr \cos \phi) \sin \phi\mathrm{d}\phi \right] \mathrm{d}k ,
\end{equation}
which reduces to
\begin{equation}\label{disp_prb_eq}
    p(\vec{r}) \propto \int_0^{\infty}k^2\left(\frac{\sin kl}{kl}\right)^n \left( \frac{\sin kr}{kr} \right) \mathrm{d}k .
\end{equation}

\begin{figure}[h!]
    \centering
    \includegraphics[width=7cm]{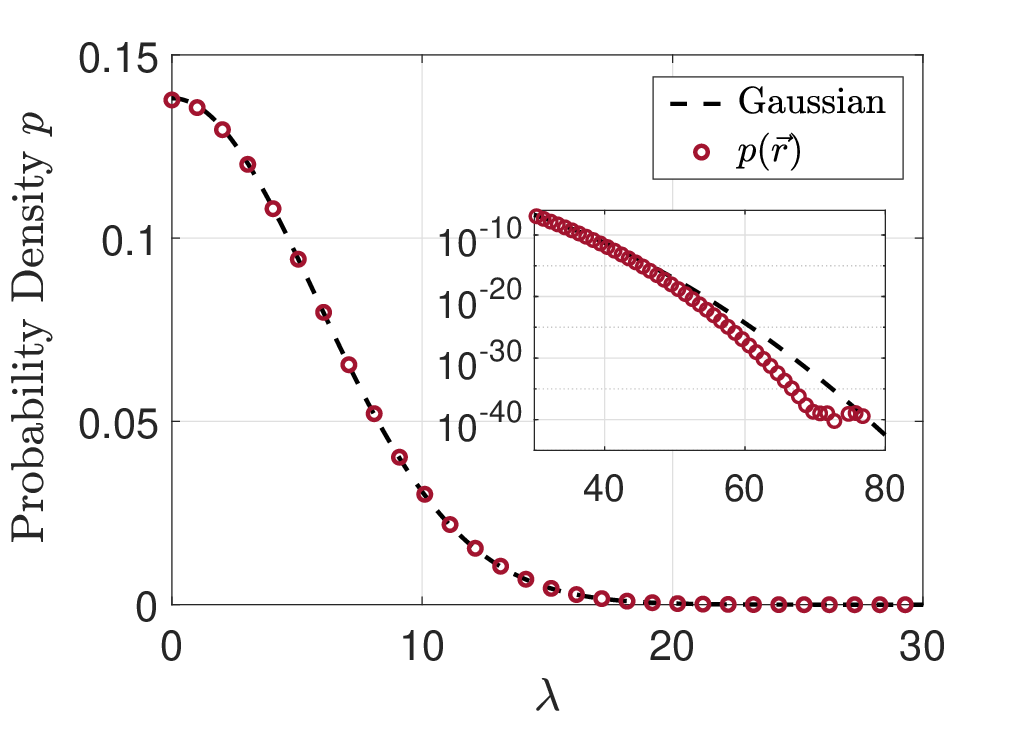}
    \caption*{(a)}
    \vspace{1pt} 
    \includegraphics[width=7cm]{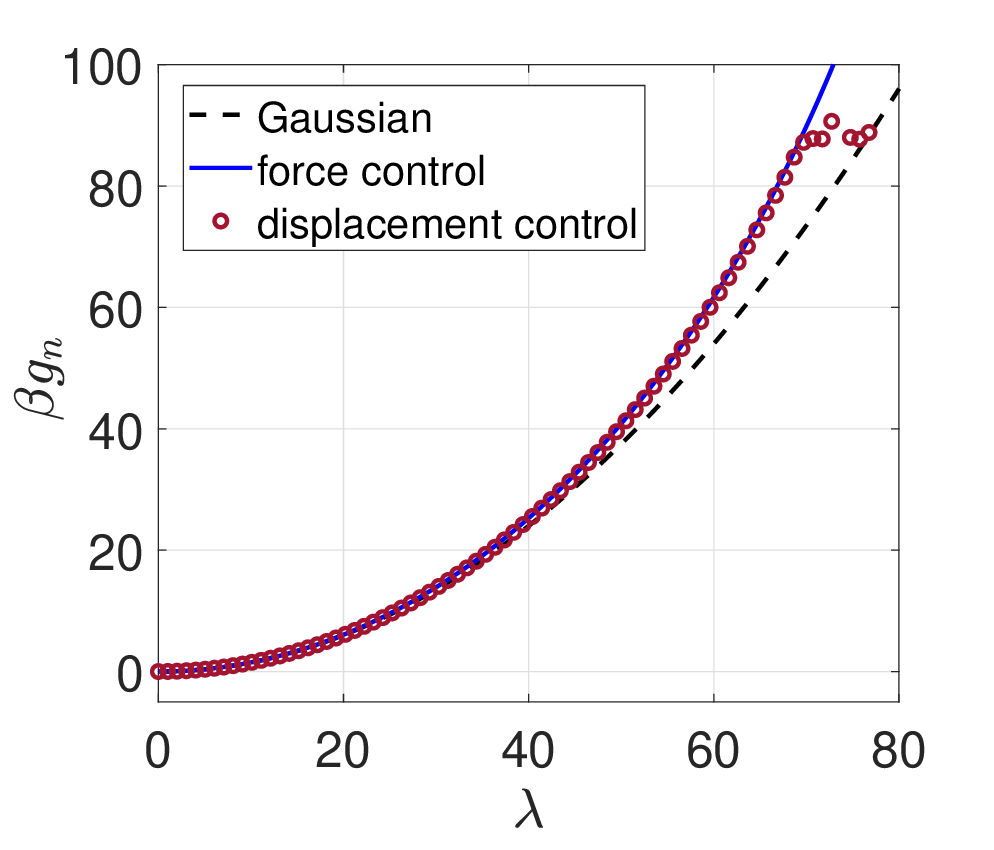}
    \caption*{(b)}
    \vspace{1pt}
    \includegraphics[width=7cm]{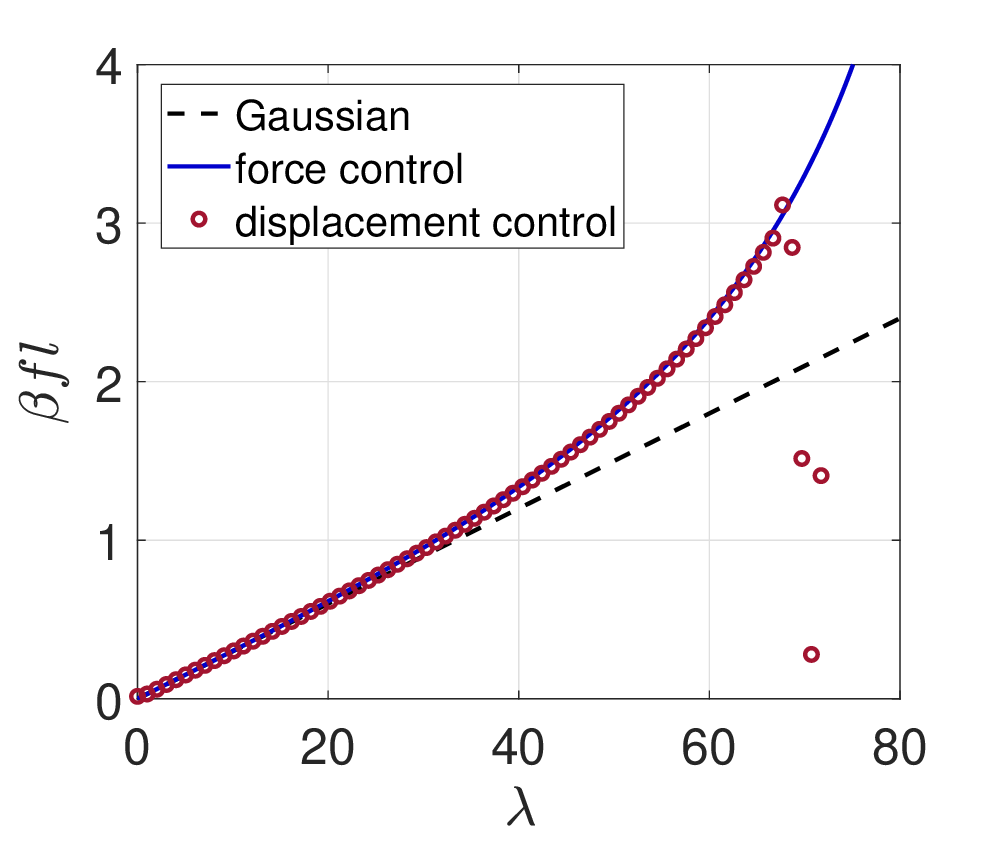}
    \caption*{(c)}

    \caption{(a) Probability density by numerically integrating Eq. \eqref{eq: disp_prb_eq_lambda} (symbols) for $n=100$, in comparison with Gaussian statistics by Eq. \eqref{eq: Gaussian} (dashed line); (b) Normalized free energy as a function of stretch, comparing Eq. \eqref{eq:gn log} to \eqref{eq:gn Gaussian} for Gaussian and \eqref{eq:free energy force} for non-Gaussian force control; (c) Normalized force-displacement diagrams, comparing Eq. \eqref{eq:force stretch} to \eqref{eq:force linear} for Gaussian and \eqref{eq:length force} for non-Gaussian force control.}
    \label{fig:probability to force}
\end{figure}

The probability density depends on two dimensionless scalars, $n$ and $\lambda = r/l$ ($r = \left| \vec{r} \right|$):
\begin{equation}\label{eq: disp_prb_eq_lambda}
    p(\vec{r}) = p(\lambda, n) \propto \int_0^{\infty} k^2\left(\frac{\sin k}{k}\right)^n 
    \left( \frac{\sin k\lambda}{k\lambda} \right)
    \mathrm{d}k .
\end{equation}
The integral on the right hand side is dimensionless and must be evaluated numerically. Figure \ref{fig:probability to force}(a) shows the calculated probability density for $n = 100$, which is very close to the Gaussian distribution (Eq. \eqref{eq: Gaussian}) except for the case of a large $\lambda$ where the probability density is extremely small (inset of Fig. \ref{fig:probability to force}a).

With the above probability density, the free energy of the chain at a prescribed end-to-end vector $\vec{r}$ can be obtained as (up to a constant)
\begin{equation}
    g_n(\Vec{r}) = g_n (\lambda) = -k_B T \log \left( p 
    (\lambda, n) \right) .
    \label{eq:gn log}
\end{equation}
The free energy depends on the end-to-end vector of the chain, $\vec{r}$, via the dimensionless scalar $\lambda$. Under the displacement control, the reaction force acting at the ends of the chain fluctuates, and the average force can be obtained as the derivative of the free energy function, namely
\begin{equation}
    \vec{f}(\Vec{r}) = \frac{\partial g_n(\vec{r})}{\partial \vec{r}},
\end{equation}
which is a vector in the direction of $\vec{r}$, and the force magnitude is
\begin{equation}
    f(r) = \frac{1}{l} \frac{\partial g_n(\lambda)}{\partial \lambda}.
    \label{eq:force stretch}
\end{equation}
Figures \ref{fig:probability to force} (b) and (c) show the calculated free energy function and the corresponding force-displacement diagram for $n = 100$. 

From Figure \ref{fig:probability to force}(a), it can be seen that the probability density $p(\lambda, n)$ is approximately Gaussian when $\lambda$ is small. 
Indeed, it can be shown that the probability density approaches Gaussian when $\lambda \ll n$ \cite{treloar1975physics, rubinstein2003polymer}. Let
\begin{equation}
    q(k)= \log \left(\frac{\sin k}{k}\right).
\end{equation}
Then, the integral in Eq. \eqref{eq: disp_prb_eq_lambda} becomes
\begin{equation}
    p(\vec{r}) \propto \int_0^{\infty} k^2 \exp{\left(nq(k)\right)} 
    \left( \frac{\sin k\lambda}{k\lambda} \right)
    \mathrm{d}k .
\end{equation}
In the limit of a large $n$, this integral is dominated by the region near the maximum of the exponent, $\exp{\left(nq(k)\right)}$ (which occurs at $k=0$). Then, expand $q(k)$ around $k=0$:
\begin{equation}
    q(k) \approx -\frac{k^2}{6} .
\end{equation}
Thus, the integral is approximately
\begin{equation}
    p(\vec{r}) \propto \frac{1}{\lambda} \int_0^{\infty} k \exp{\left(-\frac{nk^2}{6}\right)} 
    \left( \sin k\lambda \right)
    \mathrm{d}k .
\end{equation}
This integral can be calculated exactly by using the Fourier transform of a Gaussian:
\begin{equation}
    \int_{-\infty}^{\infty} \exp{\left(-a k^2 \right)} 
    \exp \left(-ik\lambda\right)
    \mathrm{d}k 
    = \sqrt{\frac{\pi}{a}} \exp \left( -\frac{\lambda^2}{4a} \right).
\end{equation}

Finally, the probability density becomes Gaussian for $\lambda \ll n$:
\begin{equation}
    p(\vec{r}) \propto \exp \left( -\frac{3\lambda^2}{2n} \right).
    \label{eq: Gaussian}
\end{equation}

The Gaussian statistics leads to a quadratic free energy function:
\begin{equation}
    g_n(\lambda) = \frac{3 k_B T}{2n} \lambda^2.
    \label{eq:gn Gaussian}
\end{equation}
Correspondingly, the force-displacement relation is linear:
\begin{equation}
    f(r) = \frac{3k_B T}{nl^2} r \quad \text{or} \quad 
    \beta fl = \frac{3\lambda}{n}  .
    \label{eq:force linear}
\end{equation}

As shown in Figure \ref{fig:probability to force}(c), the force-displacement relation calculated by Eq. \eqref{eq:force stretch} follows the linear relation in Eq. \eqref{eq:force linear} when the end-to-end distance is small compared to the fully stretched chain length ($\lambda \ll n$), but it becomes nonlinear as $\lambda$ increases. The nonlinear force-displacement relation reflects the non-Gaussian statistics at large stretch. However, the numerical calculation of the probability density by Eq. \eqref{eq: disp_prb_eq_lambda} poses a challenge: as $\lambda$ increases, the probability density $p(\lambda, n)$ becomes so small that the numerical results become inaccurate and oscillatory due to the limit of computational precision. Such a small probability density at a large stretch leads to a large free energy and a large force, which cannot be calculated accurately by the numerical integration. 
For the purpose of this study, this computational issue can be addressed by adopting the Kuhn-Grun approximation as discussed in Section 4.

The statistical distribution in form of the integral in Eq. \eqref{disp_prb_eq} was first obtained by Lord Rayleigh \cite{Rayleigh1919}. An alternative form in series expansion was obtained by Treloar \cite{Treloar1946}. 
However, calculations of both Rayleigh integral and Treloar series have been limited to short chains ($n < 20$), and the Kuhn-Grun approximation is commonly used for long chains \cite{Jernigan1969}.

\section*{Acknowledgments}
This work is partly supported by Semiconductor Research Corporation.
DEM acknowledges the US National Science Foundation for the financial support via grant CHE 2400424.


\bibliography{sn-bibliography}

\end{document}